\documentclass[aps,pre,twocolumn]{revtex4}
\usepackage{amssymb,graphicx,amsmath}
\usepackage{times}
\newcommand{\nn}{\nonumber}

\newcommand{\half}{\frac{1}{2}}
\newcommand{\ts}{\hskip0.1ex\raisebox{-1ex}[0ex][0.8ex]{\rule{0.1ex}{2.75ex}\hskip0.2ex}}

\newcommand{\fig}[2]{\includegraphics[width=#1]{./figures/#2}}
\newcommand{\Fig}[1]{\includegraphics[width=\columnwidth]{./figures/#1}}
\newcommand{\Figg}[1]{\includegraphics[width=2.065\columnwidth]{./figures/#1}}
\newlength{\bilderlength}

\newlength{\figsize}
\newcommand{\rme}{{\mathrm{e}}}
\newcommand{\rmd}{{\mathrm{d}}}

\begin{document}
\bibliographystyle{KAY}
\title{\sffamily \bfseries \Large Driven particle in a random landscape: disorder correlator, avalanche
distribution and extreme value  statistics of records}
\author{\sffamily\bfseries\normalsize Pierre Le Doussal and Kay J\"org
Wiese \vspace*{3mm}} \affiliation{ CNRS-Laboratoire de Physique
Th{\'e}orique de l'Ecole Normale Sup{\'e}rieure, 24 rue Lhomond, 75231
Paris Cedex, France. }

\date{\small\today}
\begin{abstract}
We review how the renormalized force correlator $\Delta(u)$, the function computed in the functional RG field
theory, can be measured directly in numerics and experiments on the {\it dynamics} of elastic manifolds in
presence of pinning disorder. We show how this function can be computed analytically for a particle dragged
through a 1-dimensional random-force landscape. The limit of small velocity allows to access the critical behavior at the depinning
transition. For uncorrelated forces one finds three universality classes, corresponding to
the three extreme value statistics, Gumbel, Weibull, and Fr{\'e}chet. For each class we obtain analytically the
universal function $\Delta(u)$, the corrections to the critical force, and the joint probability distribution of
avalanche sizes $s$ and waiting times $w$. We find $P(s)=P(w)$ for all three cases. All results are checked
numerically. For a Brownian force landscape, known as the ABBM model, avalanche distributions and $\Delta(u)$ can be
computed for any velocity. For 2-dimensional disorder, we perform large-scale numerical simulations to calculate
the renormalized force correlator tensor $\Delta_{ij}(\vec u)$, and to extract the anisotropic scaling exponents
$\zeta_{x}> \zeta_{y}$. We also show how the Middleton theorem is violated. Our results are relevant for the record
statistics of random sequences with linear trends, as encountered e.g.\ in some models  of global warming.  We give
the joint distribution of the time $s$ between two successive records and their difference in value $w$.
\end{abstract}
\maketitle

\section{Introduction}\label{s:Intro}

\begin{figure}[b]
{\setlength{\unitlength}{1mm}
\fboxsep0mm
\mbox{\begin{picture}(86,52)
\put(0,0){\Fig{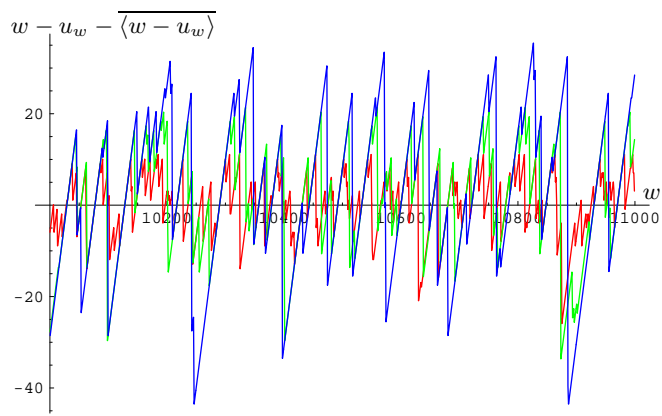}}
\put(0,52){$w-u_{w}-\overline{\left<w-u_{w}\right>}$}
\put(84,30){$w$}
\end{picture}}}
\caption{Dynamical shocks (avalanches): position of a particle $u_{w}$ pulled by a spring, of varying equilibrium position $w$, in a one dimensional random force landscape (with forces uniformly distributed between 0 and 1). The quasi-static motion shows a succession of jumps, also called shocks. Decreasing the spring constant (the mass) from $m^{2}=0.01$ (red) over $m^{2}=0.03$ (green) to $m^{2}=0.001$ (blue), the shocks become larger and larger.
}
\label{f:1}
\end{figure}%

\begin{figure}[b]
{\setlength{\unitlength}{1mm}
\fboxsep0mm
\mbox{\begin{picture}(86,50)
\put(0,0){\fig{86mm}{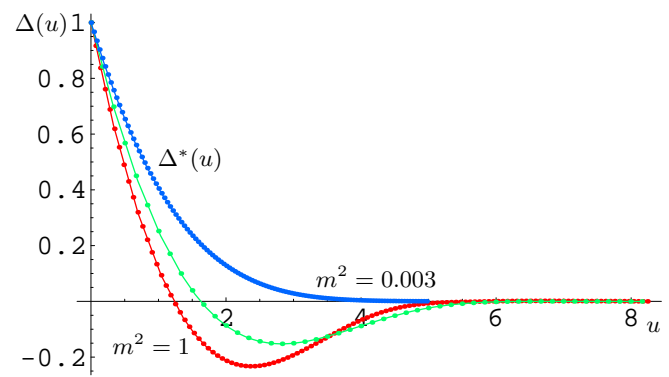}}
\put(0,48){$\Delta (u)$}
\put(84,8){$u$}
\put(13,5){$m^{2}=1$}
\put(40,14){$m^{2}=0.003$}
\put(19,30){$\Delta^{*} (u)$}
\end{picture}}}
\caption{(a): The measured correlator $\Delta (u)$ for a particle pulled in a random potential (i.e.\ RB disorder)
distributed uniformly in $[0,1]$, and rescaled such that $\Delta
(0)=1$ and $\int_{0}^{\infty} \rmd u |\Delta (u)|=1.$ From bottom
(which has $\int_{0}^{\infty} \rmd u \Delta (u)\approx 0$) to top
the mass decreases from
$m^{2} =1$ (red) to $
m^{2}=0.5$ (green) to $m^{2}=0.003$ (blue), where it has (up to small corrections) converged to the fixed point $\Delta^{*}(u)$. This demonstrates in $d=0$ the expected
crossover from
random-bond to random-field disorder (see text).
Note that the fixed point has a cusp singularity at $u=0$.}
\label{f:RBRF}
\end{figure}%
Elastic objects driven through a disordered environment are ubiquitous in nature,  including magnets
\cite{LemerleFerreChappertMathetGiamarchiLeDoussal1998,UrbachBarkhausen1995},
superconductors\cite{BlatterFeigelmanGeshkenbeinLarkinVinokur1994,NattermannScheidl2000}, density waves
\cite{Gruner1988,BrazovskiiNattermann2004}, wetting
\cite{MoulinetGuthmannRolley2002,MoulinetRossoKrauthRolley2004}, dry friction \cite{CuleHwa1998},
dislocation \cite{MorettiMiguelZaiserMoretti2004}, crack propagation \cite{PonsonBonamyBouchaud2006}, and
earthquake dynamics \cite{DSFisher1998}. These phenomena can be studied by different theoretical approaches,
including phenomenological arguments \cite{BlatterFeigelmanGeshkenbeinLarkinVinokur1994}, mean field models
\cite{DSFisher1985}, functional renormalisation group for statics
\cite{GiamarchiLeDoussal1995,Kardar1997,ChauveLeDoussalWiese2000a,Fisher1985b,DSFisher1986,BalentsDSFisher1993,ScheidlDincer2000,DincerDiplom,LeDoussalWiese2001,ChauveLeDoussal2001,LeDoussalWieseChauve2003,LeDoussalWiese2003b,LeDoussalWiese2004a,LeDoussalWiese2005a,Feldman2000,Feldman2001,Feldman2002,TarjusTissier2004,BalentsLeDoussal2002,BalentsLeDoussal2003,BalentsLeDoussal2004,LeDoussalWiese2005b,LeDoussalWiese2006b,TarjusTissier2005,TarjusTissier2006,FedorenkoLeDoussalWiese2006,FedorenkoLeDoussalWiese2006b}
 and driven dynamics
\cite{NattermanStepanowTangLeschhorn1992,NarayanDSFisher1993a,ChauveGiamarchiLeDoussal1998,ChauveGiamarchiLeDoussal2000,LeDoussalWieseChauve2002,LeDoussalWiese2002a,LeDoussalWiese2003a,LeDoussalWieseRaphaelGolestanian2004}.
They were also studied  with numerical techniques
\cite{RossoKrauth2001a,RossoKrauth2001b,RossoKrauth2002,RossoHartmannKrauth2002}. In several cases the
experimental results seem to be in reasonable agreement with the theory (see
\cite{FuchsDoyleZeldovRycroftTamegaiOoiRappaportMyasoedov1998} for vortex lattices,
\cite{ParuchGiamarchiTriscone2005} for ferroelectrics , and
\cite{LemerleFerreChappertMathetGiamarchiLeDoussal1998} for  magnetic interfaces), but some discrepancies are
still manifest, at least with the simplest theories, in some cases, e.g. the depinning of the contact line of a
fluid \cite{MoulinetGuthmannRolley2002,MoulinetRossoKrauthRolley2004,LeDoussalWieseRaphaelGolestanian2004}.

Recent theoretical progress allows not only for qualitative, but also for  quantitative tests. On one hand, for
interfaces, powerful algorithms now allow to find the exact depinning threshold and critical configuration on a
cylinder \cite{RossoHartmannKrauth2002,BolechRosso2004} and to study creep dynamics
\cite{KoltonRossoGiamarchiKrauth2006}. On the other hand the functional RG (FRG) has been extended beyond the
lowest order (one loop), and it was shown that differences between statics and depinning become manifest only at
two loops \cite{ChauveLeDoussalWiese2000a,LeDoussalWieseChauve2002}, i.e.\ to second order in an expansion in
$d=4-\epsilon$ where $d$ is the internal dimension of the manifold. Such differences appear for instance in the
roughness exponent $\zeta$. The FRG is a field theoretic tool for disordered systems, which captures the complex
glassy physics of numerous metastable states at the expense of introducing, rather than a single coupling as in
standard critical phenomena, a function, $\Delta(u)$, of the displacement field $u$, which flows to a fixed
point (FP) $\Delta^*(u)$. This FP is non-analytic, as is the effective action of the theory. The non-analyticity
is a rather unconventional feature, and the validity of the approach has been questioned: one could argue that,
although $\Delta^*(u)$ is perturbative near $d=4$, the second derivative $\Delta''(0)$ has gone to infinity,
hence we have left the domain controlled by perturbation theory. To put this criticism to rest, one first shows
how an observable of the experimental system can be defined, which is identical to the field theoretic disorder
correlator $\Delta(u)$: the idea \cite{LeDoussal2006b} is to add a quadratic confining potential to the system,
which formally acts as a mass for the elastic modes of the interface. The disorder correlator $\Delta(u)$, can
then be measured directly as the second moment of the interface displacement.

This method has been used in a numerical simulation of interfaces in a disordered magnet, to  compute
numerically the zero-temperature FRG fixed-point function $\Delta(u)$ in the statics, for interfaces ($N=1$)
using powerful exact minimization algorithms \cite{MiddletonLeDoussalWiese2006}. A variety of disorder types,
random-bond, random-field and periodic disorder were studied in various dimensions $d=0,1,2,3$. The results are
close to 1-loop predictions and deviations are consistent with 2-loop FRG. The most important feature, namely a
linear cusp in $\Delta(u)$ was clearly seen. These results come in strong support for the underlying hypothesis
of a non-analytic field theory, perturbatively accessible in a $d=4-\epsilon$ expansion.

In our Letter \cite{LeDoussalWiese2006a}, we have extended the method of
\cite{LeDoussal2006b,MiddletonLeDoussalWiese2006} to driven systems. In particular it allows to measure the FRG
fixed-point function $\Delta^*(u)$ near the depinning transition at velocity $v=0^+$. The form of this fixed
point was obtained to one loop \cite{NattermanStepanowTangLeschhorn1992,NarayanDSFisher1993a}, but the remarkable fact is that it is only
to two loop order that it differs from the static fixed point \cite
{ChauveLeDoussalWiese2000a,LeDoussalWieseChauve2002}. Shortly after, this tiny difference was measured, beyond
statistical uncertainties, together with the predicted linear cusp, in a numerical study
 \cite{RossoLeDoussalWiese2006a} of a line driven in a one dimensional medium.

In this situation, it is useful to find a simple model, which can be solved analytically, and exhibits many
features of the more complicated situation. Such a model is a particle in a random force landscape, pulled by a
moving spring. Since there is no internal degree of freedom, it is the $d=0$ limit of the depinning fixed
point for interfaces. Similar $d=0$ toy models were very useful in the study of the statics. The universality
classes there were found to be parameterized by the exponent $\zeta \geq 1$, with a (presumably unique)
universal fixed-point function $\Delta(u)=-R''(u)$ in each case, with $R(u) \sim |u|^\gamma$ at large $|u|$,
with $\gamma=4-4/\zeta$. Only in some cases this function was obtained analytically, e.g. for the case
$\zeta=4/3$ of the Sinai model \cite{LeDoussal2006b}, which corresponds to the random-field disorder class. It
is thus quite interesting to obtain the corresponding results for the depinning fixed point. Of course another
application of the $d=0$ model is  driving a fixed-size manifold over very large distance, it eventually
behaves again as a particle with some effective random force landscape.

The aim of this paper is to give a detailed account of the results summarized in \cite{LeDoussalWiese2006a}, and
to present some new ones. We summarize the basic ideas in section \ref{s:summary}, relegating details of the field-theoretic derivation in appendix \ref{app:dynamical
action}. 

In section \ref{s:particle}, we give a detailed  derivation
 of the analytical results for the $d=0$ particle model, focussing there on uncorrelated random forces. This yields
 an exponent $\zeta$ in a continuous range $0< \zeta \leq 2$ with corresponding fixed-point functions $\Delta(u)$
 easier to compute than in the statics. The result
 depends on the tail of the distribution of the local random force, i.e.\ we find the three main
 universality classes of extreme value statistics: The Gumbel, Weibull and Frechet distribution.
 
In section \ref{s:avalanches}, we calculate analytically the joint avalanche size $s$ and waiting-time $w$
distribution. The avalanche distribution was computed recently \cite{LeDoussalMiddletonWiese2008} in a
$d=4-\epsilon$ expansion using FRG, and the present results hence correspond to the $d=0$ limit. There it was
shown how the avalanche-distribution is related to the non-analyticity of the set of all cumulants of the
displacement field.

Interestingly, the problem of a particle driven in one dimension is related to the so-called record statistics
\cite{majumdar,krug}. If the particle is pulled by a spring the problem is related to record statistics for
random sequences with linear trends \cite{math}, whose interest has been revived in the context of global
warming models \cite{redner}. Translated into the language of records, we obtain in Section
\ref{s:avalanches}, the joint distribution of the time $s$ between two successive records and their difference
in value, $w$, for a sequence of i.i.d.\ random variables with trends, i.e.\ variables $Y_n = X_n + c n$ where
$X_n$ are i.i.d.\ random variables, and $c$ a drift. These results are obtained for the three classes of extremal statistics.

In section \ref{s:numerics toy}, we check some of our above results numerically; we also study numerically a
particle driven at non-zero velocity $v>0$, and find that the velocity smoothens the cusp in the force
correlator.

In section \ref{sec:abbm} we consider long-range correlated random force landscapes, specifically the case of a
Brownian force. For this model, known as the ABBM model for domain wall motion, remarkably, the stationary
distribution of instantaneous velocities can be computed \cite{abbm} for any non-zero average driving velocity
$v>0$. From that we obtain $\Delta(u)$ for any $v>0$. We also compute the quasi-static $\Delta(u)$ and avalanche
distribution. It matches with the limit $v \to 0^+$, and shows how the cusp is smoothened at $v>0$. 

In
section \ref{sec:nomass} we summarize some known results and some new ones, common to record statistics and to
the present model of a driven particle, either with no mass (fixed-force driving, symmetric records), or with a
mass (fixed-velocity driving, records with drifts). In particular we study in detail the record statistics for a
Levy-walk landscape with drift.

Finally we address the outstanding question of the depinning for systems which can move in more than one
direction, also termed ``$N>1$'', with $N$ the number of components of the displacement field $u$ (e.g. $N=2$
for a line moving in three dimension). In particular there is still no satisfactory field theoretic description
for this case based on FRG. This question was studied by Erta\c s and Kardar \cite{ErtasKardar1996}, but they made the
approximation that the disorder correlator only depends on the direction in which the system is driven.
Considering two manifolds which are driven on trajectories far apart in the transversal direction, their
renormalized disorder correlators should be independent, questioning the assumptions in the Erta\c s-Kardar
approach. We have studied this situation in the field theory \cite{fedorenkoN} (see also some study at $v>0$ in
\cite{movglass}), but consistent and {\em stable} solution of the fixed-point equations seem quite complicated
and are still lacking. In this situation it is important to have some numerical results as guide for the
analytical treatment. In section \ref{s:N-component}, we therefore discuss the changes necessary to study an
elastic manifold driven through a higher-dimensional random environment, and complement this in section
\ref{s:N=2} by a numerical study of a particle dragged through a random energy landscape. Especially, we show
numerically, that the scaling exponents (``roughness'') in the direction of the driving $\zeta_x$ and
perpendicular to it $\zeta_{y}$ are different, and satisfy $\zeta_{x}>\zeta_{y}$. We also find, that the
cross-correlator in the transversal direction (i.e.\ the force correlator between forces in the direction $x$ of
the driving, and its transversal one $y$, measured as a function of the transversal distance $u_{y}$,
$\Delta_{xy}(u_{x}=0,u_{y})$, is non-vanishing.


\section{Summary of the method}
\label{s:summary}
\subsection{General framework}
We consider the equation of motion for the over-damped dynamics of an
elastic manifold parameterized by its time-dependent displacement
field $u(x,t)$:
\begin{eqnarray}
&& \eta \partial_t u(x,t) = F_x[u(t);w(t)] \label{eq:eqmo} \\
&& F_x[u;w] = m^2 (w - u(x)) + c \nabla_x^2 u(x) + F(x,u(x)) \nonumber
\end{eqnarray}
where $F_x[u(t);w(t)]$ is the total force exerted on the manifold (we note $u(t)=\{u(x,t)\}_{x \in
\mathbb{R}^d}$ the manifold configuration, $x$ being its $d$-dimensional internal coordinate); $\eta$ is the
friction coefficient and $c$ the elastic constant. Here at the bare level, the random pinning force is
$F(x,u)=-\partial_u V(x,u)$ and the random potential $V$ has correlations $\overline{V(0,x) V(u,x')}= R_0(u)
\delta^{d}(x-x')$.  We consider first random-bond bare disorder with a short-ranged $R_0(u)$.  We have added a
harmonic coupling to an external variable $w(t)$, a given function of time (in most cases we choose it uniform
in $x$). This is the simplest generalization of the statics, where $w(t)=w$ is time-independent. It is useful to
define the fixed-$w$ energy
\begin{equation}
{\cal H}_{w}[u]= \int \rmd^d x\, \frac{m^2}{2} (u(x)-w)^2 + V(x,u(x))
\end{equation}
associated to the force $F_x[u;w]=-\frac{\delta H_{w}[u]}{\delta u(x)}$. If $w(t)$ is an increasing function of
$t$ the model represents an elastic manifold ``pulled'' by a spring.

We first describe qualitatively how to measure the FRG functions and
later justify why the relation is expected to be exact. We are
interested in the observable $w(t) - \langle \bar u(t)
\rangle$ where $\bar u(t) = L^{-d} \int \rmd^d x u(x,t)$ is the center
of mass position, and $\langle \dots \rangle$ denotes thermal
averages, i.e.\ the ground state at zero temperature. It represents
the shift between the translationally averaged displacement and the
center of the well, i.e.\ the extension of the spring. It is thus
proportional to the pulling force on the manifold, hence to the
translationally averaged pinning force minus the friction force,
i.e.\ $w(t) - \bar u(t) = m^{-2}( \eta v(t) - \int_x F(x,u(x,t)))$ (if
we use periodic boundary conditions inside the manifold). Of
particular interest are:
\begin{eqnarray}
&& \overline{ w(t) - \langle \bar u(t) \rangle } = m^{-2} f_{av}(t) \\
&& \overline{ (w(t) - \langle \bar u(t) \rangle)(w(t') - \langle \bar
u(t') \rangle) }^c = m^{-4} L^{-d} D_w(t,t') \nonumber
\end{eqnarray}
where connected means w.r.t.\ the double average $\overline{\langle
... \rangle}$. If we consider a function $w(t)$ such that $dw(t)/dt >
0$, one can also write:
\begin{eqnarray}
&& D_w(t,t')= \Delta_w(w(t),w(t'))
\end{eqnarray}
As written, the function $\Delta_w$ may in general depend on the history $w(t)$. However we expect that for
fixed $L,m$ and slow enough $w(t)$, e.g. $w(t)=vt$ with $v \to 0^+$, one has $\Delta_w(w(t),w(t')) \to
\Delta(w(t)-w(t'))$. This function $\Delta(w-w')$, which is independent of the process $w(t)$, is the one
defined in the F.T.. The derivation of this property is given in Appendix \ref{app:dynamical action} to which we
refer the reader for technical details. Note that we are discussing now $N=1$ systems (interfaces), subtleties
related to $N>1$ are discussed in section \ref{s:N-component}.

Let us now describe $T=0$ depinning. Quasi-static depinning is
studied as the limiting case where $dw/dt \to 0^+$. The quasi-static
motion can be described as follows (in the continuum model). One starts in a
metastable state $u_0(x)$ for a given $w=w_0$, i.e.\ a zero-force
state $F_{x}(u_0(x);w)=0$ which is a local minimum of
$H_{w_0}[u]$ with a positive barrier. One then increases $w$. For smooth short-scale
disorder, the resulting deformation of $u(x)$ is smooth. At some $w=w_1$, the barrier vanishes.
For $w=w_1^+$ the manifold  moves downward in energy until it is blocked again in
a metastable state $u_1(x)$ which again is  a local minimum of $H_{w_1}[u]$.
We are interested in the center of mass (i.e.\ translationally
averaged) displacement $\bar u = L^{-d} \int \rmd^d x\, u(x)$. The above
process defines a function $\bar u(w)$ which exhibits jumps at the
set $w_i$. Note that time has disappeared: evolution is only used to find the next
location. The first two cumulants
\begin{eqnarray}
&& \overline{w - u(w)} = m^{-2} f_c  \label{def-fc} \\
&& \overline{(w - u(w))(w' - u(w')}^c = m^{-4} L^{-d} \Delta(w-w')
\label{def-Delta}
\end{eqnarray}
allow a direct determination of the averaged ($m$-dependent) critical force $f_c$ and of $\Delta(w)$. Note that
$u(w)$ a priori depends on the initial condition and on its orbit but at fixed $m$ one expects an averaging
effect when $w$ is moved over a large region. This is further discussed below. Note that the definition of the
(finite size) critical force is very delicate in the thermodynamic limit \cite{FedorenkoLeDoussalWiese2006}.

Elastic systems driven by a spring and stick-slip type motion were
studied before, e.g.\ in the context of dry friction. The force
fluctuations, and jump distribution were studied numerically for a
string driven in a random potential \cite{LacombeZapperiHerrmann2001}. However, the precise
connection to quantities defined and computed in the field theory has
to our knowledge not been made. The dependence in $m$ for small
$m$ predicted by FRG, $\Delta(w) = m^{\epsilon-2 \zeta} \tilde
\Delta(w m^{-\zeta})$ is consistent with observations of
\cite{LacombeZapperiHerrmann2001} but the resulting $\tilde \Delta(w)$ has never been
measured.  Fully connected mean-field models of depinning also reduce
to a particle pulled by a spring, together with some self-consistency
condition. Ref. \cite{NarayanDSFisher1993a} discusses related issues in an
expansion around mean field. As discussed below, our main remarks here
are much more general, independent of any approximation scheme, and
provide a rather simple and transparent way to attack the problem.

Note that the manifold in the harmonic well can be approximated by $(L/L_m)^d$ roughly independent pieces with
$L_m \sim 1/m$. The motion of each piece over large distances resembles the one of a particle, i.e.\ a $d=0$
model, but with a rescaled unit of distance in the $u$ direction, $u_m \sim L_m^\zeta \sim m^{- \zeta}$.  The
``effective force'' landscape seen by each piece becomes uncorrelated on such distances, and its amplitude
scales as $F_m \sim m^2 u_m$. Hence one is in a bulk regime not dominated by extremes, i.e.\ $\Delta(w)$ probes
only motion over order one unit. It is easy to check that an arbitrary initial condition joins the common unique
orbit after about one correlation length. Hence the $d=0$ model suggests that starting the quasi-static motion
in $u_0$ and driving the manifold over $w \sim L_m^\zeta$ should then result in all orbits either converging or
having statistically identical properties. Note that if the manifold is driven over more than $L^\zeta$, a
crossover to $d=0$ behavior and extremal statistics occurs, as studied in the next Section.

The averaged critical force, defined in (\ref{def-fc}), should, for $d>0$, go to a finite limit, with
$f_c(m)=f_c^\infty + B m^{2-\zeta}$ from finite size scaling. This has been recently tested in the numerics
\cite{RossoLeDoussalWiese2006a}. Although $f_c$ is not universal and depends on short-scale details, one easily
sees that $-m \partial_m f_c(m)$ depends only on one unknown scale. We note that the definition (\ref{def-fc})
coincides with the one proposed recently as the maximum depinning force for all configurations having the same
center of mass $u_0$ \cite{FedorenkoLeDoussalWiese2006}. Since $\bar u - w$ is a fluctuating variable of order
$(L/L_m)^{-d/2}$, the two definitions should coincide in the limit where $L\to\infty$, before $m\to 0$. The one
point distribution of the critical force is obtained from the distribution of $w - u(w)$, and to one loop is
identical to the one obtained in \cite{FedorenkoLeDoussalWiese2006} provided one uses there the massive scheme.

\begin{figure}
{\setlength{\unitlength}{1mm}
\fboxsep0mm
\mbox{\begin{picture}(86,55)
\put(0,0){\fig{86mm}{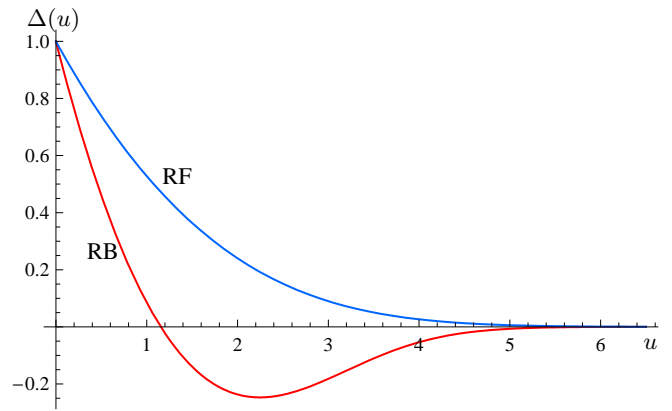}}
\put(2,51){$\Delta (u)$}
\put(84,8){$u$}
\put(10,20){RB}
\put(20,30){RF}
\end{picture}}}
\caption{Fixed-point functions $\Delta^{*} (u)$ for random-bond (RB) and random-field (RF) disorder (arbitrary scale).}
\label{f:RB-RF}
\end{figure}

Let us now recall the field-theory predictions: The FRG equation at 2-loop order for the (rescaled) force
correlator are \cite{ChauveLeDoussalWiese2000a}:
\begin{eqnarray}
- m \partial_m \Delta(u) &=& (\epsilon-2\zeta)\Delta(u) + \zeta u \Delta'(u)\nn\\
&&-\frac{\rmd^{2}} {\rmd u^{2}} \left\{ \half \left[\Delta(u)-\Delta(0)\right]^{2}\right\}\nn\\
&&+ \frac{\rmd^{2}} {\rmd u^{2}}
\left\{\Delta'(u)^{2}[\Delta(u)-\Delta(0)] {+}\lambda \Delta'(0^{+})^{2}\Delta(u) \right\}\nn\\
\label{frg2loop}
\end{eqnarray}
$\lambda=-1$ describes the statics, and $\lambda=1$ the depinning. For the statics, $\Delta(u)$ admits a
potential solution (random-bond universality class) $\Delta(u)=-R''(u)$, with $R(u)$ decaying to 0 as $u\to
\infty$, as illustrated on figure \ref{f:RB-RF}. This implies that  the integral $\int_{0}^{\infty}\rmd u \,
\Delta(u)$ remains unrenormalized. However at depinning, it flows, and no potential solution exists. For a large
class of bare disorder, the model should renormalize, as $m$ decreases, to the random-field fixed-point
solution, $\Delta_{\mathrm{RF}}^{*}(u)$, which is monotonically decaying and strictly positive. For the $d=0$
toy-model discussed in the next section, this crossover is nicely seen in our numerical simulation with
decreasing $m$, as is illustrated on figure \ref{f:RBRF}. Therefore, in the following, we can focus on the
random-field universality class, i.e.\ short-range correlated random forces.

%


%






\begin{figure}[t]
\Fig{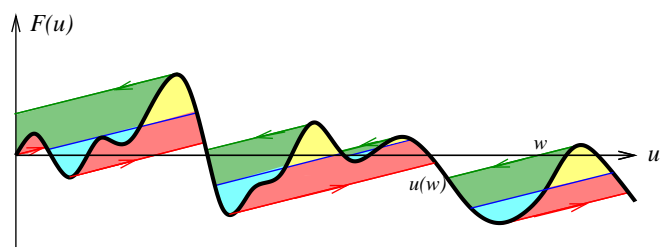}
\caption{Construction of $u(w)$ in $d=0$, for the pinning force $F(u)$ (bold black line). The two quasi-static
motions driven to the right and to the left are indicated by red and green arrows, and exhibit jumps
(''dynamical shocks''). The position of the shocks in the statics is shown for comparison, based on the Maxwell
construction (equivalence of light blue and yellow areas, both bright in black and white). The critical force is
$1/(2 M)$ times the area bounded by the hull of the construction.} \label{figgraphicalA}
\end{figure}%

\section{Particle in short-range random-force landscape ($ d=0$): exact results}
\label{s:particle}

We now study the model in $d=0$, i.e.\ a particle with equation of motion
\begin{equation}\label{d1}
\eta \partial_{t} u = m^{2} (w-u) + F (u)\ .
\end{equation}
In the quasi-static limit where $w$ is increased slower than any other time-scale in the problem, the zero force
condition $F (u) = m^{2} (u-w)$ determines $u(w)$ for each $w$, starting from some initial condition. The
graphical construction of $u(w)$ is well known from studies of dry friction \cite{charlaix}. When there are
several roots one must follow the root as indicated in Fig.\ \ref{figgraphicalA}, where $F (u)$ is plotted
versus $m^{2} (u-w)$. This results in jumps and different paths, $u^\uparrow(w)$ and $u^\downarrow(w)$
respectively for motion to the right (forward) and to the left (backward). Let us call $A$ the area of this
hysteresis loop (the area of all colored/shaded regions in Fig.\ref{figgraphicalA}). It is the total work of the
friction force when moving the center of the harmonic well quasi-statically once forth and back, i.e.\ the total
dissipated energy. The above definition of the averaged critical force (\ref{def-fc}), assuming the landscape
statistics to be translationally invariant and that one can replace disorder averages by translational ones over
a large width $M$ (which certainly holds if force correlations are short-range correlated), gives
\begin{equation}\label{d4}
f_{c} = m^{2}\,  \overline{( w-u_{w}) }^{\mathrm{tr}} = \frac{m^{2}}{M} \int_{0}^{M} \rmd w\, \left(w-u_{w}
\right)
\end{equation}
Hence, subtracting the two paths gives
\begin{eqnarray}\label{d4b}
 f_{c} &:=& \frac{1}{2} (f_c^\uparrow-f_c^\downarrow) = \lim_{M \to \infty} \frac{m^{2}}{2 M} \int_{0}^{M} \rmd
w\,
\left(u^\downarrow(w) - u^\uparrow(w) \right) \nonumber \\
& =& \lim_{M \to \infty} \frac{A}{2M}
\end{eqnarray}
where we have used $\int u \, \rmd w = \int w\, \rmd u$ and $f_c^\downarrow<0$. One can check that for $m \to 0$
this definition of $f_c$ becomes identical to the one on a cylinder, $f_d$, which for a particle ($d=0$) is $2
f_d = f_d^\uparrow-f_d^ \downarrow =\max_{u} F(u)-\min_{u} F(u)$ with $2 f_d M = \lim_{m \to 0} A(m)$. Since $A$
depends on the starting point, this definition holds after a complete tour, where the maximum (minimal) pinning
force was selected. One can also compare with the definition of shocks in the statics. There, the effective
potential is a continuous function of $w$. Therefore, when making a jump, the integral over the force must be
zero, which amounts to the Maxwell construction of figure \ref{figgraphicalA}.

\begin{figure}[t]
\Fig{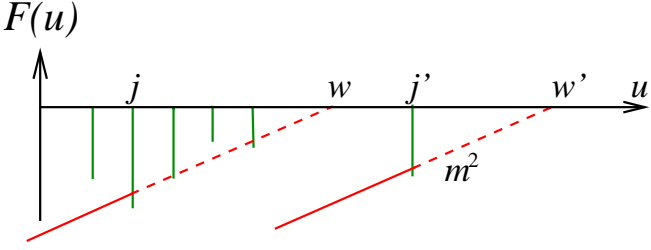} \caption{Construction of $u(w)$ in $d=0$ for the forward motion in the discretized model. The
vertical lines are the force barriers, the (red) increasing lines the spring force $m^{2}(u-w)$. A particle
moves from left to right, until it is stopped by a barrier (when the lines for spring force and barrier forces
intersect).} \label{figgraphicalB}
\end{figure}%
\subsection{Short-range correlated force: a discrete model}
\label{s:SR force: discrete model}

Let us now consider the asymptotic forward process $u(w):=u^\uparrow(w)$, defined in general as the {\it
smallest root} of the equation $F(u)=m^2(u-w)$. The set of points $(u,F(u))$ is the lightened portion of the
$F(u)$ curve, the rest being the shadow. If one starts in the shadow at $w_0$ one joins at $w_1$ the asymptotic
process for all $w>w_1$. The difference $w_1-w_0$ is finite for finite $m$ hence we will only study the
asymptotic process. Note that the area of the shadow per unit length is the critical force.

We now study short-range correlated random force landscapes. In the limit of interest, $m \to 0$, the scale of
jumps becomes large, and the finite range should be unimportant. Hence it is equivalent, and more convenient, to
consider a discrete model, $u$ being integers. The variable $w$ can be kept real. One considers a discrete
landscape $F(u)=F_i$ independently distributed with $P(F)$, and $i$ integer variable. The process $u(w)$ is then
defined on integers. Its definition is shown in Fig.\ \ref{figgraphicalB}.

Let us compute for $w' > w$ the following joint probabilities:
\begin{eqnarray}
P_w(j,F) &:=& \mbox{Prob}\big(u(w)=j \quad \text{and} \quad F_j=F\big)\nn\\
&& \\
P_{w;w'}(j,F;j',F') &:=&
\mbox{Prob}\Big(u(w)=j \text{ and }  F_j=F \nn\\
&& \text{ and }  u(w')=j' \text{ and } F_{j'}=F'\Big)
\end{eqnarray}
We define:
\begin{eqnarray}
H(F)=\int_F^{+\infty} P(f) \rmd f = 1 - \int_{-\infty}^F P(f) \rmd f
\end{eqnarray}

Since for $u(w)=j$ to hold, one must have all $F_k > m^2(k-w)$ for all $k <j$ and $F_j < m^2 (j-w)$, see Fig.\
\ref{figgraphicalB}, one has:
\begin{eqnarray}
&&P_w(j,F) = P(F) \theta(m^2(j-w) - F) \prod_{k=-\infty}^{j-1} H(m^2(k-w))\nn \\
&&P_w(j) = \left(1- H(m^2(j-w))\right) \prod_{k=-\infty}^{j-1} H(m^2(k-w))\ , \qquad
\end{eqnarray}
where the first line integrated over $F$ yields the second (and $\theta(x)$ denotes the unit step function). One
easily checks that $\sum_{j=-\infty}^{j=+\infty} P_w(j)=1$.

Next for $j'>j$ one has:
\begin{eqnarray}
&&P_{w;w'}(j,F,j',F') = P(F) \theta(m^2(j-w) - F)\nn\\
&&\quad \times  \theta(F-m^2 (j-w'))
 \prod_{k=-\infty}^{j-1} H\big(m^2(k-w)\big)  \\
&& \quad\times  P(F') \theta\big(m^2(j'-w') - F'\big)
 \prod_{k=j+1}^{j'-1} H\big(m^2(k-w')\big) \nn
\end{eqnarray}
with the convention that for $j'=j+1$, the factor $\prod_{k=j+1}^{j'-1} H(m^2(k-w'))=1$, and for $j=j'$:
\begin{eqnarray}
P_{w;w'}(j,F,j,F') &=& \delta(F-F') P(F)
\theta\big(m^2(j-w') - F\big) \nn\\
&& \times\prod_{k=-\infty}^{j-1} H\big(m^2(k-w)\big)
\end{eqnarray}
Integrating over the forces we obtain the 1-point and 2-point probability for the process $u(w)$. They read, in
compact notations with $H^w_k:=H\big(m^2(k-w)\big)$, and non-zero only for $j' \geq j$:
\begin{eqnarray}\label{d33}
P_w(j) &=& (1-H_{j}^{w}) \prod_{k=-\infty}^{j-1} H_k^{w} \\\label{d34} P_{w;w'}(j,j') &=& (H_j^{w'} - H_j^{w})
(1-H_{j'}^{w'})
\prod_{k=-\infty}^{j-1} H_k^{w} \prod_{k=j+1}^{j'-1} H_k^{w'}\nonumber\\
&& + \delta_{jj'} (1-H_{j}^{w'}) \prod_{k=-\infty}^{j-1} H_k^{w} \\\label{d35}
& =& \frac{P_w(j)}{1-H_j^w} \Big[ (H_j^{w'} - H_j^{w}) (1-H_{j'}^{w'}) \prod_{k=j+1}^{j'-1} H_k^{w'} \nn\\
&& \qquad \qquad+ \delta_{jj'} (1-H_{j}^{w'})\Big]
\end{eqnarray}
with the convention that for $j'=j$, the factor $\prod_{k=j+1}^{j'-1} H_k^{w'}=0$. Using that $\sum_{j'>j}
(1-H_{j'}^{w'}) \prod_{k=j+1}^{j'-1} H_k^{w'} = 1$ one checks the normalization $\sum_{j' \geq j} P_{w;w'}(j,j')
= P_w(j)$.

\subsection{Continuum limit for 1-point distribution}
For small $m$ the continuum limit can be taken:
\begin{eqnarray}\label{d36}
&& \!\!\!P_w(j) \approx \Big[1- H(m^2(j-w))\Big]\nn\\
&& \qquad\qquad \times \exp \left( \int_{-\infty}^j dy \ln H(m^2(y-w)\right) \\\label{d37}
&& \!\!\!=  \int_{-\infty}^{m^2(j-w)} \rmd f\, P(f)\nn\\
&& \qquad \times  \exp\left( \int_{-\infty}^j \rmd y \ln (1- \int_{-\infty}^{m^2(y-w)} P(l) \rmd l )\right) \nn
\\\label{d38} && \!\!\! \approx \int_{-\infty}^{m^2(j-w)}\rmd f\, P(f) \ \exp\!\!\left(- \int_{-\infty}^j \rmd y
\int_{-\infty}^{m^2(y-w)} P(l) \rmd l \right) .\nn
\end{eqnarray}
The last step is justified if the result is indeed dominated by the tail of $P(f)$ for $f$ negative, which is at
the heart of extremal statistics. That this is indeed true is justified a posteriori.

The quantity $m^2 [w-u(w)]\equiv m^2(w-j)$ is the ``local'', i.e.\ fluctuating critical force, and its disorder
average is $f_c(m)=m^2 \overline{[w-u(w)]}$. Its distribution can be obtained from the 1-point distribution
$P_w(j)$. To rewrite (\ref{d38}) in a simpler form we define:
\begin{eqnarray}\label{20}
 a_{w}' (j)&\equiv&  a'(j) := \int_{-\infty}^{m^2(j-w)} P(f) \rmd f\qquad \\\label{d45}
 a_{w} (-\infty) &:= 0&\ .
\end{eqnarray}
Note that $a_{w} (+\infty) = + \infty$. The 1-point distribution can thus be rewritten as:
\begin{eqnarray}\label{d39d}
P_w(j) \rmd j\, &=& \rme^{- a_{w}(j)} \rmd a_{w}(j)\ .
\end{eqnarray}
Hence one can rewrite the first moment in the form:
\begin{eqnarray}\label{d39e}
\overline{w-u(w)} &=& \int_{-\infty}^\infty \rmd j\; P_w(j) (w-j)\nn\\
& =& \int_{-\infty}^\infty \rmd j\; a'_{w}(j)  \rme^{- a_{w}(j)}  (w-j)
\end{eqnarray}
This means that the quantity $a$ has a simple exponential distribution, hence one needs to invert the relation,
i.e.\ find $j$ as a function of $a$
\begin{eqnarray}\label{d39f}
 j=j(a;w) \quad \leftrightarrow \quad a_w(j)=a\ ,
\end{eqnarray}
and use that $a$ has an exponential distribution, to get any average, e.g.
\begin{eqnarray}\label{d39h}
 \overline{(w-u(w))^p} = \int_0^\infty da \;\rme^{-a} (w - j(a;w))^p
\end{eqnarray}
for any $p$.

\subsection{Distribution of critical force : the different disorder classes}
\label{sec:criticalforce}

We now obtain the universality classes for the 1-point distribution of the process $u(w)$, i.e.\ for the
distribution of critical forces. We define $b$, a rescaled version of $a$:
\begin{eqnarray}
 b(m^2(j-w)) := m^2 a  \label{cond}
\end{eqnarray}
Using (\ref{20}), it can be written as
\begin{eqnarray}
 b(x)\equiv \rme^{- \beta(x)} = \int_{-\infty}^x \rmd y \int_{-\infty}^y P(f) d f  \label{defb}
\end{eqnarray}
The condition defines $a_w(j)$. We thus need to invert this relation to get $j$ as a function of $a$. We do this
for the three main disorder classes below.

\subsubsection{Gumbel class (class Ia)}
The first class contains distributions $P(F)$ with unbounded support and decaying exponentially fast at $F \to -
\infty$ (in some broad sense defined below). One then finds the Gumbel distribution for the critical force,
hence we call this class the Gumbel class.

Let us invert relation (\ref{cond}) and assume that the following expansion holds at small $m$:
\begin{align}\nn
 m^2 (j-w) &= \beta^{-1}(\ln \frac{1}{m^2} - \ln a) \\
& = - f_c^0(m) - \frac{\ln a}{\beta'(-f_c^0(m))}  \nn\\
&\hphantom{=}\;\;+(\beta^{-1})''(\ln \frac{1}{m^2}) \frac{(\ln a)^2}2 + \ldots \label{28} \\
 f_c^0(m) &:=  -b^{-1}(m^{2}) \equiv - \beta^{-1}(\ln \frac{1}{m^2})
 \label{29}
\end{align}
where we know that $a$ is a fluctuating number of order one with an exponential distribution,
$P(a)=\rme^{-a}\theta(a)$. That gives the distribution of the variable $m^2(j-w)$. In particular since
\begin{eqnarray}
- \int_0^\infty \rmd a\; e^{-a} \ln a = \gamma_E = 0.577216\ldots\ ,
\end{eqnarray}
we obtain the asymptotics of the (averaged) critical force as:
\begin{eqnarray}
 f_c(m)  = f_c^0(m) - \frac{\gamma_E}{\beta'(-f_c^0(m))} + \ldots
\end{eqnarray}
For this asymptotics to hold,  the ratio of successive terms has to go to zero, equivalent to
\begin{eqnarray}
 \lim_{z \to \infty} \frac{d}{dz} \ln|(\beta^{-1})'(z)| \to 0\ .
\label{defclass1}
\end{eqnarray}
This defines the Gumbel class Ia, together with the fact that the support is unbounded. An example is $\ln
P(f)=-A(-f)^\gamma$ as $f\to -\infty$. Then:
\begin{align}
\beta(x) &\approx_{x \to \infty} A (-x)^\gamma \\
\beta^{-1}(y) &\approx - (y/A)^{1/\gamma} \\
 -(\beta^{-1})'(y) & \approx\frac1{\gamma A} \left(\frac y A\right)^{\frac{1}{\gamma}-1}\ .
\end{align}
Hence class-Ia condition is satisfied for {\em any} $\gamma$, even for $\gamma<1$. Note that pre-exponential
algebraic factors to not change the result. The critical force becomes
\begin{align}
& m^2 (u(w)-w) \nn \\
& \quad=  m^2 (j-w) = - \left[\frac{\ln m^{-2}}{A}\right]^{1/\gamma} + \frac{\ln a}{A
\gamma (\frac{\ln m^{-2}}{A})^{1-\frac{1}{\gamma}}} \nn\\
&  \qquad+\, O\left( \frac{(\ln a)^2}{(\ln m^{-2})^{2-\frac{1}{\gamma}} }\right)\ , \label{fc-class1}
\end{align}
Defining the fluctuating critical force as $f_c(m)=m^2(w-u(w))$ we find
\begin{eqnarray}\label{37}
f_c(m) = f_c^0(m) + c m^2 \rho_m\ ,
\end{eqnarray}
where $c=- \ln a $ has a Gumbel distribution $P(c)=e^{-c} \exp(-e^{-c})$ on the real axis $c \in ]-
\infty,\infty[$, with $\overline{c}=\gamma_E$ and
\begin{eqnarray}
f_c^0(m) = A^{-\frac 1\gamma} (\ln m^{-2})^{\frac{1}{\gamma}}\ .
\end{eqnarray}
One also finds
\begin{equation}\label{39}
 \rho_m = \frac{(\beta^{-1})'(\ln m^{-2})}{m^{2}}= \frac1{\gamma A^{\frac 1 \gamma} m^2 (\ln m^{-2})^{1-\frac{1}{\gamma}}} \ .
\end{equation}
Since as confirmed below, $\rho_m \sim m^{-\zeta}$ is the unique scale appearing also in the second cumulant
(the disorder correlator defined in FRG), we can identify for class Ia:
\begin{eqnarray}\label{40}
 m^{- \zeta} = m^{-2} (\ln m^{-2})^{\frac{1}{\gamma}-1}\ .
\end{eqnarray}
Hence $\zeta=2$, with additional logarithmic  corrections, i.e.\ $\zeta=2^+$ for $\gamma<1$ and
 $\zeta=2^-$ for $\gamma>1$.

\subsubsection{Class Ib: bounded support with exponential singularity}

An example of this class is:
\begin{eqnarray}
 \beta(x) &=& A/(x+x_0)^\gamma \\
 \beta^{-1}(y) &=&-x_0+(y/A)^{-1/\gamma} \\
 b(x) &\sim& \rme^{-A/(x+x_0)^\gamma} \theta(x+x_0) \\
 P(f) &\sim& \rme^{-A/(f+x_0)^\gamma} \theta(f+x_0)
\end{eqnarray}
with $\gamma>0$. One sees that the condition (\ref{defclass1}) is obeyed. Hence this is still the Gumbel class,
although we introduce a distinction for convenience. A similar asymptotics can then be performed:
\begin{eqnarray}
 m^2 (j-w) = - x_0 + \left(\frac{A}{\ln m^{-2}}\right)^{\frac 1 \gamma} + \frac{1}{\gamma} A^{\frac 1 \gamma}
\frac{\ln a}{(\ln m^{-2})^{1+
\frac{1}{\gamma}}}\nn\\
\end{eqnarray}
Apart from the bound $x_0$, the result is the same as in class I with $\gamma \to - \gamma$ (in the exponents
only). Hence one finds again:
\begin{eqnarray}
f_c(m) = f_c^0(m) + c m^2 \rho_m
\end{eqnarray}
where $c=-\ln a$ has a Gumbel distribution $P(c)=e^{-c} \exp(-e^{-c})$ on the real axis $c \in ]-
\infty,\infty[$ with $\overline{c}=\gamma_E$ and:
\begin{eqnarray}
f_c^0(m) =  x_0 - A^{1/\gamma} (\ln m^{-2})^{- \frac{1}{\gamma}}
\end{eqnarray}
One also finds the characteristic scale $\rho_m \sim m^{- \zeta}$:
\begin{eqnarray}
 \rho_m =  1/(\gamma A^{-1/\gamma} m^2 (\ln m^{-2})^{1+\frac{1}{\gamma}})\ ,
\end{eqnarray}
hence  $\zeta=2^+$.

\subsubsection{Algebraic bounded support: Weibull class (class III)}
\label{WCfc} The Weibull class, or class III applies for a force distribution with bounded support (from below)
and algebraic behavior near the edge. An example is
\begin{eqnarray}
\label{49}
P(f)&=&\tilde A (f+x_0)^{\tilde \alpha} \theta(f+x_0)\\
b(x)&=&A
(x+x_0)^\alpha \theta(x+x_0)\\
\alpha&=&2 + \tilde \alpha\\
A&=&\tilde A/[(2+\tilde \alpha)(1+\tilde \alpha)]\ .
\end{eqnarray}
One must have $\tilde \alpha>-1$ hence $\alpha >1$. The box distribution corresponds to $\tilde \alpha = 0$,
i.e.\ $\alpha=2$. Here
\begin{equation}
b^{-1}(y)=-x_0 + (y/A)^{1/\alpha}
\end{equation}
 with $y \geq 0$. Hence analogously to  (\ref{28}) and (\ref{29}) we find from (\ref{cond})
\begin{eqnarray}\label{55}
m^2 (u(w)-w)
 &=&
 m^2 (j-w) = b^{-1}(m^{2}a) \nn\\
 &=&- x_0 + \left(\frac{m^2}{A}\right)^{1/\alpha} a^{1/\alpha}\ ,\qquad
\end{eqnarray}
where we recall that $a$ is a random variable with distribution $P(a)=\rme^{-a}\theta(a)$. For $\alpha \to
\infty$ one recovers the Gumbel class. Hence one finds for the fluctuating critical force
$f_c(m)=m^2(u(w)-w)=m^2(j-w)$:
\begin{eqnarray}\label{56}
f_c(m) = f_c^0(m) + c m^2 \rho_m
\end{eqnarray}
where now $c=- a^{1/\alpha}$ has a Weibull distribution $P(c)=\alpha (-c)^{\alpha-1} \exp(-(-c)^\alpha)$ with
parameter $\alpha$ on the negative real axis $c \in ]-\infty,0[$ and
\begin{eqnarray}
f_c^0(m) =  x_0\ .
\end{eqnarray}
The averaged critical force is
\begin{eqnarray}
 f_c(m) = x_0 - \left(\frac{1}{A}\right)^{1/\alpha} m^{\frac{2}{\alpha}}\,
\Gamma(1+\frac{1}{\alpha}) + \ldots\ ,
\end{eqnarray}
where we have used that $\int_0^\infty da\, \rme^{-a} a^{1/\alpha}=\Gamma(1+\frac{1}{\alpha})$. One also finds
that
\begin{eqnarray}\label{rho.m.Weibull}
 \rho_m &=&  A^{-1/\alpha} m^{-2(1-\frac{1}{\alpha})}
\\
\zeta &=& 2 - \frac{2}{\alpha}
\end{eqnarray}
with $1 < \alpha < \infty$, hence $0 < \zeta < 2$.

\subsubsection{Fr\'echet class (class II)}
The Fr\'echet class, or class II, is relevant for force distributions with large fluctuations, i.e.\ algebraic
tails on an unbounded support. An example is $P(f) \approx \tilde A (-f)^{-\tilde \alpha}\Theta(-f)$, $\tilde
\alpha > 1$. One has $b(x)=A(-x)^{- \alpha}$ with $\alpha=\tilde \alpha-2$ and $\tilde A=A(\tilde
\alpha-2)(\tilde \alpha-1)$. Since $\beta(x)=- \ln A+ \alpha \ln(-x)$ and $\beta^{-1}(y) \sim -\rme^{y/\alpha}$
one checks that the class-I condition (\ref{defclass1}) is not fulfilled. Let us first study $\tilde \alpha>2$,
i.e.\ $\alpha>0$:
\begin{eqnarray}
&& m^2 (j-w) = - A^{1/\alpha} \left(\frac{1}{m^2 a}\right)^{\!\!\frac{1}{\alpha}}
\end{eqnarray}
Hence one finds for the fluctuating critical force $f_c(m)=m^2(u(w)-w)=m^2(j-w)$:
\begin{eqnarray}\label{61}
f_c(m) = c m^2 \rho_m
\end{eqnarray}
where now $c=a^{-1/\alpha}$ has a Frechet distribution $P(c)=\alpha c^{-\alpha-1} \exp(-c^{-\alpha})$ with
parameter $\alpha>0$ on the positive real axis $c \in ]0,\infty[$ and the average critical force is:
\begin{eqnarray}
 \overline{f_c(m)} = \Gamma(1-\frac{1}{\alpha}) m^2 \rho_m
\end{eqnarray}
where we have used $\overline{c} = \int_0^\infty \rmd a\, \rme^{-a} a^{-1/\alpha}=\Gamma(1-\frac{1}{\alpha})$,
and
\begin{eqnarray}
 \rho_m =  A^{1/\alpha} m^{-2(1+\frac{1}{\alpha})}\ .
\end{eqnarray}
This corresponds to a roughness exponent
\begin{eqnarray}
\zeta = 2 + \frac{2}{\alpha}\ .
\end{eqnarray}
Note that for $\alpha<1$ ($\tilde \alpha <3$, $\zeta>4$) the average critical force is infinite.

We will see that the Frechet class is a bit pathologic in the sense that $\Delta(0)=\infty$ for $\alpha<2$. More
generally cumulants of order larger than $\alpha$ are infinite, i.e.\ they are associated to a probability
distribution with fat tails. This implies as usual that these quantities are dominated by the largest events,
hence they are sensitive to how the continuous limit is constructed from the discrete model. For $\tilde
\alpha<2$ the integral in (\ref{defb}) is divergent at its lower bound, hence undefined without a cutoff.

\subsubsection{Comparison with extremal statistics}

Until now it seems that we have recovered the standard extremal statistics classes for the distribution of the
local critical force. On one hand this is not surprising, since one expects, qualitatively, that the critical
force for two independent consecutive regions in one dimension, be the maximum of the ones for each single
region. This is certainly an exact statement for the zero-mass case recalled in Section \ref{sec:nomass}. Here,
it is quite consistent with the identification of $m^2(u(w)-w)$ as a fluctuating threshold force, a fact which
maybe was not obvious from the start. However, note that the value of the parameter $\alpha$ of the extremal
statistics classes is {\it shifted by one} from the value it takes if one models the total critical force as the
extremal one $f_c=\min_i(\{ f_c^i \})$ over $N \sim m^{-2}$ independent regions, each with its critical force
$f_c^i$ distributed with $P_c(f)$. In that case one has:
\begin{eqnarray}
{\mathrm {Proba}}(f_c > x) = \left[\int_{x}^\infty P_c(f)\right]^N \approx e^{-N \int_{- \infty}^x P_c(f)}
\end{eqnarray} which can be written equivalently as:
\begin{eqnarray}
b_c(x) := \int_{-\infty}^x P_c(f) df = \frac{a}{N}
\end{eqnarray}
where $a$ is a random variable of order one with an exponential distribution, $P(a)=\rme^{-a}\theta(a)$. A
comparison with (\ref{defb}) shows that the effective critical force in an independent region should be chosen
with a distribution with a tail:
\begin{eqnarray}
P_c(f) = \int^f_{- \infty} P(f') df'   \label{relPP}
\end{eqnarray}
for large negative $f$. Hence there is a first coarse graining which transforms the tail of $P(f)$ into the tail
of $P_c(f)$. Then one can think of the resulting critical force as the maximum over $N \sim m^{-2}$ independent
random variables distributed with $P_c(f)$. Applied to Weibull and Frechet classes, (\ref{relPP}) indeed
accounts for the shift of the index $\alpha$ by one.

\subsection{2-point probabilities and the FRG correlator $\Delta(w)$}
\label{s:2-point}

In addition to recovering the three extremal statistics classes, which, as explained above, is not surprising if
one thinks in terms of coarse grained independent random variables, there is a second, more remarkable property.
We find here that the 2-point correlation
\begin{align}
& \overline{(w-u(w))(w'-u(w'))}^c \nn\\
& =\! \int\! \rmd j\, \rmd j' P_{w,w'}(j,j') (w-j)(w'-j) - \overline{(w-j)} ~~ \overline{(w'-j')} \nn \\
& =  m^{-4}\Delta(w-w') \label{65}
\end{align}
takes,  for all three classes, the following form at small $m$:
\begin{equation}
 \Delta(w) = m^4 \rho_m^2 \tilde \Delta (w/\rho_m) \ ,  \label{formdelta}
\end{equation}
where the scale $\rho_m \sim m^{- \zeta}$ is the one identified in each case in the previous section, and the
fixed-point function $\tilde \Delta (w)$ only depends on the universality class: it is unique and identical for
all members of class I (Ia and Ib), continuously depending on $\alpha$ (hence $\zeta$) for classes II and III.

Let us now give the joint probability in the continuum limit. From (\ref{d34}) and as was done to arrive to
(\ref{d39d}) one finds for $w'>w$:
\begin{eqnarray}\label{d13}
P_{w;w'} (j,j') &=& \left[a'_{w} (j)- a'_{w'} (j) \right]a'_{w'} (j')\\&&
\quad\times \rme^{-a_{w} (j)- a_{w'} (j')+a_{w'} (j)}\theta(j'>j) \nn\\
&&  + \delta(j-j') a'_{w'} (j) \rme^{-a_{w} (j)}\ . \qquad\nn
\end{eqnarray}
Let us check the normalizations. Using that $\int_j^\infty \rmd j' a'_{w'} (j') \rme^{- a_{w'} (j')} = \rme^{-
a_{w'} (j)}$ one obtains (writing separately the two contributions):
\begin{eqnarray}
&&\!\!\!\int_{-\infty}^\infty \rmd j' P_{w;w'} (j,j') \nn\\
&& = \left[a'_{w} (j)- a'_{w'} (j) \right] \rme^{-a_{w} (j)}
+ a'_{w'} (j) \rme^{-a_{w} (j)} \nn\\
&& = P_{w}(j)
\end{eqnarray}
A similar trick yields
\begin{eqnarray}
\int_{-\infty}^\infty \rmd j P_{w;w'} (j,j') = P_{w'}(j')\ .
\end{eqnarray}
Note also that
\begin{eqnarray}
P_{w;w} (j,j') = P_{w}(j) \delta(j'-j)\ .
\end{eqnarray}
For all three classes one finds, starting from (\ref{fc-class1}), (\ref{56}) and (\ref{61}) respectively
\begin{eqnarray}\label{71}
j-w = - \frac{f_c^0(m)}{m^2} - \rho_m c(a)\ ,
\end{eqnarray}
with either $c(a)=- \ln a$ (class I), $c(a)=- a^{1/\alpha}$ (class III) or $c(a)= a^{-1/\alpha}$ (class II).
Since the constant piece in (\ref{71}) disappears when computing the connected moments, to compute $\Delta$
defined in (\ref{65}) we can simply write
\begin{eqnarray}\label{72}
j-w = - \rho_{m} c(a) \quad , \quad j'-w' = - \rho_{m} c(a')
\end{eqnarray}
and to obtain the rescaled function, $\tilde \Delta$, we can further set  $m=\rho_{m}=1$ and write:
\begin{equation}
\!\,\tilde \Delta(w-w')= \int \rmd j\, \rmd j'\, P_{w,w'}(j,j') c(a_{w}(j)) c(a_{w'}(j'))
\end{equation}

\subsubsection{Calculation of $\Delta(w)$ for class I}
We define $a_1:=a_w(j)$, $a_2:=a_{w'}(j)$, $a_3:=a_{w'}(j')$. Given the previous remark to compute the cumulants
we can set
\begin{align}
 j-w &= \ln a_1 \quad, \quad j-w'= \ln a_2 \\
 j'-w' &= \ln a_3 \quad, \quad a_2/a_1=e^{- W} \ , \label{set}
\end{align}
where we denote $W=w'-w>0$. The joint probability (\ref{d13}) then reads
\begin{eqnarray}\label{d15}
 \lefteqn{P_{w;w'} (j,j') dj dj'} \nn \\
&=& da_1 (1-\rme^{-W}) da_3 e^{ -a_1 (1-\rme^{-W}) - a_3}  \theta(j'>j) \nn \\
&& + \delta(j'-j) dj' \rme^{-W} da_1 \rme^{- a_1}\ .
\end{eqnarray}
This yields the second moment
\begin{eqnarray}\label{d17}
 \lefteqn{\!\!\!\!\! \!\!\!\!\! \!\!\!\!\!\overline{(w-j) (w'-j')}^c = \left[1-\rme^{-W} \right] \int_{0}^{\infty} \rmd a_1 \int_{a_1\,
\rme^{-W}}^{\infty} \rmd a_3}  \nn \\
&& \times \exp \left(-a_1 \left[1-\rme^{-W} \right] -a_3 \right) \ln a_1 \ln a_3 \nn \\
&& + e^{-W} (W \overline{w-j} + \overline{(w-j)^2}) - \overline{w-j}^2 \ .\qquad
\end{eqnarray}
Note the integration interval for $a_3$ which corresponds to $j'>j$ using (\ref{set}). We recall that
\begin{align}
& \overline{w-j} = - \int_{0}^{\infty} \rmd a \, \rme^{-a} \ln a = \gamma_E \\
& \overline{(w-j)^2} = \int_{0}^{\infty} \rmd a \, \rme^{-a} (\ln a)^2 = \gamma_E^2 + \frac{\pi^2}{6}\ .
\end{align}
Thus we obtain:
\begin{eqnarray}
\label{d18a}
 \tilde \Delta(W) &=&   (1-\rme^{- W})  \int_{0}^{\infty}
\rmd a_1 e^{-a_1 (1-\rme^{- W})} \ln a_1  \nn \\
&& \times
\int_{a_1\, \rme^{- W }}^{\infty} \rmd a_3\, e^{-a_3} \ln a_3\nn \\
&& + \rme^{- W } ( \gamma_E^2 + \frac{\pi^2}{6} + W \gamma_E) - \gamma_E^2
\end{eqnarray}
The calculation is performed in Appendix \ref{app:integral}. The final result for the fixed-point function of
class I is:
\begin{equation}\label{c97}
\tilde \Delta (w) =\frac{w^2}{2}+\text{Li}_2\left(1-e^w\right)+\frac{\pi ^2}{6}
\end{equation}
where $\mbox{Li}_n(z)=\sum_{k=1}^\infty z^k/k^n$. One can also use the alternative formula (\ref{c96}). Another
equivalent compact expression for the result is:
\begin{eqnarray}\label{largew0}
 \tilde \Delta (w) &=& \sum_{n=1}^{\infty} \frac{1+nw }{n^{2}}\rme^{-n w}
\end{eqnarray}
A numerical test is performed in section \ref{s:numerics toy}, see figure \ref{f:data2}.

The behavior of the fixed-point function at small $w>0$ is:
\begin{eqnarray}\label{p1}\nn
 \tilde \Delta (w) &=& \frac{\pi
   ^2}{6}-w+\frac{w^2}{4}-\frac{w^3}{36}+\frac{w^5}{3600}-\frac{w^7}{211680}+O(w^{9})\end{eqnarray}
hence we confirm that {\it there is a cusp}, and a power series expansion in $|w|$. The behavior at large $w$ is
easier obtained from (\ref{largew0}) and reads:
\begin{eqnarray}\label{largew}
 \tilde \Delta (w)
 &=&
 (w+1) e^{-w}+\frac{1}{4} (2 w+1) e^{-2 w}+\frac{1}{9} (3 w+1) e^{-3 w} \nn \\
&& +O\left(e^{-4 w}\right)
\end{eqnarray}
It is characteristic of short-ranged correlations in the force with an exponential decay.

\subsubsection{Calculation of $\Delta(w)$ for class III}
\label{KN1} As for class I, we define
 $a_1=a_w(j)$, $a_2=a_{w'}(j)$, $a_3=a_{w'}(j')$ with (attention: $j$ is ahead of $w$)
\begin{eqnarray}
 j-w &=& a_1^{1/\alpha}\\
   j- w' &=& a_2^{1/\alpha} \\
 j'-w' &=& a_3^{1/\alpha}   \ .
\label{class3def-b}
\end{eqnarray}
One must distinguish the cases $j>w'$ (then $j'>j$ is equivalent to $a_3>a_2$ and $W=w'-w=a_1^{1/\alpha} -
a_2^{1/\alpha}$) and $j<w'$, in which case $a_2=0$. There are thus two pieces for the part with $j'>j$:
\begin{eqnarray}
d_{1}&:=& \overline{(j-w)(j'-w')}|_{j'>j}  \\
&=& \int_{w}^{w'} \rmd j\, \frac{\rmd a_1}{\rmd j\,} \rme^{-a_1} \int_0^\infty \rmd a_3\, \rme^{-a_3} (a_1
a_3)^{\frac 1 \alpha}\nn \\
&& + \int_{w'}^\infty \rmd j\, \frac{\rmd(a_1-a_2)}{\rmd j\,} \rme^{- (a_1-a_2)} \int_{a_2}^\infty \rme^{-a_3}
(a_1 a_3)^{\frac 1 \alpha} \nn
\end{eqnarray}
In the first integral the integration bounds over $a_1$ is from $0$ to $W^\alpha$ and in the second the relation
$w'-w=a_1^{1/\alpha} - a_2^{1/\alpha}$ holds. Hence one obtains
\begin{eqnarray}
 d_1 &=& \Gamma\left(1+\frac{1}{\alpha}\right) \int_0^{W^\alpha} \rmd a_1\,
a_1^{1/\alpha} \rme^{-a_1} \\
&& + \int_{W^\alpha}^\infty \rmd a_1 \left(1 - \frac{\rmd a_2}{\rmd a_1}\right) a_1^{1/\alpha} \rme^{-a_1+a_2}
\Gamma\left(1+\frac{1}{\alpha},a_2\right)\nn
\end{eqnarray}
The second contribution ($j=j'$) is:
\begin{eqnarray}
 d_2 &:=& \overline{(j-w)(j'-w')}|_{j'=j} \nn\\
 & =& \int_{w'}^{\infty} \rmd j\,
\frac{\rmd a_2}{\rmd j\,} \rme^{-a_1} (a_1 a_2)^{1/\alpha} \nn\\
&=& \int_{W^\alpha}^\infty \rmd a_1 \frac{\rmd a_2}{\rmd a_1} \rme^{-a_1} (a_1 a_2)^{1/\alpha}
\end{eqnarray}
One has for the disorder correlator:
\begin{eqnarray}
 \tilde \Delta(W)=d_1 + d_2 - \left(\int_0^\infty \rmd a_1 \rme^{-a_1} a_1^{1/\alpha}\right)^{\!\!2}
\end{eqnarray}
Using $a_2=(a_1^{1/\alpha}-W)^\alpha$, $\rmd a_2/\rmd a_1=(1 - w a_1^{- 1/\alpha})^{\alpha-1}$ and the variable
$y=a_2^{1/\alpha}$, $a_1^{1/\alpha}=y+W$, one finds:
\begin{eqnarray}
\lefteqn{\tilde \Delta(W)=-\Gamma \left(1+\frac{1}{\alpha}\right)
\Gamma\left(1+\frac{1}{\alpha},W^\alpha\right) }\\
&& + \alpha \int_0^\infty \rmd y (y+W)
\rme^{-(y+W)^\alpha} \nn\\
&& \qquad~~ \times \big[y^\alpha+ \rme^{y^\alpha} \Gamma(1+\frac{1}{\alpha},y^\alpha)
\left((y+W)^{\alpha-1}-y^{\alpha-1}\right) \big] \nn
\end{eqnarray}
Integration by part of the last term finally yields:
\begin{eqnarray}\label{Delta-Weibull}
\tilde \Delta(w)&=&-\Gamma\left(1+\frac{1}{\alpha}\right) \Gamma\left(1+\frac{1}{\alpha},w^\alpha\right) \nn\\
&&+ w\, \Gamma\left(1+\frac{1}{\alpha}\right)
\rme^{-w^\alpha} \nn\\
&& + \int_0^\infty \rmd y\, \rme^{-(y+w)^\alpha + y^\alpha} \Gamma\left(1+\frac{1}{\alpha},y^\alpha\right) \
.\qquad
\end{eqnarray}
We recall that
\begin{eqnarray}\label{109}
\Gamma(a,x) = \int_{x}^\infty \rmd z\, z^{a-1} \rme^{-z}\ .
\end{eqnarray}
Hence we find a fixed-point function continuously dependent on $\alpha$, and $\zeta=2-2/\alpha$ yielding a
unique form for each value of $0<\zeta<2$.

The value at $w=0$ has a simple expression:
\begin{eqnarray}
 \tilde \Delta(0)=\Gamma\left(1+\frac{2}{\alpha}\right)- \Gamma\left(1+\frac{1}{\alpha}\right)^2
\end{eqnarray}
We find that the function $\tilde \Delta(w)$ has a cusp with
\begin{equation}\label{94}
 - \Delta'(0^+)=\frac{\Gamma\left(1+\frac{1}{\alpha}\right)}{\alpha}\ .
\end{equation}
Since there are a priori $w^\alpha$ terms (we recall $\alpha>1$), we want to understand at which order the
expansion in $|w|$ breaks down.

More explicit expressions can be obtained in special cases. For the box-distribution, i.e.\ $\alpha=2$, we find
\begin{eqnarray}\label{tDWalpha=2}
\tilde \Delta(w)&=&\frac{e^{-w^2}} {4
   w}{ \left[2 w-e^{w^2} \sqrt{\pi } \left(2 w^2+1\right) \text{erfc}(w)+\sqrt{\pi }\right]}\nn\\
   &&+\frac{1}{2}  \sqrt{\pi } \left[w\, e^{-w^2} - \Gamma \left(\frac{3}{2},w^2\right)\right]\ ,
\end{eqnarray}
which has a power series expansion in $|w|$ around $w=0$:
\begin{eqnarray}
\tilde \Delta(w)&=& \left(1-\frac{\pi }{4}\right)-\frac{\sqrt{\pi } w}{4}+\frac{w^2}{3}-\frac{\sqrt{\pi }
   w^3}{24}-\frac{w^4}{30}+\frac{\sqrt{\pi } w^5}{120}\nn\\
   && +\frac{w^6}{210}-\frac{\sqrt{\pi }
   w^7}{672}-\frac{w^8}{1512}+\frac{\sqrt{\pi } w^9}{4320}+\frac{w^{10}}{11880}\nn\\
   && +O\left(w^{11}\right)
\end{eqnarray}
(\ref{tDWalpha=2}) together with (\ref{formdelta}) is checked numerically in section \ref {s:numerics toy}, see
figure \ref{f:data2}.

However for $\alpha$ non-integer the situation is more complicated. Despite the presence of $w^\alpha$ terms (we
recall $\alpha>1$), the one sided second derivative at $\Delta''(0^+)$ seems to exists. Numerically one finds
that for $\alpha$ non-integer, but close to 1, e.g.\ $\alpha=3/2$, the third derivative at 0 exists, but does
not go to zero with a finite slope. Analytically, one obtains an expansion around $\alpha=1$, as
\begin{equation}
\tilde \Delta(w) = \Delta_{0}(w) + (\alpha-1) \Delta_{1}(w) + O(\alpha-1)^{2}
\end{equation}
with
\begin{align}
\Delta_{0}(w)&=\rme^{-w}\\
\Delta_{1}(w)&= (w-2) \Gamma (0,w)+e^{-w} (-(w+2) \log (w)-4) \label{101}
\end{align}
with the incomplete $\Gamma$-function defined in (\ref{109}). Note that $$
\partial_{w}^{2}\,\Delta_{1}(w) =-e^{-w} (w \log (w)+1) $$
and the Taylor expansion of (\ref{101}) around 0 is then:
\begin{align}
\Delta_{1}(w)=&\, (-4+2 \gamma )+(2-\gamma ) w-\frac{w^2}{2}\nn\\& +\left[\frac{11}{36}-\frac{\log
(w)}{6}\right]
   w^3+\left[\frac{\log (w)}{12}-\frac{13}{144}\right] w^4\nn\\
   &+O\left(w^5\right)
\end{align}
whose logarithmic part can be summed up as:
\begin{align}
\Delta_{1}(w)=&\, \left[2-w- e^{-w} (w+2) \right] \log (w) \nn\\
&+ \mbox{analytic in }|w|
\end{align}
These expansions confirms that for the Weibull class at non-integer $\alpha$ the second derivative at $0^+$,
$\Delta''(0^+)$, exists, but not the third one.

Finally let us note that if $\rho_m^{\mathrm{III}}$ is scaled as $\alpha$ one should recover class I from the
large-$\alpha$ limit of class III, i.e.\ one can indeed check from (\ref{set}) and (\ref{class3def}) that
$\lim_{\alpha \to \infty} \alpha^2 \tilde \Delta^{\mathrm{III},\alpha}(w/\alpha) = \tilde
\Delta^{\mathrm{I}}(w)$. (the indices refer to the class). An example of these limit procedure is given below.

\subsubsection{Calculation of $\Delta(w)$ for class II (Frechet)}
We define again $a_1=a_w(j)$, $a_2=a_{w'}(j)$, $a_3=a_{w'}(j')$ with
\begin{eqnarray}
 j-w &=& - a_1^{- 1/\alpha}\\
   j- w' &=& -a_2^{- 1/\alpha} \\
 j'-w' &=& - a_3^{- 1/\alpha}    \label{class3def}
\end{eqnarray}
One sees that $j$ can only vary in the interval $[-\infty, w]$, and that $j'>j$ is equivalent to $a_3>a_2$. As
for Weibull, there are two pieces:
\begin{eqnarray}
d_{1}&:=& \overline{(j-w)(j'-w')}|_{j'>j}  \\
&=&  \int_{-\infty}^w \rmd j\, \frac{\rmd (a_1-a_2)}{\rmd j\,} \rme^{- (a_1-a_2)} \int_{a_2}^\infty \rmd a_3\,
\rme^{-a_3} (a_1 a_3)^{\frac{-1}{\alpha}} \nn
\end{eqnarray}
where the relation $W=w'-w=a_2^{-1/\alpha} - a_1^{-1/\alpha}$ holds. This yields:
\begin{eqnarray}
 d_1 &=&  \int_{0}^\infty \rmd a_1 \left(1 - \frac{\rmd a_2}{\rmd a_1}\right) a_1^{-1/\alpha} \rme^{-a_1+a_2}
\Gamma(1-{\textstyle\frac{1}{\alpha}},a_2)\qquad~~
\end{eqnarray}
The second contribution is:
\begin{eqnarray}
 d_2 &:=& \overline{(j-w)(j'-w')}|_{j'=j} \nn\\
 & =& \int^{w}_{-\infty} \rmd j\,
\frac{\rmd a_2}{\rmd j\,} \rme^{-a_1} (a_1 a_2)^{- 1/\alpha} \nn\\
&=& \int_{0}^\infty \rmd a_1 \frac{\rmd a_2}{\rmd a_1} \rme^{-a_1} (a_1 a_2)^{- 1/\alpha}
\end{eqnarray}
One has for the disorder correlator:
\begin{eqnarray}
 \tilde \Delta(W)=d_1 + d_2 - \left(\int_0^\infty \rmd a_1 \rme^{-a_1} a_1^{- 1/\alpha}\right)^{\!\!2}
\end{eqnarray}
Using the variable $y=a_1^{-1/\alpha}$, $a_2^{-1/\alpha}=y+W$, one finds:
\begin{eqnarray}\label{114}
 \tilde \Delta(W) &=&  \alpha \int_0^\infty \rmd y\, y (y+W)^{-\alpha} e^{-y^{- \alpha}}
\nn \\
&& + \alpha \int_0^\infty \rmd y \,y (y^{-1-\alpha} - (y+W)^{-1-\alpha})
\nn \\
&& \qquad ~~~\times e^{-y^{- \alpha} + (y+W)^{-\alpha}} \Gamma(1-{\textstyle \frac{1}{\alpha}},(y+W)^{-\alpha}) \nn\\
&& - \Gamma(1-{\textstyle \frac{1}{\alpha}})^2
\end{eqnarray}
As for the Weibull class, see (\ref{Delta-Weibull}), we can integrate by part the second term into:
\begin{align}\label{115}
&- \lim _{\Lambda\to \infty} \int\limits_0^ \Lambda \rmd y \,
e^{-y^{- \alpha} + (y+W)^{-\alpha}}\frac{\rmd }{\rmd y}
\left[ y \Gamma(1-{\textstyle \frac{1}{\alpha}},(y+W)^{-\alpha}) \right]\nn\\
& \qquad \qquad + \Lambda \Gamma(1-\textstyle \frac1\alpha )\ ,
\end{align}
where the last term comes from the upper bound in the partial integration, after putting $y\to \Lambda\to
\infty$ there. Rewriting it as $\int _{0}^{\Lambda} \rmd y \,\Gamma(1-1/\alpha)$, and using the fact that the
first term in (\ref{114}) cancels,  we arrive at the simple final expression for the Frechet class:
\begin{align}\label{116}
& \tilde \Delta(w) = - \Gamma(1-{\textstyle \frac{1}{\alpha}})^2\\
& + \int\limits_0^ \infty \rmd y\, \Gamma({\textstyle 1-\frac1\alpha})- e^{-y^{- \alpha} + (y+w)^{-\alpha}}
\Gamma(1-{\textstyle \frac{1}{\alpha}},(y+w)^{-\alpha}) \nn
\end{align}
We find a fixed-point function, which depends continuously on $\alpha$, hence on $\zeta=2+2/\alpha$ with a
unique form for each value of $\zeta$, in the domain $0<\zeta<2$.

\begin{figure}[tb]
\Fig{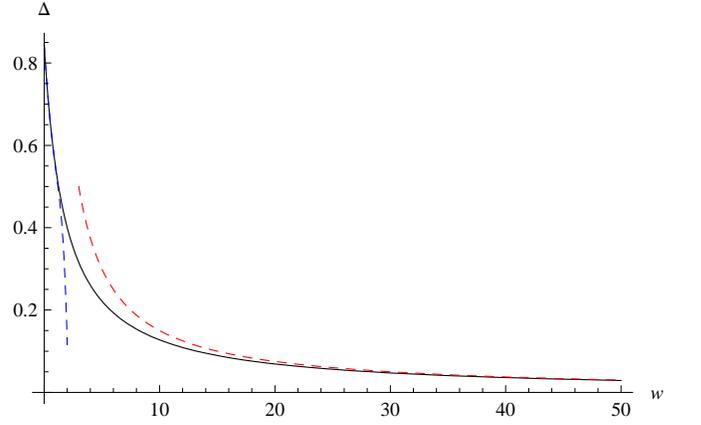} \caption{$\tilde \Delta(w)$ from eq.\ (\ref{116}) for $\alpha=3$ (bold). One clearly
sees the long tail for $w\to \infty$. The asymptotic behavior for large $w$ from eq.\ (\ref{118}) is shown
(dashed/red) as the small-$w$ expansion (blue/dashed).} \label{f:DeltaFrechetalpha=3}
\end{figure}

We can obtain the small-$w$ behavior easily from (\ref{114}). The value at zero,
\begin{eqnarray}
\tilde \Delta(0) = \Gamma(1 - {\textstyle \frac{2}{\alpha}}) - \Gamma(1-{\textstyle \frac{1}{\alpha}})^2\ ,
\end{eqnarray}
is found consistent with the results from the previous section on the distribution of the critical force,
\begin{eqnarray}
\overline{(w-u(w))^n} = \rho_m^n \Gamma(1 - {\textstyle \frac{n}{\alpha}})\ .
\end{eqnarray}
One sees that $\tilde \Delta(0)$ is defined for $\alpha>2$ and diverges as $\alpha \to 2^+$. As discussed in the
previous section this is because in the Frechet class the distribution of the critical force has algebraic tails and an
infinite $n$-th moment for $\alpha \leq n$. We thus consider $\alpha>2$. The fixed-point function has a cusp, with from (\ref{116})
\begin{equation}\label{94frechet}
 - \tilde \Delta'(0^+)=\frac{\Gamma\left(1-\frac{1}{\alpha}\right)}{\alpha}\ ,
\end{equation}
and a well defined Taylor expansion in $|w|$:
\begin{eqnarray}
\tilde \Delta(w)&=& \Gamma \left({\textstyle \frac{\alpha -2}{\alpha }}\right)- \Gamma(1-{\textstyle
\frac{1}{\alpha}})^2 +\frac{\Gamma
   \left(-\frac{1}{\alpha }\right) w}{\alpha
   ^2}+\frac{\alpha  w^2}{4 \alpha +2}\nn\\
   && -\frac{(\alpha +1)
   \Gamma \left(\frac{1}{\alpha }\right) w^3}{36 \alpha
   +24}+\frac{\alpha ^3 \Gamma \left(3+\frac{2}{\alpha
   }\right) w^4}{48 (2 \alpha  (4 \alpha
   +9)+9)}\nn\\
   &&+\frac{\alpha ^2 (\alpha +1) \left(\alpha
   ^2-4\right) \Gamma \left(2+\frac{3}{\alpha }\right)
   w^5}{240 \left(15 \alpha ^2+32 \alpha
   +16\right)}\nn\\
   &&+O\left(w^6\right)
\end{eqnarray}
The large-$w$ behavior of the fixed-point function is quite different from the other classes. Indeed  $\tilde
\Delta(w) $ decays to zero rather slowly for large $w$. This can be seen by writing
\begin{align}\label{117}
 \tilde \Delta(w) &= t_{1}(w)+t_{2}(w)- \Gamma(1-{\textstyle \frac{1}{\alpha}})^2\\
t_{1}(w)&=  \int_{0}^ \infty \rmd y\,  e^{-y^{- \alpha} + (y+w)^{-\alpha}} \int_{0}^{(y+w)^{-\alpha}}\rmd t\,
t^{-\frac 1 \alpha}\rme^{-t}\nn\\
t_{2}(w)&=  \int_0^ \infty \rmd y\, \Gamma({\textstyle 1-\frac1\alpha}) \left[1- e^{-y^{- \alpha} +
(y+w)^{-\alpha}}  \right] \nn
\end{align}
The leading term  for large $w$ comes  from $t_{1}(w)$, via a series of approximations:
\begin{align}
t_{1}(w)\approx& \int_{0}^ \infty \rmd y\,  e^{-y^{- \alpha} + (y+w)^{-\alpha}}
\int_{0}^{(y+w)^{-\alpha}}\rmd t\, t^{-\frac 1 \alpha}\nn\\
 =&\frac{\alpha }{\alpha -1} \int_{0}^ \infty \rmd y\,  e^{-y^{- \alpha} + (y+w)^{-\alpha}} (w+y)^{1-\alpha } \nn\\
 \approx &\frac{\alpha }{\alpha -1} \int_{0}^ \infty \rmd y\,  (w+y)^{1-\alpha } \nn\\
 =& \frac{w^{2-\alpha } \alpha }{(\alpha -2) (\alpha -1)}\ ,
 \label{118}
 \end{align}
whereas $t_{2}(w)$ is of order $1/w^{\alpha}$ plus a constant, since one can simply expand the exponential
function for large $w$. For $\alpha=3$ the function $\tilde \Delta(w)$ and the asymptotics (\ref{118}) are
plotted on figure \ref{f:DeltaFrechetalpha=3}.

\section{Avalanche-size and waiting-time distributions}
\label{s:avalanches}
\subsection{Avalanche-size distribution}
\label{ss:avalanches}
\begin{figure}[tb]
\begin{center}
\Fig{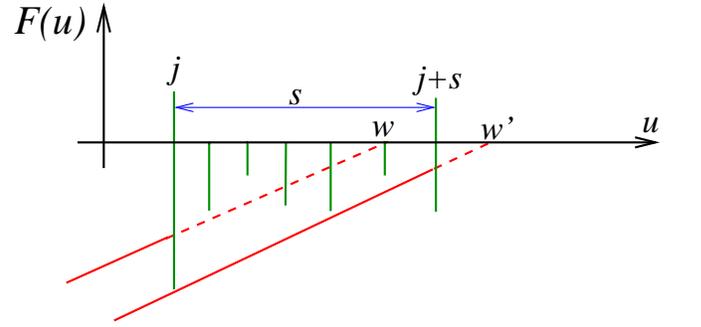} \caption{Geometrical construction for the size of an avalanche} \label{f:avalanche}
\end{center}
\end{figure}
Successive avalanches, or jumps, occur at a discrete set of $w=w_i$ such that $u(w_i^-)=j_i$ and
$u(w_i^+)=j_i+s_i$ where $s=s_i>0$ is the {\em size of the avalanche}. The {\em waiting time} between
consecutive avalanches is denoted $W=w_{i}-w_{i-1}$. It is not properly a time, but we will term it here loosely
waiting time since for a driving with a constant velocity it is the waiting time $t_w=W/v$, with here $v=0^+$
(in that limit the jump time is negligibly shorter). Here we compute the joint distribution of avalanche sizes
and waiting times. In section \ref{s:numerics toy} we discuss an algorithm to generate the sequence of
avalanches and the Markov-chain property.

There are two useful probabilities for which the general expressions for the discrete model are easy to write.
The first is, for $w'>w$:
\begin{align}
&P_w(j;w') := \\
&~~ \mbox{Proba}( u(w)=j ~ \text{and next avalanche is in} ~ [w' , \infty] )\nn
\end{align}
When $u(w)=j$ the next avalanche occurs at $w''>w$ such that $m^2 (j-w'')=F_j$. Thus to realize $w''>w'$ we need
$F_j<m^2 (j-w')$. Hence one has:
\begin{align}
 &P_w(j;w') = (1- H(m^2 (j-w'))  \prod_{k=-\infty}^{j-1}
H(m^2(k-w))\ .
\end{align}
For $w=w'$,  $P_w(j;w)=P_w(j)$ defined in (\ref{d33}). Of interest is the differential waiting time distribution
\begin{align}
&P_w(w') \rmd w'= - \rmd w' \partial_{w'} \sum_{j=-\infty}^{\infty} P_w(j;w')\\
&~~= \mbox{Proba}( \text{given $w$, the next avalanche is in} ~ [w' , w'+\rmd w'] ) \nn
\end{align}
The joint (integrated) probability is more involved:
\begin{align}
P_w(j,w',S) \rmd w' : =& \nn\\
\mbox{Proba}( u(w)=j ~ &\text{and} ~ \text{next avalanche is in} ~ [w',w'+\rmd w'] ~ \nn\\ &\text{and} ~
\text{of size} ~ s>S ~  )
\end{align}
One has:
\begin{align}
 P_w(j,w',S) &= \prod_{p=1}^{S} H(m^2(p+j-w')) (- \partial_{w'} P_w(j;w')) \label{jointdiscrete}
\end{align}
and of course $P_w(j,w',S=0)=- \partial_{w'} P_w(j;w')$, i.e.\ the probability that $u(w)=j$ and that the next
avalanche occurs at $w'$ (per unit $dw'$). Of particular interest is:
\begin{align}
 P_w(w',s>S)&:= \sum_j P_w(j,w',S)  \\
& \,\,= \mbox{Proba}(\text{next avalanche is in} ~ [w',w'+dw'] ~ \nn\\
&\text{\hspace{2cm} and} ~ \text{of size} ~ s>S ~  )\nn
\end{align}
This is illustrated on figure \ref{f:avalanche}.

We now consider the limit $m \to 0$ in which the continuum limit can be used, and extract the distributions of
waiting times and avalanche sizes. We find that they depend on the same single scale $\rho_m$ as defined in the
previous Sections. The waiting time distribution is denoted below $P(w)$ (and sometimes $P(W)$), and has thus
the scaling form:
\begin{eqnarray}
P(w)=\rho_m^{-1} \tilde P(w/\rho_m)
\end{eqnarray}
However, to simplify notations, unless specified, we drop below the tilde on $\tilde P$ and formally replace
$\rho_{m} \to 1$ (formally, since all statements below are about scaling forms in the limit $m \to 0$). When no
confusion is possible we use the symbol $w$ for either the argument of $u(w)$ and the waiting time
$w=w_i-w_{i-1}$, and use $W$ for the waiting time when confusion is possible. Similarly the avalanche size
distribution is loosely noted with the same symbol:
\begin{eqnarray}
P(s)=\rho_m^{-1} \tilde P(s/\rho_m)
\end{eqnarray}
and we perform the same simplification in notations, and similarly for the joint distribution. The dependence on
$\rho_{m}$ can be restored by replacing in the final formulas
\begin{eqnarray}
&& j \to \frac{j}{\rho_{m}}  \\
&& w \to  \frac{w}{\rho_{m}} \quad , \quad s \to  \frac{s}{\rho_{m}}
\end{eqnarray}
similarly for $S$ and $W$, and correcting the normalizations of probabilities to $1$.

\subsection{Waiting-time distribution}
Using the method introduced in the previous section, see especially Eq.~(\ref{20})~ff., one finds that the
continuum limit for the probability $P_w(j;w')$ is
\begin{equation}
P_w(j;w') \rmd j = \rmd a_{w'}(j) e^{-a_w(j)}\  , \label{solu1}
\end{equation}
and we use again the notation $a_1=a_w(j)$ and $a_2=a_{w'}(j)$.

\subsubsection{Class I (Gumbel)}
We recall (\ref{71}) for class I
\begin{eqnarray}
 j-w &=& - f_c^0 +  \ln a_1 \nn \\
 j-w' &=& - f_c^0 +  \ln a_2   \label{rel0}
\end{eqnarray}
where $f_c^0$ is a non-fluctuating constant, and as in (\ref{set}) $a_2=e^{- W} a_1$. Thus one finds:
\begin{eqnarray}
P_w(j;w') \rmd j = e^{-W} \rmd a_1\, \rme^{-a_1}
\end{eqnarray}
Integrating over $a_1$ in $[0,\infty[$ one obtains
\begin{eqnarray}\label{127}
P_w(w') =  \theta(w'-w)\, \rme^{w-w'}\ .
\end{eqnarray}
\begin{figure}
\Fig{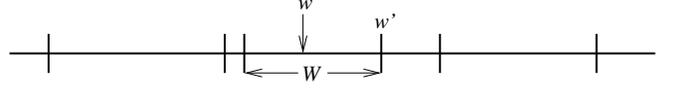} \caption{The probability  $P_{w}(w')$ to have a point $w$ in the interval of size $W$
preceding $w'$.} \label{f:explainPwwp}
\end{figure}%
From (\ref{127}) we can infer the distribution $P(W)$ of waiting times $W$.
Since the probability that a uniformly chosen $w$ on the real axis falls in an interval of size $W$ is $W
P(W)/\int \rmd\,W W P(W)$ (see figure \ref {f:explainPwwp}), and that then the probability of of $w'-w$ is
uniform, i.e.\ $ \theta(0 < w'-w < W)/W$, multiplying the two expressions one finds the general relation:
\begin{eqnarray}
 \frac{\int_{w'-w}^\infty \rmd W\, P(W)}{\int_0^{\infty} \rmd W\, W P(W)}= P_w(w') \ .  \label{relW}
\end{eqnarray}
This is true only in the small-$m$ limit (because of uniform measure assumption). For any member of class I we
find the distribution of ''waiting time'' $W$ (restoring the dependence on $\rho_{m}$):
\begin{eqnarray}\label{wtI}
 P(W) = \rho_m^{-1} e^{- W/\rho_m}
\end{eqnarray}

\subsubsection{Class III (Weibull)}
Let us consider now class III. One has, following the notation in section \ref{KN1}:
\begin{eqnarray}
 j-w &=& - f_c^0 + a_1^{1/\alpha} \\
 j-w' &=& - f_c^0 + a_2^{1/\alpha} \ .
\label{rel1}
\end{eqnarray}
Using the variable $y>0$ such that $a_2^{1/\alpha}=y$ and $a_1^{1/\alpha}=y+W$ one now finds from (\ref{solu1}):
\begin{equation}
P_w(j;w') \rmd j =  \alpha \, y^{\alpha-1} e^{-(y+W)^\alpha}\, \rmd y\,
\end{equation}
Integrating w.r.t. $y$, taking $-\partial_{w'}$ and integrating by part we obtain:
\begin{equation}
 P_w(w') =  \alpha (\alpha-1) \int_0^\infty \rmd y\, y^{\alpha-2} e^{-(y+w'-w)^\alpha}
\end{equation}
We recall that $\alpha>1$ hence the integral is always convergent. From (\ref{relW}) we obtain:
\begin{equation}
 \langle W \rangle^{-1} = P_w(w) = \alpha \, \Gamma\!\left(2-\frac{1}{\alpha}\right)
\end{equation}
and, taking $-\partial_{w'}$ on both sides of (\ref{relW}) we obtain for class III:
\begin{equation}\label{134}
 P(W) =  \frac{\alpha(\alpha-1)}{\Gamma(2-\frac{1}{\alpha})} \int_0^\infty \rmd y \;y^{\alpha-2} (y+W)^{\alpha-1} e^{-(y+W)^\alpha}
\end{equation}
If $\alpha>2$ it can be integrated by part into
\begin{equation}\label{135}
P(W) =  \frac{(\alpha-1)(\alpha-2)}{\Gamma(2-\frac{1}{\alpha})} \int_0^\infty \rmd y\, y^{\alpha-3}
e^{-(y+W)^\alpha}  \ .
\end{equation}
For completeness let us give the result for $\alpha=2$, which corresponds to a  box distribution for the forces:
\begin{eqnarray}\label{136}
 P_w(w') &=&  \sqrt{\pi} \,\text{erfc}( W) \\
 \label{137}
 P(W) &=& \frac{2}{\sqrt{\pi}} e^{- W^2} \theta(W)
\end{eqnarray}
which is a simple one-sided Gaussian.

\subsubsection{Class II (Fr\'echet)}
Let us consider now class II (Frechet). One has:
\begin{eqnarray}
j-w &=& -  a_1^{-1/\alpha} \\
j-w' &=& -  a_2^{-1/\alpha} \label{rel1-b}
\end{eqnarray}
Using the variable $y>0$ such that $a_1^{-1/\alpha}=y$ and $a_2^{-1/\alpha}=y+W$ one now finds from
(\ref{solu1}):
\begin{eqnarray}
P_w(j;w') \rmd j =  \alpha  (y+W)^{-\alpha-1} \rme^{- y^{-\alpha}}\rmd y
\end{eqnarray}
Integrating w.r.t. $y$, taking $-\partial_{w'}$ we obtain:
\begin{eqnarray}
 P_w(w')
=  \alpha (\alpha+1) \int_0^\infty \frac{\rmd y}{(y+W)^{2+\alpha}}\, \rme^{- y^{-\alpha}}
\end{eqnarray}
We recall that $\alpha>0$ hence the integral is always convergent. From (\ref{relW}) we get
\begin{eqnarray}
 \langle W \rangle^{-1} = P_w(w) = \alpha \, \Gamma\! \left(2+\frac{1}{\alpha }\right)\ .
\end{eqnarray}
Deriving again (\ref{relW}) w.r.t.\ $W$, we obtain
\begin{equation}\label{144}
P(W) =  \frac{(\alpha+1)(\alpha+2)}{\Gamma(2+\frac{1}{\alpha})}\int_0^\infty \frac{\rmd y}{(y+W)^{3+\alpha}}\,
\rme^{- y^{-\alpha}}    \ .
\end{equation}
One easily checks that (\ref{144}) is correctly normalized.

\subsection{Joint avalanche-size and waiting-time distribution}
The continuum limit of Eq.~(\ref{jointdiscrete}) can be written as
\begin{align}
P_w(j,w',S) \,\rmd j =  [ - \partial_{w'} ( \rmd a_2\, \rme^{-a_1} ) ] \rme^{-a_3+a_2}\ ,
\end{align}
where we use the same notations $a_w(j)=a_1$, $a_{w'}(j)=a_2$ as above and in addition $a_{w'}(j'=j+S)=a_3$. We
remind that in order to include $\rho_{m}$, $S$ gets rescaled as $j$ and $w$.

\subsubsection{Class I (Gumbel)}
Consider now class I. In addition to (\ref{rel0}) one has
\begin{eqnarray}
 j+S-w' = - f_c^0 +  \ln a_3\ .
\end{eqnarray}
Hence we have $a_2 = e^{-W} a_1$ and $a_3 = e^{-W - S} a_1$ where
\begin{align}
 P_w(j,w',S) \rmd j &=  [ - \partial_{w'} e^{-W} ( \rmd a_1\, \rme^{-a_1} ) ] \rme^{-a_1 \rme^{-W} (\rme^{S}-1)} \nn \\
&= e^{-W} \rmd a_1\, \exp\big( -a_1 [1 + e^{-W} (e^{S}-1)] \big)
\end{align}
Integrating over $j$ and $a_1$ respectively, we obtain the joint distribution:
\begin{equation}
 P_w(w',s>S) = \theta(w'-w)\, \frac{{1}}{e^{w'-w} + e^{S} - 1}
\end{equation}
Again this is the probability that if one observes the system at $w$, the next avalanche occurs at $w'$ {\it
and} has size $s>S$. This allows to find the joint probability $P^+(W,s>S)$ that $w_{n+1}-w_n=W$ and the {\em
next} avalanche is $s_{n+1} >S$ (see Appendix \ref{s:numerics toy} for further notations and definitions).
Indeed one has, by the same reasoning as above \footnote{Note one could also imagine to divide by
$\int_0^{\infty} dW W P^+(W,s>S)$ but this is wrong.}:
\begin{equation}
 \frac{\int_{w'-w}^\infty \rmd W\, P^+(W,s>S)}{\int_0^{\infty} \rmd W\, W P^+(W,0)}= P_w(w',s>S)
\end{equation}
This yields for class I:
\begin{equation}
P^+(W,s>S) = \frac{ \rme^{W} }{(\rme^{W} + e^{S} - 1)^2}
\end{equation}
Integrating over $W$ this yields the cumulative joint waiting-time and avalanche-size distribution for class I:
\begin{equation}\label{151}
P(w>W,s>S) =  \frac1{\exp(W)+\exp(S)-1}\ .
\end{equation}
Setting $W=0$ gives the (rescaled) avalanche-size distribution for class I:
\begin{equation}\label{152}
P(s) =   e^{- s} \ , \qquad P(s>S) = e^{- S}\ .
\end{equation}
Setting $S=0$, and deriving w.r.t.\ $W$ reproduces the waiting-time distribution (\ref{wtI}). We thus obtain
that the avalanche exponent, such that $P(s) \sim s^{- \tau}$ at small $1 \ll s \ll \rho_m$, is here $\tau=0$.
We can give the lowest moments:
\begin{align}
\left< w s \right> &= \frac{\pi^{2}}6\approx 1.64493\\
\left<w \right>&=\left<s \right> = 1\\
\left<w^{2} \right>&=\left<s^{2} \right> =2\ .
\end{align}
in units of $\rho_m$. We also note that the relation:
\begin{eqnarray}
\frac{\left<s ^{2}\right>}{\left< s \right>} = -2\Delta'(0^{+})\ .
\end{eqnarray}
is obeyed, using $\tilde \Delta'(0^{+})=-1$ from (\ref{p1}). This relation between the cusp and the second
moment holds quite generally \cite{LeDoussalMiddletonWiese2008} and is used here as a useful check.

\subsubsection{Class III (Weibull)}
Consider now class III, i.e.\ (\ref{rel1}) and
\begin{equation}
 j+S-w' = - f_c^0 + a_3^{1/\alpha}
\end{equation}
This leads to
\begin{eqnarray}
 P_w(j,w',S) \rmd j &=&  \left[- \partial_{w'}\frac{\rmd a_{w'}(j)}{\rmd j}\rmd j\, \rme^{-a_{w(j)}}  \right] \nn\\
 && \times e^{-a_{w'}(j+S)+a_{w'}(j)}
\end{eqnarray}
Now setting as before $y=a_{2}^{1/\alpha} = j-w'$, the only derivative in the bracket to be taken is of the
variable $y$, i.e.\ $-\partial_{w'} y =1$. In short-hand then gives
\begin{eqnarray}
 P_w(j,w',S) \rmd j &=&  - \partial_{w'} \alpha y^{\alpha-1} \rmd y\, e^{-(y+W)^\alpha - (y+S)^\alpha  + y^\alpha} \nn\\
 &=&  \alpha (\alpha-1) y^{\alpha-2} \rmd y\, e^{-(y+W)^\alpha - (y+S)^\alpha  + y^\alpha} \nn\\
\end{eqnarray}
Integrating over $y$ (i.e.\ $j$) it yields
\begin{equation}
 P_w(w',s>S) = \alpha(\alpha-1) \int_0^\infty \rmd y\,
y^{\alpha-2} e^{-(y+W)^\alpha - (y+S)^\alpha  + y^\alpha}\ .
\end{equation}
The final result for the joint probability for class III takes the form:
\begin{eqnarray}\label{156}
&& \!\!\!P^+(W,s>S) \\
&&\!\!\!\quad = \frac{\alpha}{\Gamma(1-\frac{1}{\alpha})}  (-\partial_{W}) \int_0^\infty \rmd y\, y^{\alpha-2}
e^{-(y+W)^\alpha - (y+S)^\alpha  + y^\alpha}\nn\ .
\end{eqnarray}
Integrating over $W$ yields the (rescaled) cumulative distribution for class III:
\begin{eqnarray}\label{156b}
&& \!\!\!P(w>W,s>S) \\
&&\!\!\!\quad = \frac{\alpha}{\Gamma(1-\frac{1}{\alpha})}  \int_0^\infty \rmd y\, y^{\alpha-2} e^{-(y+W)^\alpha
- (y+S)^\alpha  + y^\alpha}\nn\ .
\end{eqnarray}
Setting $W=0$, we obtain the avalanche-size distribution.
\begin{equation}\label{157}
 P(s>S) =   \frac{\alpha}{\Gamma(1-\frac{1}{\alpha})} \int_0^\infty \rmd y\,  y^{\alpha-2} e^{- (y+S)^\alpha }
\end{equation}
On the other hand, setting $S=0$ in  (\ref{156b}) gives back the waiting-time distribution (\ref{134}).

Let us comment these results. Since (\ref{157}) can be Taylor expanded in $S$ around $S=0$, it is clear that for
class III also the avalanche exponent is again $\tau=0$. For large $S$, the decay of (\ref{157}) is $P(s>S) \sim
P(S) \sim \exp(-S^{\alpha})$, i.e.\ a stretched exponent decay with exponent $\delta=\alpha$. Next, one checks
that the general relation involving the cusp is obeyed:
\begin{align}
\frac{\left<s ^{2}\right>}{\left< s \right>} &= \frac{\Gamma \left(2+\frac{1}{\alpha }\right)}{\alpha } \rho_m =
-2 \Delta'(0^{+})\ .
\end{align}
using (\ref{94}), a useful check on our calculations.

One can also obtain simple expressions for the lowest moments (in units of $\rho_m$):
\begin{align}
\left<s \right> &= \left<w \right> = \frac{1}{\alpha  \Gamma \left(2-\frac{1}{\alpha }\right)}
\\
\left< s w \right> & = \frac{\Gamma (1+\frac1\alpha) H_{-\frac{1}{\alpha }}}{\Gamma \left(-\frac{1}{\alpha
}\right)}
\end{align}
as a function of the harmonic number $H_n=\sum_{k=1}^n 1/k$, i.e.\ $H_{-\frac1\alpha}:= \gamma_{\mathrm E} +
\psi(1-1/\alpha)$, $\psi(x):=\Gamma'(x)/\Gamma(x)$. From these one can construct a fully universal dimensionless
ratio:
\begin{align}
\frac{\left<s w \right>}{\left< s \right>\left< w \right>} &= -\frac{\pi  (\alpha -1)^2 H_{-\frac{1}{\alpha
}}}{\alpha
   ^2  \sin \left(\frac{\pi }{\alpha }\right)}
\end{align}

Finally one can check that for $\alpha \to \infty$, one recovers class I distribution (\ref{151}). More
precisely:
\begin{eqnarray}
&& \lim _{\alpha \to \infty} P(w> W/\alpha, s>S/\alpha) \\
&& = \lim _{\alpha \to \infty} \frac{1}{\Gamma(2-\frac{1}{\alpha})}  \int_0^\infty \rmd x\,
x^{-\frac 1 \alpha}  \nn \\
&& \qquad \times \exp\left( -x \left[ \left(1 +\frac{W}{\alpha x^{\frac1\alpha}}\right)^{\!\!\alpha} +\left(1
+\frac{S}{\alpha x^{\frac1\alpha}}\right)^{\!\!\alpha}  -1\right] \right)\nn
\\
&& = \int_0^\infty \rmd x\,\exp\left( -x \left[   \rme^{W}+\rme^{S}  -1\right] \right) = (\ref{151})\ ,
\end{eqnarray}
where we have used $x=y^{\alpha}$ as variable and $\lim _{\alpha \to \infty} (1+W/\alpha)^{\alpha} = \exp(W)$
and that the factors of $x^{1/\alpha}$ could be dropped.

\subsubsection{Class II (Fr\'echet)}
Consider now class II, i.e.\ (\ref{rel1}) and
\begin{align}
 j+S-w' = - \rho_m a_3^{-1/\alpha}
\end{align}
hence $S=a_2^{-1/\alpha} - a_3^{-1/\alpha}= y + W - a_3^{-1/\alpha}$. This leads to:
\begin{align}
P_w(j,w',S) \rmd j &=  [ - \partial_{w'} \alpha (y+W)^{-\alpha-1} \rmd y\, e^{- y^{-\alpha}} ]\nn \\
&\qquad \times e^{(y+W)^{-\alpha} - (y+W-S)^{-\alpha}} \nn\\
&= \alpha(\alpha+1) (y+W)^{-\alpha-2} \rmd y\, e^{- y^{-\alpha}} \nn \\
& \qquad\times e^{(y+W)^{-\alpha} - (y+W-S)^{-\alpha}}
\end{align}
The domain of integration depends on whether $W>S$ or $S>W$, and can be expressed as
\begin{align}
&P_{w}(w',s>S)\nn\\
&\qquad =\alpha(\alpha+1) \int_{\max(0,S-W)}^{\infty} \frac{\rmd y}{(y+W)^{\alpha+2}}\nn\\
&\qquad  \qquad\qquad\qquad\times e^{- y^{-\alpha}+(y+W)^{-\alpha} - (y+W-S)^{-\alpha}}\\
&\qquad =\alpha(\alpha+1) \int_{\max(W,S)}^{\infty} \frac{\rmd y}{y^{\alpha+2}}\nn\\
&\qquad  \qquad\qquad\qquad\times e^{- (y-W)^{-\alpha}-(y-S)^{-\alpha} + y^{-\alpha}}
\end{align}
This translates with the same arguments as for (\ref{156b}) into
\begin{align}\label{173}
&P(w>W,s>S)\nn\\
&\qquad =\frac{\displaystyle~\int \limits_{\max(W,S)}^{\infty} \frac{\rmd y}{y^{\alpha+2}}\, e^{-
(y-W)^{-\alpha}-(y-S)^{-\alpha} + y^{-\alpha}}} {\displaystyle\int\limits_{0}^{\infty} \frac{\rmd
y}{y^{\alpha+2}}\, e^{-  y^{-\alpha}}}
\end{align}
As for the Weibull and Gumbel universality classes, this expression is symmetric in $S$ and $W$. We therefore
conclude that for all three classes:
\begin{equation}
P(w)=P(s)\ .
\end{equation}
This property is proved with slightly more general argument in Appendix \ref{app:details}. It is of course valid
only for the {\it rescaled distributions} in the limit $m \to 0$ in the sense described above (at the level of
the discrete model $w$ is a continuous variable while $s$ is discrete).

The resulting distribution of (rescaled) avalanche size thus reads:
\begin{equation}\label{144s}
P(s) =  \frac{(\alpha+1)(\alpha+2)}{\Gamma(2+\frac{1}{\alpha})}\int_0^\infty \frac{\rmd y}{(y+s)^{3+\alpha}}\,
\rme^{- y^{-\alpha}}    \ .
\end{equation}
with (in units of $\rho_m$):
\begin{eqnarray}
&& \langle s \rangle^{-1}  = \alpha \, \Gamma\! \left(2+\frac{1}{\alpha }\right)\ . \\
&& \langle s^2 \rangle/ \langle s \rangle = \frac{- 2 \Gamma(-1/\alpha)}{\alpha^2} = - 2 \Delta'(0^+)
\end{eqnarray}
the last equality being a check, using (\ref{94frechet}). Note that $P(s)$ has power law decay for {\it large}
$s$, i.e.\ $P(s) \approx s^{-(2+\alpha)}$. The avalanche exponent however is still $\tau=0$, since it is related
to {\it small} avalanches (i.e.\ $s \ll \rho_m$). The Frechet class yields to a cutoff for large avalanches (i.e.
$s \gg \rho_m$) which is itself a power law.

\begin{figure}
\Fig{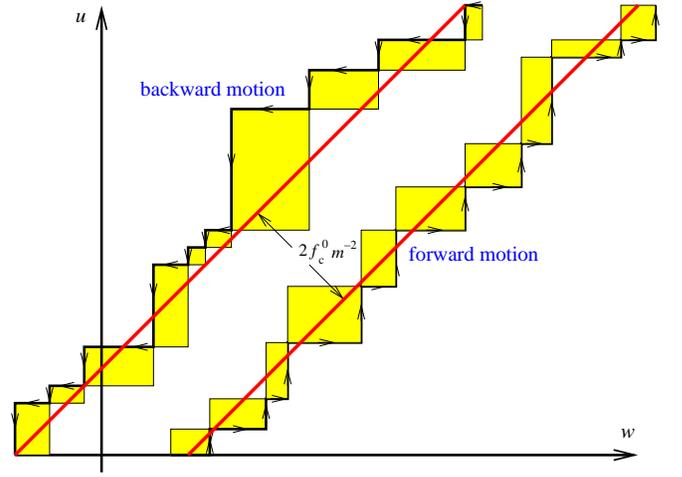} \caption{The plot shows
$u(w)$ for the forward motion with $m^2 \left<w-u\right>= f_{c}$, and the backward motion with $m^2
\left<u-w\right>= f_{c}$. Both trajectories are indicated with arrows.  The world line $u(w)$ fluctuates around
$w \pm f_{c}^{0} m^{-2}$. These fluctuations have a geometrical interpretation  as the yellow/shaded region,
whose area is related to the expectation $\left<w s \right>$. } \label{f:ExpUW}
\end{figure}

\subsubsection{Local fluctuations of the area of the hysteresis loop }

Figure \ref{f:ExpUW} illustrates a typical hysteresis loop for uncorrelated random force landscape, for
convenience assumed to have a symmetric distribution $P(-f)=P(f)$. The plot shows $u(w)$ for the forward motion
with $m^2 \left<w-u\right>= f_{c}>0$ (disorder or translational averages) and the backward motion $m^2
\left<u-w\right>= f_{c}$. Both curves $u(w)$ fluctuate around $w \pm f_{c}^{0}/m^2$, and the enclosed area $A$
of the hysteresis loop exhibits a uniform part $\bar A = 2 |f_c^0(m)| w $, computed in Section
\ref{sec:criticalforce} for each class, plus a fluctuating part $\tilde A$:
\begin{eqnarray}
 A = \bar A + \tilde A \ , \qquad \tilde A = m^2 \sum_i w_i s_i\ .
\end{eqnarray}
It has a geometrical interpretation as  the area of the yellow/shaded region. Its
average value per unit length over a large sample $w \in [0,M]$ is
\begin{eqnarray}
\frac{1}{M} \langle \tilde A \rangle = m^2 \frac{\left< w s \right>}{\left< s \right>}\ ,
\end{eqnarray}
since the number of avalanches is $\sum_i = N$, $\sum_i s_i = M$ hence $\left< s \right> = M/N$. Note en passant
that one has also $\sum_i w_i = M$ hence $\left< s \right> = \left< w \right>$, i.e.\ the first moments of $s$
and $w$ are always equal. (We neglect boundary contributions which for uncorrelated disorder scale
subdominantly). Similarly one can consider the moments of the local hysteresis area:
\begin{eqnarray}
 \frac{1}{M} \sum_i (w_i s_i)^p = \frac{\left< (w s)^p \right>}{ \left< s \right>}\ .
\end{eqnarray}
They can be obtained from the moments of the variable $a = w s$. Hence it is useful to compute the distribution of
this variable for the three classes. This is performed in Appendix \ref{app:details}.

Let us give the result for the (rescaled) distribution of the Gumbel class:
\begin{align}\label{PA3d}
P(a> A)&=
  \int\limits_{0}^{\infty}\rmd x  \, \frac{e^{x}}{\left(e^{x}+e^{A/x}-1\right)^{2}}\nn\\
  &=
  \int\limits_{1}^{\infty}\rmd y  \, \frac{1}{\left(y+e^{A/\ln(y)}-1\right)^{2}}
\end{align}
from which we give some moments:
\begin{eqnarray}
&& \left< a \right> = 1.64493 \quad , \quad
\left<a^{2}\right> = 18.3995 \ , \nn \\
&& \left<a^{3}\right> = 547.343 \quad , \quad  \left<a^{4}\right> =  30764.6\ ,
\end{eqnarray}
measured in units of $\rho_m^2$.

\section{Numerics for the toy model }\label{s:numerics toy}
\subsection{Basic definitions}
\begin{figure}
{\setlength{\unitlength}{1mm} \fboxsep0mm \mbox{\begin{picture}(86,54) \put(0,0){\fig{86mm}{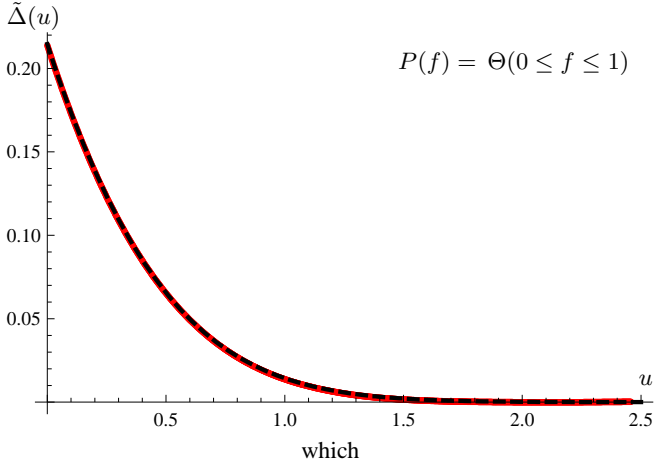}}
\put(0,54){$\tilde \Delta (u)$} \put(84,6){$u$} \put(52,48){$P (f) = \, \Theta(0\le f\le 1)$}
\end{picture}}}
which
\caption{Numerically calculated $\tilde \Delta (u)$  as a function of $u$ (red, fat) for the box distribution
$P(f)=\Theta(f)\Theta(1-f)$, and comparison with analytical result (black, dashed) from equation (\ref
{tDWalpha=2}); $\tilde A=1$, $\tilde \alpha =0$, $A=1/2$, $\rho_{m}=\sqrt 2/m$, $m^{2}=10^{-5}$; there is  no
adjustable parameter.} \label{f:data1}
\end{figure}
\begin{figure}
{\setlength{\unitlength}{1mm} \fboxsep0mm \mbox{\begin{picture}(86,54) \put(0,0){\fig{86mm}{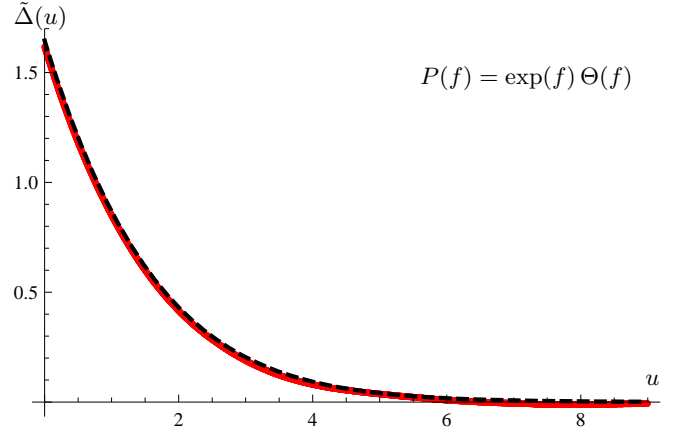}}
\put(0,54){$\tilde \Delta (u)$} \put(84,6){$u$} \put(54,46){$P (f) = \exp(f)\,\Theta(f)$}
\end{picture}}}
\caption{$\tilde \Delta (u)$ as a function of $u$ for the exponential distribution (red, fat), and comparison
with analytical result from equation (\ref{c97}) (black, dashed); $\rho_{m}^{-1}=m^{2}=0.003$, no adjustable
parameter.
} \label{f:data2}
\end{figure}

\begin{figure}
{\setlength{\unitlength}{1mm} \fboxsep0mm \mbox{\begin{picture}(86,48) \put(0,0){\fig{86mm}{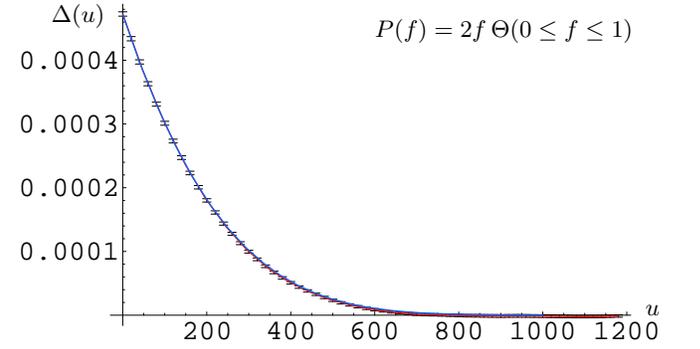}}
\put(5,47){$\Delta (u)$} \put(84,8){$u$} \put(48,45){$P (f) =2 f \, \Theta(0\le f\le 1)$}
\end{picture}}}
\caption{$\Delta (u)$ versus $u$ for $\alpha =3$ (Weibull class); $m^{2}=0.0001$. No adjustable parameter. Red
$=$ data with error-bars. Blue $=$ analytic solution (\ref{Delta-Weibull}) for $\alpha=3$. The other parameters
used in (\ref{49}) ff.\ are: $\tilde A = 1/\gamma$ and $\tilde \alpha = \frac{1}{\gamma}-1$. We have chosen
$\gamma =1/2$ as an example, hence $\tilde \alpha =1$, $\alpha =3$, $\tilde A=2$, $A = \frac{1}{3}$. We use
$m^2=0.0001$, with $10^{7}$ disorder points. One finds $f_{c}/m^{2}$: $-599.8$ (numerics) versus   $-597.79$
(analytic). The parameter-free numerical result for $\Delta(u)$ is compared with the analytical one, and found
in excellent agreement.} \label{f:alpha=3}
\end{figure}

\begin{figure}
{\setlength{\unitlength}{1mm} \fboxsep0mm \mbox{\begin{picture}(86,54) \put(0,0){\fig{86mm}{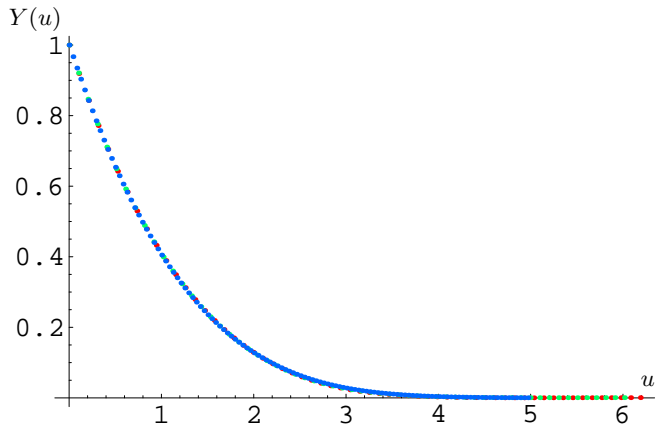}}
\put(0,54){$Y (u)$} \put(84,6){$u$}
\end{picture}}}
\caption{{\it Rescaled} disorder correlator: The function $Y(u)$ is the rescaled version of $\Delta(u)$, s.t.\
$Y(0)=1$ and $\int_{u} Y(u)=1$. The data is the same as on figures \ref{f:data1}, \ref{f:data2} with in addition
RB disorder (i.e.\ random potential): Blue is the box distribution in [0,1], red the exponential, green is RB.
This shows that for $m=0.003$ the different microscopic disorders yield very similar rescaled correlators,
although the unrescaled ones are different. For the difference between rescaled disorders see figure
\ref{f:Delta-diffs}.} \label{f:data3}
\end{figure}\begin{figure}[t]
{\setlength{\unitlength}{1mm} \fboxsep0mm \mbox{\begin{picture}(86,48) \put(0,-4){\fig{86mm} {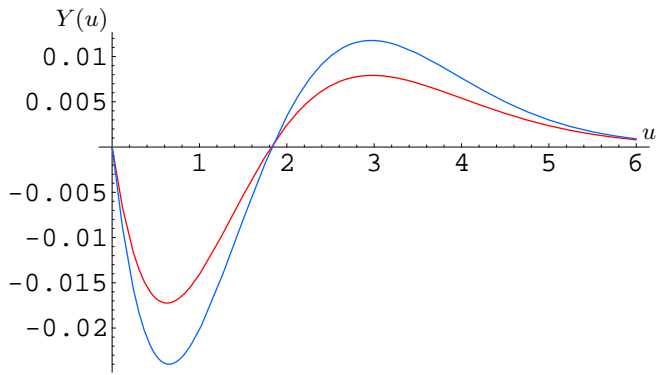}}
\put(7,46){$Y (u)$} \put(85,31){$u$}
\end{picture}}}
\caption{Comparison of the analytical results for the various classes. The function $Y(u)$ is the rescaled
version of $\Delta(u)$, s.t.\ $Y(0)=1$ and $\int_{u>0} Y(u)=1$. Plotted here are the (small) differences
$Y_{\mathrm{I}}(x)-Y_{\mathrm{III,2}}(x)$ (red) and $Y_{\mathrm{I}}(x)-Y_{\mathrm{III,3}}(x)$ (blue). The index
refers to class, and $\alpha$. This explains why the various classes are very close on figure \ref{f:data3}.}
\label{f:Delta-diffs}
\end{figure}

\begin{figure}
{\setlength{\unitlength}{1mm} \fboxsep0mm \mbox{\begin{picture}(86,54) \put(0,0){\fig{86mm}{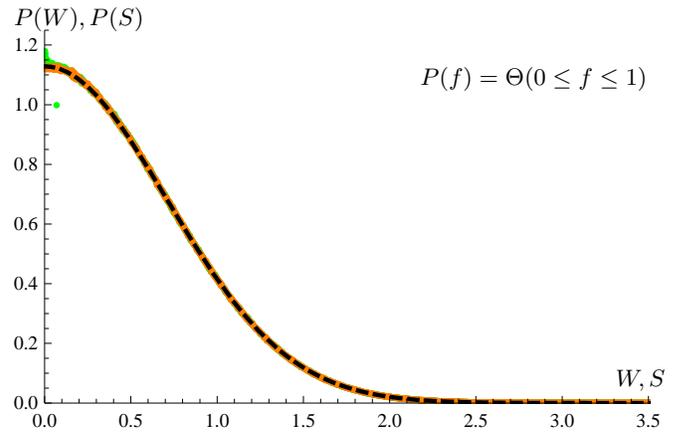}}
\put(0,54){$P(W),P(S)$} \put(80,6){$W,S$} \put(54,46){$P (f) = \Theta(0\le f\le 1)$}
\end{picture}}}
\caption{$P(W)$ (orange dots) and $P(S)$ (green dots) as a function of $W$ and $S$ respectively, for the  box
distribution $P(f)=\Theta(f)\Theta(1-f)$, and comparison with analytical result (black, dashed) from equation
(\ref {tDWalpha=2}); $\tilde A=1$, $\tilde \alpha =0$, $A=1/2$, $\rho_{m}=\sqrt 2/m$, $m^{2}=10^{-5}$; there is
no adjustable parameter. One sees that the two distributions are identical, with $P(W)$ plotted on top of
$P(S)$. However for small avalanche sizes $S$, the numerics has not yet converged against the analytical result,
i.e.\ finite-size corrections are present and visible on the plot.
} \label{f:PwPs}
\end{figure}

In this section we study numerically the discrete model of the last section. We consider four different disorder
distributions:

\leftline{\underline{Box}:}
\begin{equation}\label{k6}
P_{1} (f) = 1 \mbox{ if } f\in [0,1], \mbox{ 0 and else.}
\end{equation}

\leftline{\underline{Exponential}:}
\begin{equation}\label{k7}
P_{2} (f) = \exp (f) \mbox{ if }f>0\mbox{ and 0 else.}
\end{equation}

\leftline{\underline{RB (random bond)}:}

short-ranged correlated potential. The resulting force at site $i$ is $f_{i} = e_i - e_{i+1}$ where the energies
$e_i$ are uncorrelated random variables distributed with the box distribution $P_1(e)$. We call this distribution $P_3(f)$. 

\leftline{\underline{Class III with $\alpha=3$}:}
\begin{equation}\label{k8}
P_{4}(f)=2 f\, \Theta(0\le f\le 1) \ .
\end{equation}
Note that power law-distributed forces can be generated by defining $f := x^{\gamma}$, with $x \in [0,1]$
uniformly distributed. This yields
\begin{equation}\label{k3a}
P (f) =\frac{1}{\gamma} f^{\frac{1}{\gamma}-1} \Theta(0\le f\le 1)\ ,
\end{equation}
with $\gamma=1/2$ for (\ref{k8}).

We integrate numerically the equation of movement (\ref{d1}), first at $\eta=0$. This is the discrete model
defined in section \ref{s:SR force: discrete model}. In practice, for given $w$, we move the particle as long as
the force. i.e.\ the r.h.s.\ of equation (\ref{d1}) is positive. The point at which we stop defines $u(w)$. We
then update $w\to w+1$. This is an approximation to the process defined in section \ref{s:SR force: discrete
model}, but since jumps as well as waiting times diverge when $m\to 0$, the scaling limit is the same. A second
algorithm, described in Appendix \ref{app:markov}, was used to independently compute $P(w)$ and $P(s)$ (not
shown), and check the present results.

\subsection{$\Delta(w)$}
We have shown on figures \ref{f:data1}, \ref{f:data2}, and \ref{f:alpha=3},  comparisons between the numerically
computed functions $\Delta(u)$, and the analytical predictions. The corresponding analytical results are
referenced in the corresponding captions. Hence there is {\it no adjustable parameter} in figures \ref{f:data1}
to \ref{f:alpha=3} and the  agreement is excellent.

An general important question at depinning is whether the random bond class (i.e.\ uncorrelated potentials)
flows to the random field one (uncorrelated forces). This appears clearly in figure \ref{f:data3}, where we plot
the rescaled (as explained in the caption) $\Delta(u)$, for the 3 disorders $P_{1}(f)$,  $P_{2}(f)$, and
$P_{3}(f)$, defined in (\ref{k6}) ff.
The cross-over as the mass decreases from RB disorder to RF disorder can also nicely be seen in our simulations,
presented on figure \ref{f:RBRF}. One expects that the random short-ranged energy model, i.e.
$f_i=e_{i}-e_{i+1}$ with $e_i$ distributed with $P_{\mathrm{RB}}(e)$ should flow to the random force model with
$P(f)=\int_e P_ {\mathrm{RB}}(f+e) P_ {\mathrm{RB}}(e)$, i.e.\ the convolution of  $P_ {\mathrm{RB}}(e)$ and $ P_
{\mathrm{RB}}(-e)$. This is because the rare large forces are isolated and become uncorrelated. This predicts
that the box distribution for $e$ should flow to the $\alpha=3$ class III.

We have seen on Fig.\ \ref{f:data3} that the {\it shapes} of the correlator functions $\Delta(u)$ (i.e.\ their
rescaled form as explained in the caption) are rather similar for the various universality classes, while their
unrescaled forms are very different. These rescaled forms obtained from the analytical calculations are compared
in Fig.\ \ref{f:Delta-diffs}.

\subsection{Shocks and Avalanches}

In figure \ref{f:1} we have shown the avalanches, also called dynamical shocks: As a function of $w$, we plot
$w-u_{w}$ (minus its average), for different masses. First consider the smallest mass, $m^{2}=0.03$. We see that
$w-u_{w}$ is growing linearly with $w$, before it jumps. The linear parts are those, at which the particle is
localized by a large disorder force, before jumping (in zero time at $\eta v=0$) to a new position (vertical
parts).  When decreasing the mass, we see that the linear parts, i.e.\ the ``time'' (i.e.\ distance in $w$)
between jumps, as well as the jumps itself become larger, while sharing parts of their trajectories. This can be
interpreted as merging of the (dynamical) shocks.

In figure \ref{f:PwPs} we show the avalanche-size and waiting-time distributions for $P(f)$ being the box
distribution. It clearly shows that avalanche-size and waiting-times distributions are identical, as follows for
all disorder classes from (\ref{177}). Moreover, the result is in agreement with the parameter free prediction
of eq.\ (\ref{157}), using $\rho_{m}$ defined in equation (\ref{rho.m.Weibull}).

\subsection{Finite velocity}
In this section, we consider the equation of motion at finite velocity $v$, or rather at finite $\eta  v$, since
the latter is the parameter entering all equations.

The algorithm for finite $\eta v$ is as follows: We generalize the position $u_{t}\equiv u (w_t)$ of the point
to now take non-integer values. It follows the Langevin equation
\begin{equation}\label{D4a}
\eta \frac{\rmd}{\rmd t} u_{t} = f_{[u_{t}]}  + m^{2}(u_{t}-w_{t})
\end{equation}
where
\begin{equation}\label{D5}
[u] := \mbox{largest integer} \le u\ .
\end{equation}
Condition (\ref{D5}) reflects that the disorder only changes at discrete values of $u$.

\begin{figure}[t]
{\setlength{\unitlength}{1mm} \fboxsep0mm \mbox{\begin{picture}(86,50) \put(0,0){\fig{86mm}{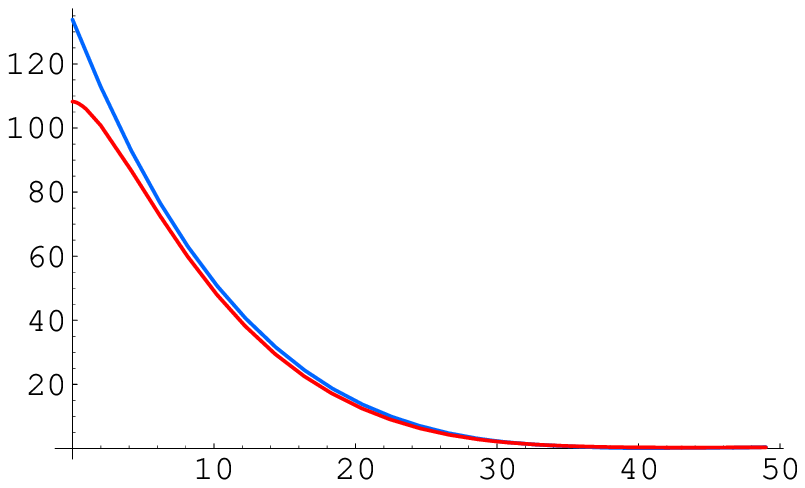}}
\put(0,50){$\Delta (u)$} \put(84,6){$u$} \put(14,16){$\eta v=0.2$} \put(16,36){$\eta v=0$}
\put(27,48){$\Delta' (u)$} \put(85,50){$u$} \put(32,18){\mbox{\fig{55mm}{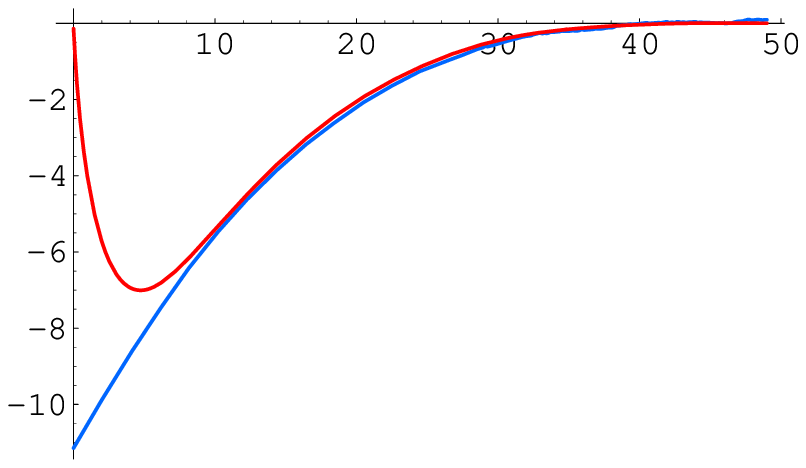}}}
\end{picture}}}
\caption{Rounding of $\Delta(u)$ through a finite velocity; $\eta v=0.2$. Inset: $\Delta'(u)$. The disorder
distribution is $P(f)=\Theta(0\le f\le 1)$, $m^{2}=0.003$.} \label{f:data4}
\end{figure}

In practice, we discretize (\ref{D4a}) with a step-size $\delta t=1/ 100$, integrating this discretized equation
of motion in time following the It\^o scheme, and using $w_{t}=v t$.
\begin{equation}\label{D4b}
\eta \left[u_{t+\delta t}-u_{t}\right] = \delta t \left[ f_{[u_{t}]}  + m^{2}(u_{t}-w_{t}) \right]
\end{equation}
To guarantee that our time discretization is fine enough, we  report ``maxslide'', the maximum of (\ref{D4b})
encountered in a simulation. For the simulation shown on figure \ref{f:data4}, this was 0.05. The figure shows
$\Delta(u)$ at $\eta v=0$, calculated with the discrete algorithm used in the previous sections, and $\Delta(u)$
at $\eta=1$, and $ v=0.2$. The microscopic disorder is a box distribution for the force, given by eq.\
(\ref{k6}). The result clearly shows a {\it rounding of the cusp} by the non-zero velocity.

\section{Long-range correlated forces}

\label{sec:abbm}

\begin{figure}[b]
{\setlength{\unitlength}{1mm}
\fboxsep0mm
\mbox{\begin{picture}(86,52)
\put(0,0){\fig{86mm}{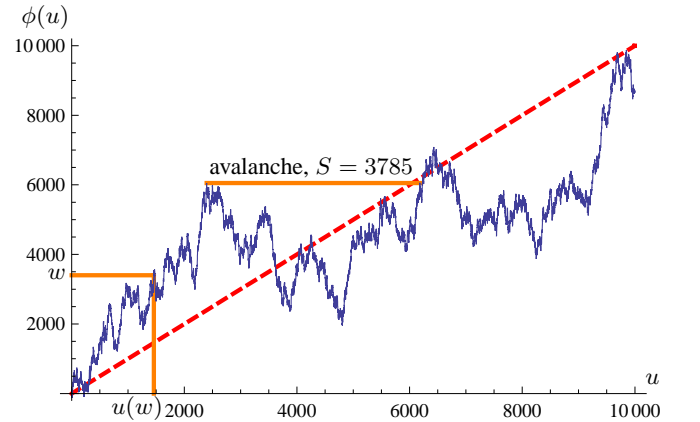}}
\put(1,53){$\phi(u)$}
\put(84,5){$u$}
\put(4.5,19){$w$}
\put(13,1){$u(w)$}
\put(26,33){avalanche, $S=3785$}
\end{picture}}}
\caption{$\phi(u)$ defined in (\ref{199}) for $f$ a random walk with $D=1.67 \times 10^{3}$ (blue). The dashed (red) line  denotes $\phi(u)=u$. We show explicitly the graphical construction of $u(w)$ for $w=3400$, as well as a larger avalanche of size $S=3785$. The system is discretized with stepsize 1 in $u$ direction.}
\label{f:f=RW}
\end{figure}
In addition to the three universality classes for short-range (SR) correlated forces, there is also a family of universality classes for long-range (LR) correlated forces. Consider a gaussian distributed force landscape with no bias $\overline{F(u)-F(u') }=0$ and second moment:
\begin{eqnarray}
\overline{[F(u)-F(u')]^2 } = 2 \sigma |u-u'|^\gamma   \label{corr1}
\end{eqnarray}
We focus on  $\gamma=1$, i.e.\ a Brownian-force landscape, but one expects a continuously varying fixed point as a function of $\gamma$. Although in most cases random-force landscapes at depinning have short-ranged correlations, these more exotic LR landscapes exhibit some interesting properties. Note that {\it in the statics} the (very) LR correlated random potential landscape corresponding to the case $\gamma=1$ was studied by Sinai \cite{sinairandomforce}. It was found that shocks are dense there.

Quite remarkably, exact results can be obtained for this model for any non-zero velocity $v>0$,
as first noticed by Alessandro, Beatrice, Bertotti and Montorsi \cite{abbm} who introduced this model, hence referred to as ABBM model, as a
realistic description of the Barkhausen effect in metallic ferromagnets, and compared the results to experiments.
The mass term originates from the magnetostatic fields: the demagnetizing field (resulting from the effective monopoles sitting at the end of the sample) provides a long-ranged restoring force which
acts as a spring, precisely as in the model considered
in this paper (see also discussions in \cite{ZapperiCizeauDurinStanley1998} and \cite{UrbachBarkhausen1995}).

We first obtain some results on the quasi-static model, then we recall some results of the
ABBM analysis at $v>0$, and obtain from it the renormalized correlator at non-zero velocity.
We then discuss how these results match at $v \to 0^+$.

\subsection{Quasi-static motion}
The forward process $u=u(w)$ is defined as the {\it smallest} root of
\begin{eqnarray}\label{199}
\phi(u) = u - m^{-2} F(u) = w\ .
\end{eqnarray}
For $\gamma=1$ the process $\phi(u)$ is a Brownian motion (BM) of diffusion constant $D=\sigma/m^4$ and upward
drift $b=1$. The time of the Brownian is $t\equiv u$. From Fig.\ \ref{f:f=RW} one sees that it becomes
a first-passage-time problem, i.e.\ $u=u(w)$ is the first time the process $\phi(u)$ reaches altitude $w$.
Conversely the process $w(u) = \max_{u' \leq u} \phi(u')$, is the maximum position over all previous times
reached by a brownian motion. To avoid pathologies we assume a cutoff which makes $F(u)$ smooth at very short
scales; equivalently we can discretize in $u$-direction, as was done to generate figure \ref{f:f=RW}.
In Appendix \ref{app:passage} we have collected some useful properties of first-passage times and maxima
of the Brownian motion that we now use extensively. We refer to this Appendix for all details.

\subsubsection{Avalanche distribution}
From Fig.\ \ref{f:f=RW} one sees that the avalanche distribution $P(s)$ identifies with the return probability to the origin of the BM with a drift $b=1$ and diffusion constant $D=\sigma/m^4$.
\begin{eqnarray}
 P(s) &\approx& P(s;W_0) \nn\\
 &=& \frac{s_0^{1/2}}{2 \sqrt{\pi} } s^{-3/2}
\exp\left( - \frac{s}{4 s_m} - \frac{s_0}{4 s} + \sqrt{\frac{s_0}{4 s_m}} \right)\ , \nn \\
\label{distribava}
\end{eqnarray}
where $u=s$, $P(u;W)$ is defined in (\ref{firstp}),
$b=1$ and $D=\sigma/m^4$. $W_0$ is a
non-universal short-distance scale. We have defined $s_0:= W_0^2/D$ and $s_m:=D$ the
short-scale and large-scale cutoffs for the avalanche size. In the limit of small $m$ one has $s_0 \ll s_m$, allowing to drop the last term in (\ref{distribava}).
There are many small avalanches of the order of $s_0$, i.e.\ the distribution is concentrated at $s_{0}$.
However the moments $\langle s^p \rangle$ for $p>1/2$ are dominated by large avalanches. For $s \gg s_0$ one has:
\begin{eqnarray}
 P(s) &=& \frac{\langle s \rangle}{2 \sqrt{\pi s_m}} \frac1{s^{3/2}} e^{- s/(4 s_m)} \\
 s_m &=& D=\sigma m^{-4}
\end{eqnarray}
and $\langle s \rangle = \sqrt{s_m s_0}$. This is exactly the distribution found in the mean-field theory of
sandpiles \cite{meanfield} and of the random-field Ising model, and was recently shown to hold,
using FRG \cite
{LeDoussalMiddletonWiese2008,us_inprep}, for elastic manifolds in $d=4$. The random-walk picture goes back to the so-called {\em Galton process} \cite{galton} for survival of family names (see \cite{fisher_review98} for a recent
discussion in the context of depinning) which exhibits the same mean-field power-law
behavior at the threshold.

\subsubsection{2-point conditional distribution}
It turns out that the 1-point-probability and critical-force distribution is a subtle issue for this model, due to the long-range nature of the landscape and the choice of boundary conditions. We do not discuss it in details here, but some considerations are given in the Appendix \ref{app:E}. A full solution requires a separate study.

We can still offer some simple remarks. If we know, e.g. by observation
in a numerical simulation or an experiment, that the process is such that
$u(w_1)=w_1$, then one can easily compute, from the Markov property of Brownian motion,
the probabilities of all {\it future} events, i.e.\ the {\it conditional probability} for
$u(w_2),\ldots,u(w_n)$ with $w_1 < w_2 <\ldots <w_n$:
\begin{eqnarray}
&& \!\!\!P_{w_2,..w_n}(u_2,..,u_n|u(w_1)=u_1) = \\
&& P(u_2-u_1;w_2-w_1) \times \ldots \times P(u_n-u_{n-1};w_n-w_{n-1})  \label{resmarkov} \nn
\end{eqnarray}
where $P(u;w)$ is the first-passage-time probability defined in (\ref{firstp}). Computing the moments one finds:
\begin{eqnarray}
 \overline{u(w_n)-w_n} &=& u(w_1)-w_1 \\
 \overline{[u(w_n)-w_n - u(w_p)-w_p]^2}^c
 &=& 2 D |w_n-w_p| \nn\\
 &=&  \frac{2 \sigma}{m^4}  |w_n-w_p|\qquad  \label{second}
\end{eqnarray}
for any $n,p \geq 1$. This defines the renormalized correlator:
\begin{eqnarray} \label{refs}
\Delta(0) - \Delta(w) =  \sigma w  \ ,
\end{eqnarray}
i.e.\ it is exactly the bare disorder correlator. Note that although this result was derived from a conditional
probability, it is independent of the choice of $u_1$ and $w_1$ {\it provided} all points $w_i$ to which it
applies are {\it larger} than $w_1$. Note also that although the 2-point correlator is the bare one, the higher
cumulants are different: they are non-trivial and can be easily computed from (\ref{resmarkov}). Hence the system
flows to a non-trivial fixed point. Remarkably, one can check that (\ref{refs}) is an exact fixed point of the
2-loop FRG equation for $\Delta'(u)$, i.e.\ the derivative of (\ref{frg2loop}) \footnote{To obey
(\ref{frg2loop}) one should take $\Delta(0)=+\infty$ from the long-range character, or consider a box-size in
$u$ space, which modifies the FRG analysis.} using the value $\zeta=\epsilon=4-d$ (here $\zeta=4$ which is the
correct value for the present model).

We can also check that (\ref{refs}) satisfies the general relation
\begin{equation}
 -\frac{ 2 \Delta'(0^+)}{ m^{4}} = 2 \frac{\sigma}{ m^{4}}
 = 2 s_m = \frac{\langle s^2 \rangle}{ \langle s \rangle}\ .
\end{equation}
In fact, we can also check some of the general relations discussed in
\cite{LeDoussalMiddletonWiese2008,us_inprep} for the so-called higher Kolmogorov cumulants, defined there: 
\begin{eqnarray}  \label{Kolmo}
 G(\lambda)&: =& \overline{\exp(\lambda [u(w)-w-u(w')-w')] - 1}  \nonumber \\
&=& \exp\left([w-w'][ \frac{1 - \sqrt{1-4 \lambda s_m}}{2 s_m}-\lambda]\right)-1 \nn   \\
&=& \left(\frac{1 - \sqrt{1-4 \lambda s_m}}{2 s_m} - \lambda\right) (w-w') + O((w-w')^2) \nonumber\\
\end{eqnarray}
for $w>w' \geq w_1$, using formula (\ref{laplace}) for the Laplace transform of the first passage time
probability with $b=1$ and $D=s_m$. This is related to the fact that the full distribution of avalanche sizes at
the tree level (i.e.\ in mean field) in the field theory coincides with the distribution of return times of the 1-dimensional Brownian motion.

\subsection{Motion at finite velocity $v>0$}
Let us now consider the case $w(t)= v t$, i.e.\ a particle pulled by a spring at constant velocity:
\begin{eqnarray}
\partial_t u(t) = F(u(t)) + m^2 (v t - u(t))\ .
\end{eqnarray}
For simplicity,  we set $\eta=1$(it can be restored by setting $t \to t/\eta$ and $v \to \eta v$). For $v>0$ and
since it is an over-damped equation of motion (no overshoot), the instantaneous velocity $v_t:=\partial_t u$ is
positive (possibly after a short transient), hence one can write $v_t={\sf v}(u(t))$, which satisfies:
\begin{eqnarray}
 \partial_t v_t &=& F'(u(t)) v_t + m^2 (v  - v_t) \label{first} \\
 \partial_u {\sf v}(u) &=& m^2 \left[\frac{v}{{\sf v}(u)}-1\right] + F'(u)
\end{eqnarray}
Since for the Brownian-force landscape $F'(u)$ is a white noise, the second equation is a standard Langevin
equation, hence the probability to observe ${\sf v}$ at $u$, given that the velocity is ${\sf v}_0$ at $u_0$
satisfies the Fokker-Planck equation:
\begin{eqnarray}
 \partial_u P &=& \partial_{\sf v} \left[ \sigma \partial_{\sf v} P + (\partial_{\sf v} E({\sf v})) P\right] \\
 E({\sf v}) &=& - m^2 v \ln {\sf v} + m^2 {\sf v}
\end{eqnarray}
with delta function initial condition. For $u \to \infty$ it converges to the equilibrium measure:
\begin{eqnarray}
 P_{\mathrm{eq}}({\sf v}) &=& Z^{-1} e^{- E({\sf v})/\sigma}\nn \\
 &=& \frac{(m^2/\sigma)^{m^2 v/\sigma}}{v \Gamma(m^2 v/\sigma)} {\sf v}^{m^2 v/\sigma} e^{- m^2 {\sf v}/\sigma}\qquad \label{distribP}
\end{eqnarray}
One can also directly work with (\ref{first}), rewriting it as a stochastic equation \cite{abbm}:
\begin{eqnarray}
 \rmd v_t = \rmd F(t) + m^2 (v  - v_t) \rmd t \label{first-2}\ .
\end{eqnarray}
$\overline{dF(t)^2} = 2 \sigma v_t dt $ is a Brownian motion up to a time reparametrization, and the factor $v_t$ can
be seen by writing $\int \rmd u (\frac{\rmd F(u)}{\rmd u})^2 = \int \rmd t v_t^{-1} (\frac{\rmd F(t)}{\rmd t})^2$. In It\^o
prescription this yields the Fokker-Planck equation for the probability $Q(v_t={\sf v}, t|v_0
,t_0)$  of velocity ${\sf v}$: 
\begin{eqnarray}
 \partial_t Q  &=& \partial_{\sf v} \left[ \sigma \partial_{\sf v} ({\sf v} Q) + m^2 ({\sf v} - v) Q \right] \label{eqQ} \\
& = &\partial_{\sf v} \left[ \sigma \partial_{\sf v} ({\sf v} Q) + {\sf v} (\partial_{\sf v} E({\sf v})) Q \right]
\end{eqnarray}
Hence the steady-state solution for $t \to \infty$ is $Q_{\mathrm{eq}} \sim {\sf v}^{-1} P_{\mathrm{eq}}$, the velocity factor
originating from the change of variable from $u$ to $t$. One has \cite{abbm}:
\begin{equation}
 Q_{\mathrm{eq}}({\sf v}) = \frac{(m^2/\sigma)^{m^2 v/\sigma}}{\Gamma(m^2 v/\sigma)} {\sf v}^{-1+ m^2 v/\sigma} e^{-
m^2 {\sf v}/\sigma} \label{distribQ}
\end{equation}
which yields the average velocity $\overline{{\sf v}}^Q=v$, as expected, and the connected expectation of the square of the velocity $\overline{{\sf v}^2}^{Q,c}=v
\frac{\sigma}{m^2}$. Note that the average velocity using $P$ is $\overline{{\sf v}}^P=v+\frac{\sigma}{m^2}$,
hence it does not even vanish as $v \to 0^+$: this is because most (in fact, as $m \to 0$, all) of the $u$
segments belong to avalanches, yielding a finite average velocity if weighted by $\rmd u$, but that the fraction of
time spent on them goes to zero, consistent with $v \to 0^+$.

ABBM also noted that the correlation function
\begin{equation}
 C(t-t_0):=\int \rmd v_t \rmd v_0 (v_t - v)(v_0 - v) Q(v_t, t|v_0 ,t_0) Q_{\mathrm{eq}}(v_0)
\end{equation}
satisfies the very simple equation (see also \cite{abbm2}):
\begin{equation}
 \partial_t C = - m^2 C  \label{simple}
\end{equation}
obtained by multiplying (\ref{eqQ}) by $(v_t-v)(v_0-v) Q_{\mathrm{eq}}(v_0)$, integrating over $v_t={\sf v}$ and $v_0$ and
using the fact that the current $J=\sigma \partial_{\sf v} {\sf v} Q + m^2 ({\sf v} - v) Q$ vanishes at ${\sf
v}=0$ and ${\sf v}\to \infty$. After integration by parts one obtains (\ref{simple}).

From this we can now obtain the renormalized disorder $\Delta$ at $v>0$ and discuss the crossover. One first
notes that from the definition (\ref{def-fc}), inserting $w= v t$ and $w'=v t'$ and taking two derivatives, one
has
\begin{equation}
 C(t-t')= \overline{(v_t-v)(v_{t'}-v)} = - v^2 m^{-4} \Delta''(v(t-t'))\ .
\end{equation}
Since from (\ref{simple}) and above, 
\begin{equation}
 C(t)= v \sigma m^{-2} e^{- m^2 t}\ ,
\end{equation}
we obtain
\begin{equation}
 \Delta''(w) = - \frac{\sigma}{v} m^2 e^{- m^2 w/v}\ .
\end{equation}
Integrating twice we finally get:
\begin{equation}
 \Delta(0)-\Delta(w) = \sigma w - \frac{\sigma v}{m^2} (1 - e^{- m^2 w/v})\ , 
\end{equation}
a formula valid for any $v>0$. The integration constant has been fixed by either of the two equivalent conditions: (i)  no cusp at $w=0$; (ii) the large-$w$ behavior is the same as in the bare model, and as in the statics, i.e.\ in the limit $v\to 0$, given by  (\ref{refs}). A non-zero velocity $v>0$ thus smoothens the
cusp in a boundary layer of size $w \sim v/m^2$, but the function remains non-analytic: there is a {\it subcusp},
i.e.\ a non-zero $\Delta'''(0^+)$; indeed one has at small $w$:
\begin{equation}
\Delta(0)-\Delta(w) = \frac{m^2 \sigma}{2 v} w^2 - \frac{m^4 \sigma}{6 v^2} w^3 + O(w^4)
\end{equation}
This indicates continuity of $u(w(t))$ but jumps in its derivative, the velocity. It remains to be understood
whether this feature is more general or if it is tied to the long-range nature of the random force landscape.

The distribution of avalanches times (in $t$) and sizes (in $u$) at $v>0$ can be extracted by studying the
returns ``near'' the origin of the process , i.e.\ the return to the origin inside the potential well
$E({\sf v})$. This is given by  (\ref{first}) or equivalently by 
\begin{equation}
 \rmd {\sf v} = m^2 \left(\frac{v}{\sf v}-1\right) \rmd u + \sqrt{2 \sigma}\, \rmd B\ ,
\end{equation}
where $\rmd B(u)$ is a standard unit Brownian motion.
 Near ${\sf v}=0$ we can first ignore the drift term $-m^2 \rmd u$. Define the change of variables $y=2
\sigma u$, then  ${\sf v}(u)=r(y=2 \sigma u)=\sqrt{\sum_{i=1}^{\tilde d} x_i^2(y)}$ is the norm of a
$\tilde d$-dimensional BM in the variable $y$ \cite{jpprivate2}, which satisfies:
\begin{eqnarray}
 \rmd r &=& \frac{\tilde d-1}{2 r} \rmd y + \rmd B(y) \\
\tilde d &=& 1 + \frac{m^2 v}{\sigma}\ .
\end{eqnarray}
This yields the avalanche-size exponent $\tau=2-\tilde d/2$, for $\tilde d \leq 2$, from the power law
decay of first return probabilites  of a Brownian near the origin, see appendix \ref{app:G}
\begin{eqnarray}
\tau = \frac{3}{2} - \frac{m^2 v}{2 \sigma}
\end{eqnarray}
below a critical velocity $v \leq \sigma/m^2$. This result was anticipated in Ref. \cite{durin,jpprivate}. Note
that the definition of avalanches at $v>0$ is not clear cut and requires a small velocity cutoff noted ${\bf
v}_0$. In Appendix \ref{app:avalanchedistrib} it is shown that:
\begin{eqnarray}
 P(s) &=& \frac{1}{\Gamma(\tau-1)} \frac{1}{s} \left(\frac{s_0}{s}\right)^{\tau-1} e^{-s_0/s} \\
 s_0 &=& {\sf v}_0^2/(4 \sigma)  \label{cutoff}
\end{eqnarray}
for ${\sf v}_0 \ll v$. Since the drift is neglected (\ref{cutoff}) holds only for $s \ll s_m=\sigma/m^4$ the
large scale cutoff. The drift provides a large velocity cutoff ${\sf v}_{>} = \sigma/m^2$ in (\ref{distribP}),
(\ref{distribQ}) and a large relaxation time cutoff $t_{>} = m^{-2}$, with $s_m = {\sf v}_{>} t_{>}$. For larger
velocity $v > \sigma/m^2$ the behavior changes qualitatively. It corresponds to $\tilde d>2$, see appendix \ref{app:G}, and the most probable velocity
 in $Q_{\mathrm{eq}}({\sf v})$ is no longer near ${\sf v}=0$.

More details and the solution including the drift term, are given in Appendix \ref{app:avalanchedistrib} for the
various regimes. In particular it is shown that one recovers the quasi-static size-distribution obtained above
in the limit $v=0^+$ \footnote{Note however that these results are for a pure Brownian landscape. If the force
landscape is smooth at short scale $u_0$, (\ref{cutoff}) holds only for $s_0 \gg u_0$. For $s_0 \leq u_0$ the
short-scale cutoff function for $P(s)$ is more complicated, but this has no consequence for the behavior for
$s_0 \ll s \ll s_m$ still given by (\ref{cutoff}). In the small $v$ limit, there is a further crossover velocity
$v_c = \sqrt{4 \sigma u_0}$ below which the short-scale cutoff function progressively becomes the one of the
quasi-static limit.}. For $v \leq \sigma/m^2$ the random walk ${\sf v}(u)$, in the continuum limit, comes back
infinitely often near the origin (i.e.\ near ${\sf v}_0$), hence the role of the drift term is mainly to cut off
the rare large avalanches, very much like in the statics (see e.g.\ the discussion  in
\cite{LeDoussalMiddletonWiese2008,us_inprep}). For $v
> \sigma/m^2$ the random walk in velocity space is still certain to come back near the origin
but only because of the drift. There are then two types of avalanches. In a fraction of them (computed in 
Appendix \ref{app:avalanchedistrib}) the instantaneous velocity ${\bf v}$ does not reach $v$: these avalanches are still described by the model
without the drift (first-return ``time'', conditioned to return) and lead to power law distributions. In the rest,
the velocity reaches $v$ and equilibrates in the well $E({\bf v})$; the ``time'' between two returns at
small ${\bf v}_0 \ll v$ can then be estimated as $\sim ({\bf v}_0)^{- m^2 v/\sigma}$ proportional to the
inverse equilibrium probability (either $P_{\mathrm{eq}}$ or $Q_{\mathrm{eq}}$ depending on whether one is interested in avalanche size or
duration). Typically there will be a bunch of small avalanches of the first kind separated by one of the second
kind. Eventually at larger velocities returns to the origin ${\bf v}_0 \ll v$ become very rare events and there is no
real sense in which one can talk about avalanches.

\section{Depinning and extreme statistics of records}
\label{sec:nomass}
\subsection{Model without a mass: records without drifts}
For a particle pulled through a random-force landscape it is also possible to consider the problem without a
parabola. The problem is easier to solve, but the correspondence with the FRG calculations is much less clear.
Let us give here some elementary results.

We now have to solve for the smallest root $u(f)$ of
\begin{eqnarray}
&& f = - F(u)
\end{eqnarray}
where $f$ is the applied force. We study the case where the force is continually increased. The process $u(f)$
then has jumps from $u_i$ to $u_{i+1}$ as the force crosses the values $f_i$, which form an increasing sequence.
These values are called the {\it record values} for the process $F(u)$, and the $u_i$ the {\it record times}.
Statistics of records thus naturally occurs in the physics of depinning. The problem is to find the running
maximum (i.e.\ the record) of an {\it unbiased} process, while in the case of a mass it had a drift. In the
absence of a drift the only scale in the problem is the system size $M$.

\subsubsection{Uncorrelated forces}
Let us start with the discrete model of uncorrelated forces studied in Section \ref{s:SR force: discrete model},
characterized by a force distribution $P_{f}(F)$ for each site. A similar problem was studied in \cite{laloux}
(section IV-B). There the probability distribution of the full record value sequence
$(f_1<f_2,\ldots<f_n,\ldots)$ was obtained for a semi-infinite line. It can be mapped onto a sum of independent
variables as follows: The sequence distribution can be obtained from
\begin{eqnarray}
 \Phi(f_n) = \sum_{i=0}^n a_i
\end{eqnarray}
for any $n>1$, where the $a_i$ are independent positive random variables,
each with an exponential distribution $P(a) \rmd a=e^{-a} \rmd a$. The function:
\begin{eqnarray}
 \Phi(f) = - \ln \int_{-\infty}^f  \rmd F\, P_{f}(F)
\end{eqnarray}
describes the tail of the distribution, here the smallest $f$. For stretched exponential tails, as in class I,
the growth is $f_n \sim n^{1/\gamma}$, while for power law tails, as in class II, the growth is exponential in
$n$.

Another set of results, remarkably universal, is known \cite{majumdar,krug} for
the probability $P(N|M)$ of the
number of records $N$, here equal to the number of jumps, for a system of size $M$ (notations are inverted as
compared to \cite{majumdar}). Then for an uncorrelated sequence of $F_i$ it was shown \cite{nevzorov} that at
large $M$
\begin{eqnarray}
 N = \ln M + \xi \sqrt{\ln M}\ ,
\end{eqnarray}
where $\xi$ is a univariate gaussian random variable  \cite{redner,math}. Hence the translationally-averaged avalanche size in absence of a mass should be
\begin{eqnarray}
 \frac{1}{N} \sum_{i=1}^N s_i = \frac{M}{N} \to \frac{M}{\ln M}\ ,
\end{eqnarray}
i.e.\ it is the typical avalanche size $\frac{M}{\overline{N}}$. In the language of records the avalanche sizes $s_i=u_{i+1}-u_i$ are the time
intervals between successive records, also called {\it record ages}. The translational average grows unboundedly with system size. Hence there
are very few avalanches and they are almost as large as the system. Note that, at variance with the results on
the sequence $f_i$, this result is {\it independent} of the distribution $P_{f}(F)$ for continuous distributions.

\subsubsection{Forces correlated as a random walk}

\begin{figure}[b]
{\setlength{\unitlength}{1mm}
\fboxsep0mm
\mbox{\begin{picture}(86,52)
\put(0,0){\fig{86mm}{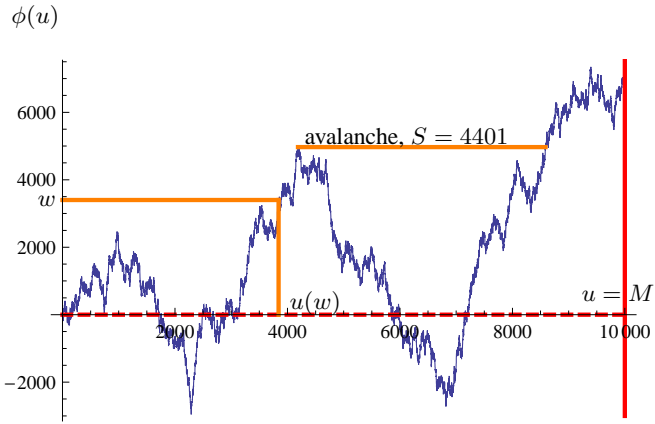}}
\put(1,53){$\phi(u)$}
\put(76.8,16){$u=M$}
\put(4.5,28.7){$w$}
\put(38,15){$u(w)$}
\put(40,37){avalanche, $S=4401$}
\end{picture}}}
\caption{$\phi(u)$ defined in (\ref{199}) for $f$ a random walk with $D=1.67 \times 10^{3}$ (blue). The dashed (red) line  denotes $\phi(u)=0$ (walk with no drift). We show explicitly the graphical construction of $u(w)$ for $w=3400$, as well as a larger avalanche of size $S=4401$. The system is discretized with stepsize 1 in $u$ direction.}
\label{f:f=RWunbias}
\end{figure}

In the case of a landscape obtained as a discrete-time random walk, $F_i=F_{i-1}-\eta_i$ where $\eta_i$ are
uncorrelated random variables drawn from the same {\it symmetric} continuous distribution $P(-\eta)=P(\eta)$, it was recently
obtained in Ref. \cite{majumdar} that
\begin{eqnarray}
 \sum_{M=N-1}^\infty P(N|M) z^M = \frac{(1-\sqrt{1-z})^{N-1}}{\sqrt{1-z}} \label{exact}\ .
\end{eqnarray}
Equivalently:
\begin{eqnarray}
 P(N|M) &=& 2^{-2 M + N -1}  \left( ^{2M-N+1}_{~~ M} \right)\nn\\
 & \sim&_{M \to \infty}
\frac{1}{\sqrt{\pi M}} e^{-N^2/(4 M)} \nonumber\ .
\end{eqnarray}
Hence for large sizes $M$, the average number of jumps behaves as \cite{majumdar}:
\begin{eqnarray}
 \overline{ N } \approx \frac{2}{\sqrt{\pi}} \sqrt{M}
\end{eqnarray}
There are also results for the jump sizes $s_i$, named {\em record ages} $l_i$ in Ref.\ \cite{majumdar}. The typical
jump size is $s_{\mathrm{typ}} = M/\overline{N} = \sqrt{\pi M}/2$ while the average maximal jump size is
$\overline{s_{\mathrm{max}}} = 0.626508 M$ and the average minimal jump size is $\overline{s_{\mathrm{min}}} = \sqrt{M/\pi}$
\cite{majumdar}.

\subsection{Model with a mass: records with a drift}

The usual problem of records with drifts \cite{krug} consists in studying the sequence:
\begin{eqnarray}
Y_i = X_i + c i
\end{eqnarray}
with $c>0$ where the $X_i$ are symmetric random variables. One way to present the correspondence to the
depinning model with a mass \footnote{One can also set $X_i=-F_i/m^2$ and $c=1$ as in Section \ref{sec:abbm}.} is
that
\begin{eqnarray}
X_i = - F_i \ , \qquad c = m^2\ .
\end{eqnarray}
The set of (upper) records $Y_{i_p}$, $p=1,\ldots, N$, $i=1,\ldots,M$, i.e.\ successive highest values, are the values $m^2
w_p$ at which a jump from $u_p=i_p$ to $u_{p+1}=i_{p+1}$ occurs in the process $u(w)$.

\subsubsection{Short-range correlations}

In the case of i.i.d.\ random variables with a drift it was shown that the total number of records $N$ up to time
$M$ grows linearly as $N \sim r(c) M$ with (normal) fluctuations which were characterized \cite{krug,math}.
However obtaining analytic results, even for $r(c)$, for a general distribution was found difficult and
some results were obtained only for special distributions $P(F)$ \cite{krug,math}. The function $r(c)$ is
related to the avalanche density $1/\left<s\right>=r(c)$, which is finite in presence of a mass, and is
computed here for small $m^2=c$. We solved the problem for arbitrary distributions $P(F)$ and found universality
in the small $c$-limit, with three classes. In addition we obtained the joint distribution $P(w,s)$ of (i) the
time $s$ between one record and the next; (ii) the difference in value $w$ with the previous record
\footnote{With the same method one can obtain the difference in value $w$ with the next record.}. These results
were given in  in Section \ref{s:avalanches}.

\subsubsection{Long-range correlations}

Let us now extend   the discussion of Ref.\ \cite{majumdar} to records with drift, i.e.\ depinning with a mass.
Again we consider the random walk $X_i=X_{i-1}+\eta_i$ with i.i.d.\ random variables $\eta=-F$ of distribution
$P_{f}(\eta)$. The alert reader will note that $P_{f}(\eta)$, $P_{\mathrm{ava}}(s)$ and $P(s)$ below denote three different
probabilities and functions. $P_{f}(\eta)$ produces a correlated sequence $X_n - X_0 = \sum_{i=1}^n \eta_i$. For $n
\geq 1$ we set:
\begin{eqnarray}
 P(n) &=& \mbox{Prob}(X_n < X_0 - c n) = \mbox{Prob}(Y_n < Y_0) \nonumber \\
 Q(n) &=& \mbox{Prob}(X_i < X_0 - c i, i=1,\ldots,n) \\
 &=& \mbox{Prob}(Y_i < Y_0 , i=1,\ldots,n)
\end{eqnarray}
The Sparre-Andersen theorem \cite{sparre1,sparre2,bauer} states that
\begin{equation}
Q(z):=\sum_{n=0}^\infty Q(n) z^n = \exp \Big( \sum_{n=1}^\infty \frac{P(n)}{n} z^n \Big) \ ,\label{sparretheo}
\end{equation}
setting $Q(0)=1$ by convention. We denote $F(n):=Q(n-1)-Q(n)$ the first passage probability that $Y_n$ crosses
$Y_0$ between steps $n-1$ and $n$. As in \cite{majumdar} the joint distribution of record ages (jump sizes)
$s_i$ and number $N$ of records is:
\begin{equation}
 P(\{s\}, N|M) = F(s_1) F(s_2) \cdots F(s_{N-1}) Q(s_N) \delta_{\sum_i^N s_i=N}
 \label{distribs}
\end{equation}
While for $c=0$, $P(n)=1/2$ independent of $n$, leading to (\ref{exact}) and the very universal results of
\cite{majumdar} quoted above, for $c>0$ the sequence $P(n)$, hence $Q(n)$, usually depends on the details of the
distribution $P(\eta)$. Hence apart from the asymptotic behaviour at large $n$ (hence $M$), one expects less
universality.

The following formula are still valid: The generating function for the probability to have $N$  records given $M$, $P(N|M)=\sum_{\{s\}} P(\{s\},N|M)$ can be
written as
\begin{equation}\label{247}
\sum_{M=N-1}^\infty z^M P(N|M) = F(z)^{N-1} Q(z)   \ ,
\end{equation}
where $F(z):=\sum_{n=1}^{\infty}F(n) z^{n}=1 - (1-z) Q(z)$. For instance the generating function for the average number of jumps is obtained by multiplying (\ref{247}) by $N$, and summing over $N$:
\begin{equation}\label{248}
\sum_{M \geq 0} z^M \overline{N}(M) = \frac{1}{(1-z)^2 Q(z)} \ .
\end{equation}
Similar results hold  for higher moments.

If one considers $P_{f}(\eta)$ with a finite second moment, the $X_i$ are in the universality class of the
Brownian motion and one should recover the results of Section \ref{sec:abbm}, using that \cite{bauer}
\begin{equation}
 P(n) \sim n^{-1/2} e^{- n S(c)} \quad , \quad Q(n) \sim n^{-3/2} e^{- n S(c)}
\end{equation}
with a common function $S(c)=O(c^2)$ at small $c$. For instance, if $P_{f}(\eta)$ is a univariate gaussian,
$P(n)=\frac12 \left[1 - \mbox{erf}(c \sqrt{n/2}) \right]\approx (2 \pi c^2 n)^{-1/2} e^{-c^2 n/2} $ for large $n$. We will not study the Brownian case
in detail, since it was already discussed in Section \ref{sec:abbm}, and we refer to Ref.\ \cite{bauer} for a
detailed asymptotic analysis (as well as a nice proof of (\ref{sparretheo})).

Of course we expect that stable distributions play a special role. Here we detail one example of a Levy-type random-force landscape, for which the results for the records with drift are particularly simple, and present
a nice generalization of Ref.\ \cite{majumdar}, although  they may not be as universal. 
Consider the Cauchy distribution, 
\begin{equation}
 P_{f}(\eta) =\frac{a}{\pi (\eta^2 + a^2)}\ ,
\end{equation}
such that the distribution of $X_n-X_0$ is also Cauchy with parameter $a \to n a$. Quite extraordinarily, 
$$P(n)=\int_{c n}^{\infty}\frac{na \,\rmd x}{\pi[x^{2}+(na)^{2}]} = 
\int_{c }^{\infty}\frac{a \,\rmd x}{\pi[x^{2}+(a)^{2}]} =:p
$$ is independent of $n$, with $0<p=\arctan(a/c)/\pi<1/2$ for $c>0$. Hence $Q(z)=(1-z)^{-p}$ and $F(z)=1 -
(1-z)^{1-p}$, and 
\begin{eqnarray}
\label{251}
 Q(n)&=&\frac{\Gamma(n+p)}{\Gamma(1+n) \Gamma(p)} \\
 F(n)&=& (1-p) \frac{\Gamma(n+p-1)}{\Gamma(1+n) \Gamma(p)}\ .
 \label{252}
\end{eqnarray}
Using (\ref{248}) one finds the average number of records (i.e.\ of jumps):
\begin{equation}
 \overline{N} = \frac{\Gamma(2+M-p)}{\Gamma(1+M) \Gamma(2-p)}  \sim_{M \to \infty}  \frac{M^{1-p}}{\Gamma(2-p)}
\end{equation}
which grows as a power law of the size. Higher moments grow with the same scale:
\begin{equation}
 \overline{N^2}  = - \overline{N} + \frac{2 \Gamma(3+M-2p)}{\Gamma(1+M) \Gamma(3-2 p)}\ .
\end{equation}
Hence at large $M$ the connected fluctuations are
\begin{equation}
 \overline{N^2} - \overline{N}^2 =  \left[ \frac{2}{ \Gamma(3-2p)} - \frac{1}{\Gamma(2-p)^2} \right] M^{2(1-p)}\ ,
\end{equation}
and in all cases the results of Ref.\ \cite{majumdar} are recovered for $p=1/2$, the case without drift.
The full distribution takes a scaling form at large $M$:
\begin{equation}
P(N|M) \approx M^{p-1} g_p(N M^{p-1}) \label{scalingp}
\end{equation}
Summing (\ref{247}) with this scaling ansatz at large $M$, i.e.\ $x:=-\ln z \approx 1-z$ small yields:
\begin{eqnarray}
 \int_0^\infty \rmd M\, M^{p-1} e^{- M x} g_p(N M^{p-1}) \approx x^{-p} e^{- N x^{1-p}} \nonumber
\end{eqnarray}
with $g_{1/2}(y)=e^{-y^2/4}/\sqrt{\pi}$.

From (\ref{distribs}) one sees that the distribution of avalanche sizes (i.e.\ record ages) is $P_{\mathrm{ava}}(s)=F(s)$
for $s=1,2,\ldots$. For fixed $p$ and large $s$ it decays from (\ref{252}) as a power law with $\tau=2-p$:
\begin{equation}
P_{\mathrm{ava}}(s)= F(s) \approx \frac{ s^{-(2-p)}}{-\Gamma(p-1)}\ .
\end{equation}
This leads to a simple interpretation in terms of a directed random walk with traps of independent
random release times $s_i$, distributed as $P(s) \sim s^{-(1+\mu)}$ and   $1/2 \leq \mu=1-p < 1$.   $M$ is the
total time $t$ and $N$ the distance $x$ traveled. As is well known, for $\mu<1$, $x \sim t^\mu$ and the
distribution of $z=t/x^{1/\mu}=M/N^{1/(1-p)}$ is a Levy stable distribution $L_\mu(z)$ with positive support,
which is indeed the solution of (\ref{scalingp}), $g_p(y)=\mu^{-1} y^{1-1/\mu} L_{\mu}(y^{-1/\mu})$.

Although the strong universality of the symmetric case does not hold, we expect that all processes in the class
of the Cauchy process remain critical even with a drift which has a power law distribution of avalanches given
above, and a continuously varying exponent. For stable processes intermediate between Cauchy and Gaussian,
avalanches should be cut at a finite scale, which diverges with different exponents as $c\to 0$. The
situation of stable processes broader than the Cauchy distribution remains open.

\section{General considerations about an $N$-component displacement field}\label{s:N-component}
\begin{figure}[b]
\begin{center}
\Fig{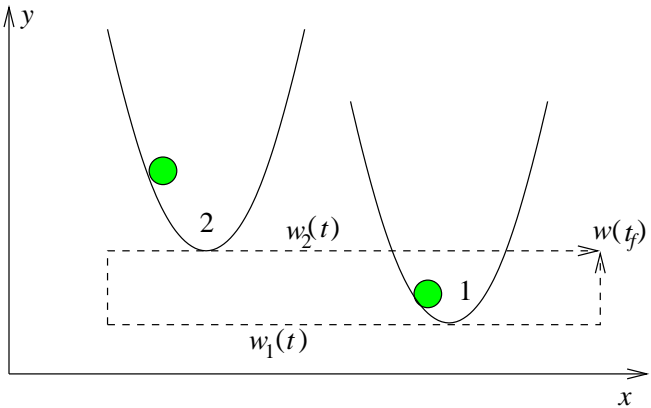} \caption{2 particles dragged through a random energy landscape by parabolic potentials (springs)
whose centers have identical starting and final positions but follow different paths $\{w_{1}(t)\}$ and
$\{w_{2}(t)\}$.} \label{f:w1w2}
\end{center}
\end{figure}
\begin{figure}[t]
\begin{center}
\Fig{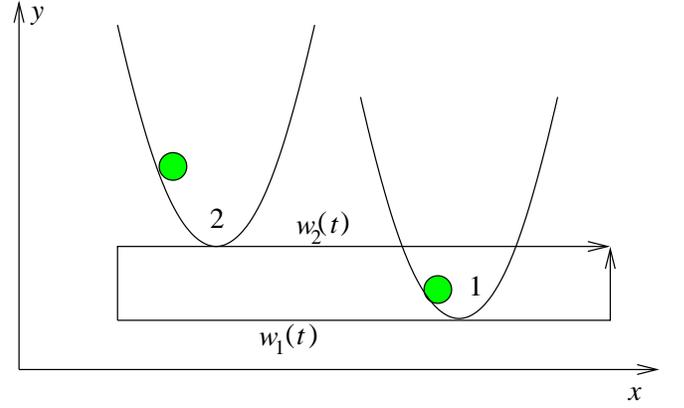} \caption{2 particles dragged through a random energy
landscape by parabolic potentials (springs) whose centers follow parallel straight lines as described in the
text} \label{f:14}
\end{center}
\end{figure}
\begin{figure*}
{\setlength{\unitlength}{1mm}
\fboxsep0mm
{\begin{picture}(179,35)
\put(0,-2){\Figg{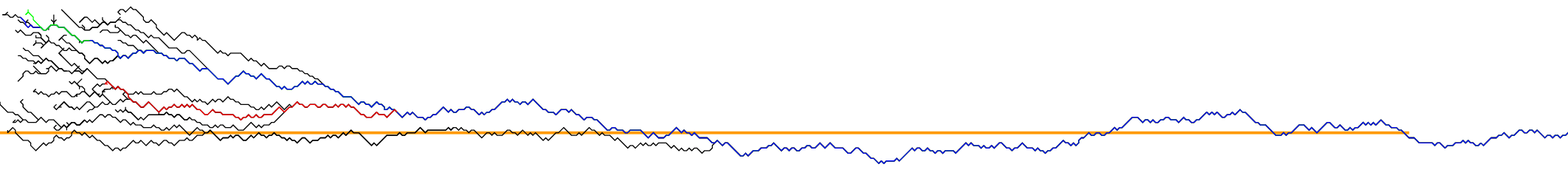}}
\put(148,8){\fig{3cm}{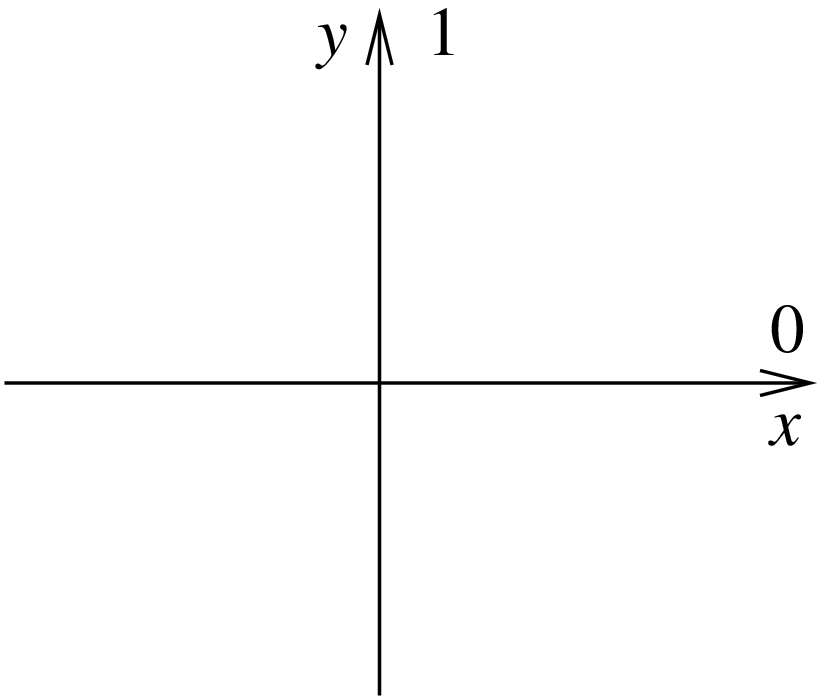}}
\put(59,11.1){\fbox{\fig{8cm}{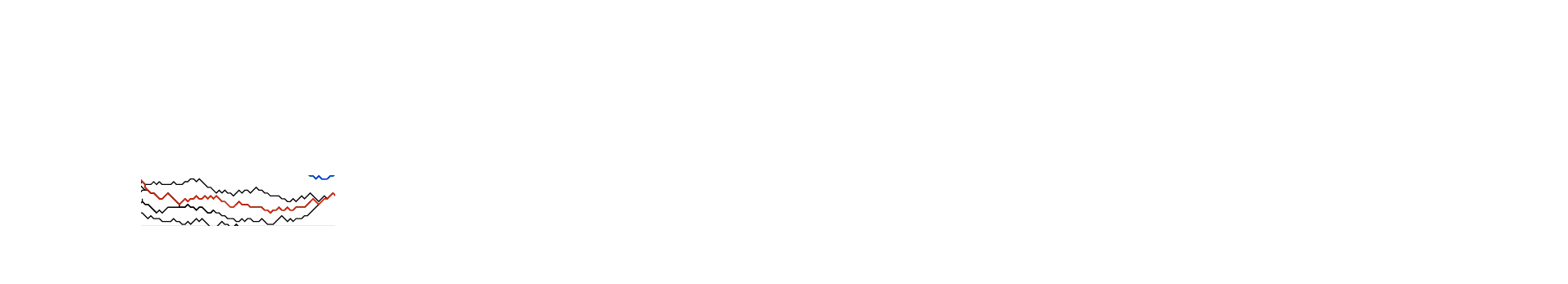}}}
\put(16,4.8){\fbox{\rule{0mm}{5.5mm}\rule{21.8mm}{0mm}}}
\put(16,10.2){\vector(2,1){43}}
\end{picture}}}
\caption{Trajectories of particles dragged from left to
right by a parabolic well. Trajectories starting
at different random initial positions all
converge towards the same trajectory, with the same position at a
given time (which is not visible on the plot). The minimum position of
the parabola in $y$-direction is indicated by a straight (orange/grey)
line. At the right, we show our coordinate system.
The inset at the top is a blowup of part of the curve. It shows a deviation from the no crossing property which holds for $N=1$ (Middleton theorem \cite{Middleton1992}): Trajectories which are together can split up, even if later on they join again (see text).}
\label{k9}
\end{figure*}%
\begin{figure*}
\Figg{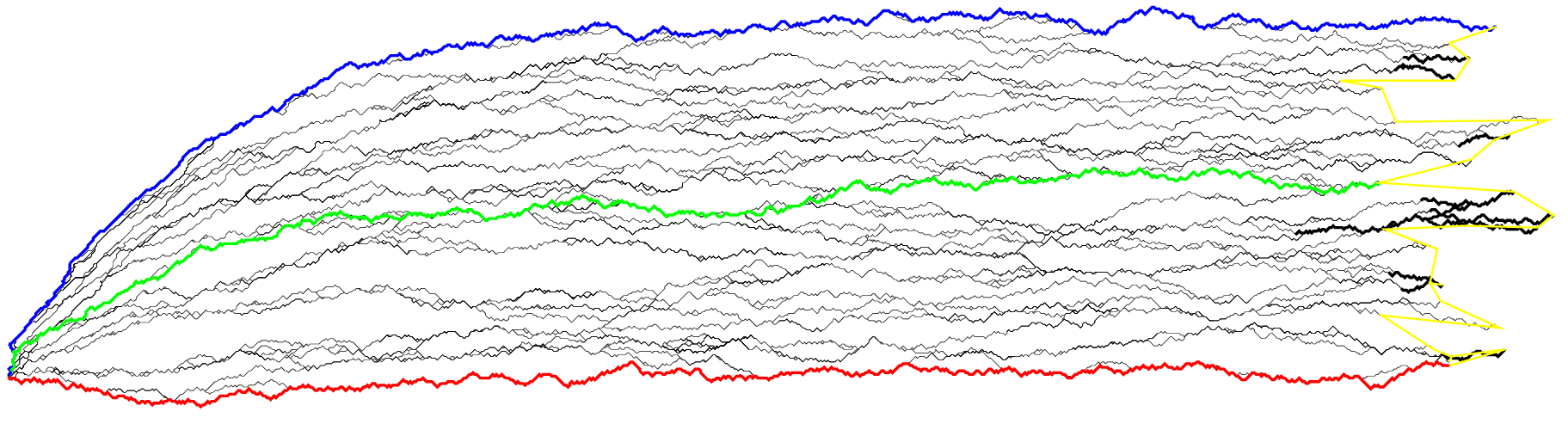}
\caption{Trajectories of 200 particles dragged from left to
right. All particles start at (0,0). However, they sit in different
potential wells (as described in the main text). As can be seen, they
mostly move together on preferred trajectories, before separating
again. The endpoints are joined by yellow lines. $m^{2}=0.003$. The
particles in the two outer wells, as the center one, are marked in
colors (red/blue, green). The well was moved for a total of 1000
steps.
}
\label{k10}
\end{figure*}

Up to now, we have considered particles, and more generally elastic objects and manifolds, whose position is a 1-component function. We now consider particles or elastic objects embedded in higher dimensions $N>1$. For simplicity, we focus on a particle, but the considerations in this short section can be extended to $d$-dimensional elastic manifolds. In the next section we consider an application to a particle driven in a 2-dimensional random energy landscape.

Consider two particles, which see the same random energy landscape, but which sit in different parabolas,
labeled 1 and 2. These parabola are chosen with the same curvature $m^2$ but their centers differ, and can have
very different trajectories, which we call $\{w_{1}(t)\}$ and $\{w_{2}(t)\}$. An interesting case is when the
trajectories differ but the endpoints coincide $w_1(t_{i,f})=w_2(t_{i,f})$ as in the example of figure
\ref{f:w1w2}. In that example it is clear that
\begin{equation}
|\overline{u_{w_{1}}(t_{f}) -   u_{w_{2}}(t_{f})}| \approx \sqrt{2} \frac{f_{c}}{m^{2}}\ .
\end{equation}
This is because there should be a non-zero critical force, $f_{c}$, and that each particle lags behind each parabola center $w_{i}(t)$, up to fluctuations, roughly in minus the direction of drift. Thus the process $u_{w}$ depends on the trajectory $w(t)$, and to define a single valued function $u_{w}$, we have to restrict to a single well-defined trajectory $w(t)$ in a quasi-static limit. Consider now  figure \ref{f:14}.
Both parabolas move with the same velocity $v$ in
$x$-direction. They are completely characterized by their position
$\vec w=\left\{x =x_{0}+v t,\vec y \right\}$, with $\vec w\in
\mathbb{R}^{N}$, $\vec y\in \mathbb{R}^{N-1}$. Especially note that
without loss of generality, $x_{0}$ can be put to 0. Again we
integrate the Langevin-equation (\ref{eq:eqmo}), to define $\vec u
(\vec w)$.

A more difficult question is whether $\vec u (\vec w)$ depends on the initial
condition $\vec u (t_{0})$, and since we have set $x_{0}$ to 0,
implicitly on the starting time
$t_{0}$. One expects (see next Section how it occurs) that the dependence of $\vec u (\vec w)$ on
$t_{0}$ disappears in the limit of $t_{0}\to -\infty$, and this is the
limit we are interested in. This could be checked, similar to the {\em
exact sampling} method, see e.g.\ \cite{SMAC}, by starting at time
$t_{0}$, and checking that at the time of interest $t$, all
trajectories from all possible initial conditions have converged
towards a single one. If not, one starts at an earlier time $t_{-1}$,
and checks again, repeating this procedure until all trajectories have
converged. This defines a function $\vec u (\vec w)$, which is
now independent of the initial time and conditions. In the next
section it will be checked numerically
that for a particle driven through a two dimensional
bounded random energy landscape, all trajectories indeed converge, see
figure \ref{k9}. It is also found there that the so-called no crossing property (Middleton theorem \cite{Middleton1992}) does not hold for $N=2$: although violations appear to be rare there are some instances of two trajectories splitting up. This results from a second particle (more properly, a second trajectory of the same particle with a different initial condition) arriving at a later time on the same site: it then feels a stronger drive from the parabola and may jump forward and pass the first particle. An example is shown in figure \ref{k9}.

Having given an unambiguous definition of $\vec u (\vec w)$, we can calculate connected
correlation functions of its moments, which again define $\Delta$,
(now a tensor), and higher cumulants. For this tensor the driving direction $x$ will play a special role (for $N >2$ we expect isotropy in the other $N-1$ directions. The calculation is done for a particle in the next section. Other definitions of $\vec u (\vec w)$ could of course be given. The simplest one is to pick a fixed but different driving direction. From statistical isotropy of the disorder the results should be the same up to the rotation. We defer the study of more complicated driving processes to future work.

\section{A particle dragged in two dimensions: $d=0$, $N=2$}
\label{s:N=2}
We now
study particles dragged through a 2-dimensional random-energy
landscape.

The algorithm works as follows: We generate a random-energy landscape
on a square lattice.  A particle in addition sees a parabolic
well. The total energy is the sum of both.  We will mostly use a
box distribution for the energy of a site, uniform in $[0,1]$.  Energies on different sites are uncorrelated. We then update
all particle positions: If a particle can move in a direction s.t.\ it
will lower its potential energy, it will do so. If there are several
such directions, it will choose the one with the lowest final
energy. We allow moves to the eight nearest neighbors numbered from $1$ to $8$ (starting at the
center 0):
$$
\fig{3cm}{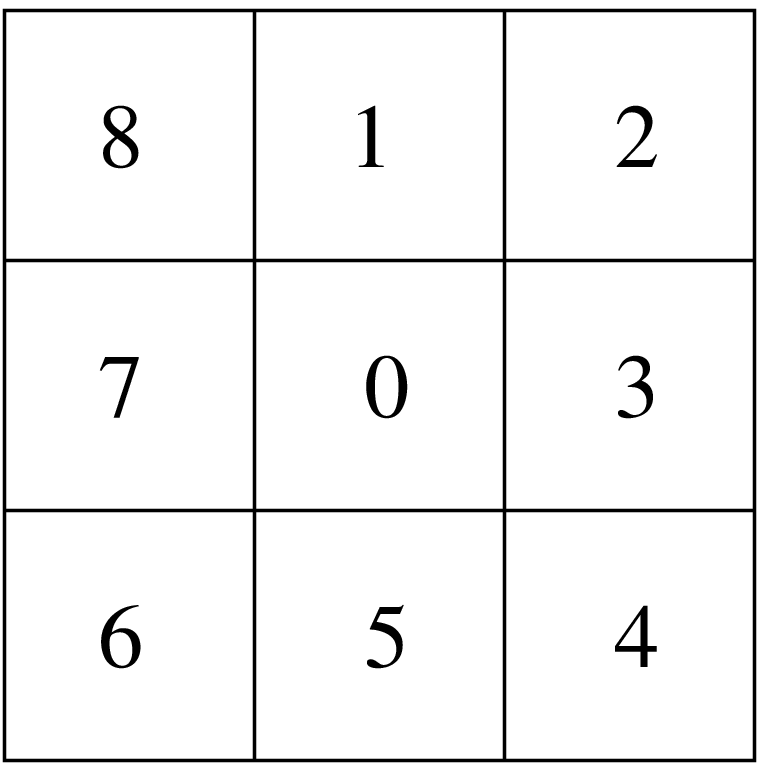}
$$
If a move is possible, we perform it and then try other moves again, until the particle finally gets stuck. If several moves are possible, we take the one which results in the largest descent in energy, i.e.\ we go into the direction of the maximum force. Only then, we update the
position of the parabola, by moving it from $w$ to $w+\rmd w = w+ v \rmd t$. We record the particle position
as a function of time $t$.

We first show numerically that there is a unique attractor trajectory (see figure \ref{k9}). Start particles at random positions, but in the
same parabolic well. Then move the parabola in a given direction (here always to the right, also denoted the $x$-direction).
One sees that trajectories converge, and particles will have the same
position at a given time (not visible on figure \ref{k9}, which only shows the world-lines.) This convergence can be understood from the fact that if two ``particles'' (in fact these are the same particle but with different initial conditions) meet at a site at the same time, their future evolution is identical. Hence the deep sites with low energies where the particle gets temporarily stuck act as sinks where the trajectories merge.  Clearly, the particle needs to be trapped long enough for the process to be efficient.

Our proper simulation is done with many particles (here 200), each sitting in a parabola which are
displaced by one unit (or in general by $\rmd y$)
 to the top. This gives us data-points in the $y$-direction. In the $x$-direction
(in which we move the parabolas), we use that after some time $t$ the parabola has been displaced by
a distance $v t$. We denote the minimum of the parabola $w_{t,y}$ and its $i$th component $w_{t,y}^{i}$ as
\begin{equation}
w_{t,y}:= (v t , y)\ , \qquad w_{t,y}^{0}:= v t \ , \qquad w_{t,y}^{1}:= y
\end{equation}
The particle sitting in this parabola will have position $u_{w_{t,y}}$, with components $u_{w_{t,y}}^{i}$.
We then define $\Delta^{ij}(t v,y)$ as
\begin{equation}
\Delta^{ij}\big((t'-t) v,y\big):= m^{4 } \, \overline {(u^{i}_{w_{t,0}}-w^{i}_{t,0})(u^{j}_{w_{t',y}}-w_{t',y}^{j})}^c \label{deldij} 
\end{equation}
The connected symbol $c$ indicates that we have subtracted the critical force.
$\Delta^{ij}(x,y)$ has the following symmetry properties
\begin{eqnarray}
\Delta^{ii}(x,y) &=&  \Delta^{ii}(x,-y) \\
\Delta^{01}(x,0) &=& 0\\
\Delta^{01}(0,y) &=& - \Delta^{10}(0,y)\ .
\end{eqnarray}
This is a consequence of the relabeling symmetry in (\ref{deldij}) $\Delta^{ij}(x,y)=\Delta^{ji}(-x,-y)$ and obvious covariance under the parity symmetry $y \to - y$. For smaller and smaller masses, there will be more and more data-points.  Steps in the $x$-direction are necessarily discretized,
of size $\rmd w = v \rmd t $.
This poses an additional problem not present for $N=1$: there choosing a $\rmd w$ too large results in a loss  in precision (since some smaller jumps may be overrun) but does not have dramatic consequences for large jumps, especially does not change the endpoint $u(w)$, due to Middleton's theorem \cite{Middleton1992}. In contrast, for $N=2$, if the parabola is not moved adiabatically , the particles will see a strong force forward, and therefore be more
likely to move forward, instead of sideward, thus embarking on a different trajectory. This may alter the whole trajectory over a much larger region. In practice, we decided to never move the parabola by more than one unit, before checking whether a move could be made. It may be a possible source for finite-size
corrections. These will disappear {\em if, and only if} the critical force scales to zero for $m\to 0$, since the energy gain for an elementary move is
\begin{equation}\label{221}
\frac{m^{2}}2 \left[ (u+1-w)^{2}-(u-w)^{2} \right] = m^{2} \left(u-w+\frac12\right) \approx  f_{c}
\end{equation}
However, $f_{c}$ goes slowly towards 1, by which it is bounded.
Unfortunately we find e.g. $f_{c}(m^{2}=0.01)=0.464602$,
$f_{c}(m^{2}=0.0001)=0.785073$.
This might indicate that the step-size we have used is still too large. We have not attempted to use a smaller step-size due to the enormous computing powers needed. We nevertheless believe that the results are valid for the following reason: $f_{c}$ measures the time average of $u_{w}-w$, but we have to know the forward force exerted by the spring, when the particle {\em arrives} at the trap, Clearly, this must be much smaller, otherwise in a few steps the force would have increased by 1, which is sufficient to overcome {\em any} barrier for the box-distributed random energies, and the particle would not remain pinned for a long time. However we see diverging trapping times in the simulations, thus the argument using eq.\ (\ref{221}) is not valid.

\begin{figure}[t]
\Fig{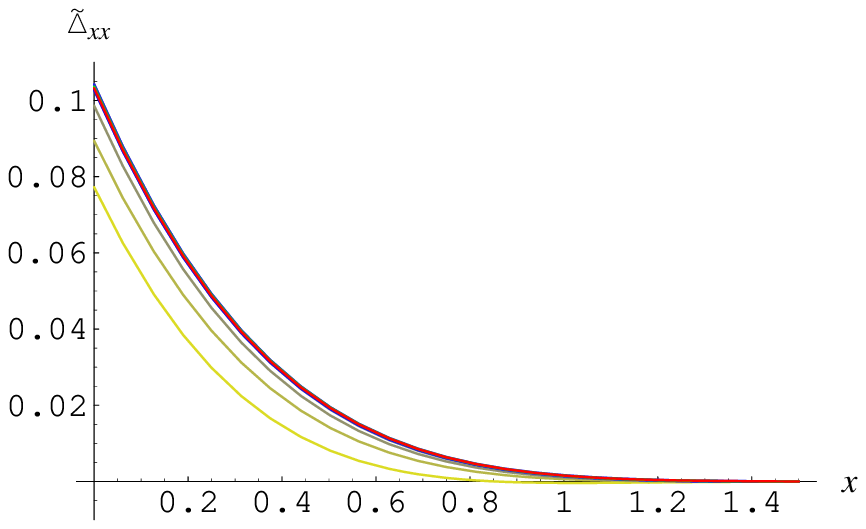}\hspace{-3cm}\raisebox{20mm}{\fig{3cm}{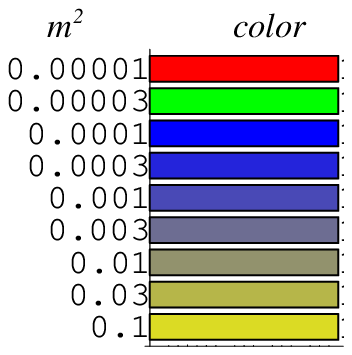}}
\caption{Scaling collapse for $\tilde \Delta_{xx}(x,0):= m^{-4+2 \zeta_{x}} \Delta_{xx}(x\, m^{\zeta_{x}},0)$, with $\zeta_{x}=1.595$. The scaling collapse is perfect except for the two largest masses.}
\label{f:D1}
\end{figure}

\begin{figure}[t]
\Fig{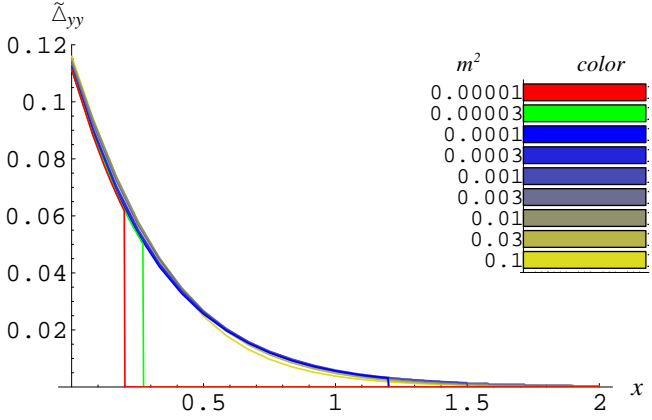}\hspace{-3cm}\raisebox{20mm}{\fig{3cm}{color-edited}}
\caption {Scaling collapse for $\tilde \Delta_{yy}(x,0):= m^{-4+2 \zeta_{y}} \Delta_{yy}(x\, m^{\zeta_{x}},0)$, with $\zeta_{x}=2$ and $\zeta_{y}=1$. The descending lines indicate that no data has been collected for larger values, due to an insufficient choice of parameters. Good scaling collapse.}
\label{f:D2}
\end{figure}

\begin{figure}[t]
\Fig{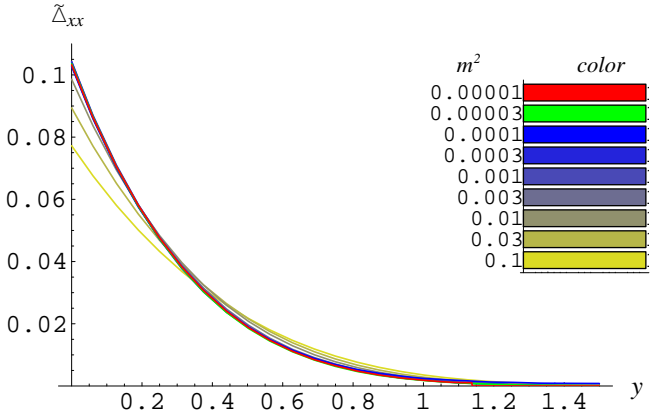}\hspace{-3cm}\raisebox{20mm}{\fig{3cm}{color-edited}}
\caption{Scaling collapse for $\tilde \Delta_{xx}(0,y):= m^{-4+2 \zeta_{x}} \Delta_{xx}(0,y\, m^{\zeta_{y}})$, with $\zeta_{x}=1.595$ and $\zeta_{y}=1$. The scaling collapse is excellent for all but the three largest masses.}
\label{f:D3}
\end{figure}

\begin{figure}[t]
\Fig{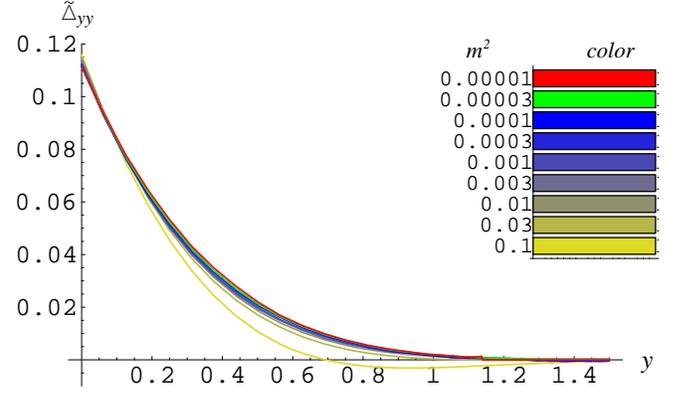}\hspace{-3cm}\raisebox{20mm}{\fig{3cm}{color-edited}}
\caption{Scaling collapse for $\tilde \Delta_{yy}(0,y):= m^{-4+2 \zeta_{y}} \Delta_{xx}(0,y\, m^{\zeta_{y}})$, with $\zeta_{y}=1$. The collapse is good, except for the 3 largest masses.}
\label{f:D4}
\end{figure}

\begin{figure}[htbp]
\Fig{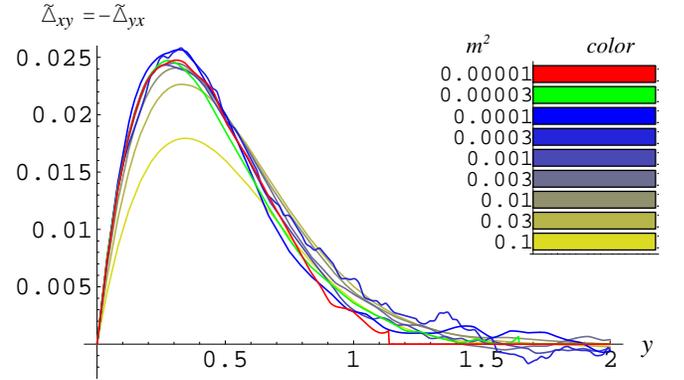}\hspace{-3cm}\raisebox{20mm}{\fig{3cm}{color-edited}}
\caption{Scaling collapse for $\tilde \Delta_{xy}(0,y):= m^{-4+\zeta_{x}+\zeta_{y}} \Delta_{xy}(0,y\, m^{\zeta_{y}})$, with $\zeta_{x}=1.25$ and $\zeta_{y}=1$. The signal-to-noise ratio is rather big. In order to improve the statistics, we have used (after numerical verification) that $\tilde \Delta_{xy}(0,y)=-\tilde \Delta_{yx}(0,y)$, to plot  $\frac12[\tilde \Delta_{xy}(0,y)-\tilde \Delta_{yx}(0,y)]$. Fair scaling collapse, except for the 2 largest masses.}
\label{f:D5}
\end{figure}

We now present data for the force-force correlators in figures \ref{f:D1} to \ref{f:D5}, for masses ranging from $m^{2}=0.1$ to $m^{2}=10^{-5}$, descending in half-decades. A first and important qualitative conclusion to be drawn is that all correlators not only depend on $x$, but also on $y$. This is in contradiction to the fixed-point structure used by Erta\c s and Kardar \cite{ErtasKardar1996}, whose $\Delta_{ij}$ depends only on $x$ but not on $y$.

Our aim is to determine the scaling exponents $\zeta_{x}$ and $\zeta_{y}$ from the finite-mass scaling-ansatz, suggested by the FRG equations for this problem \cite{fedorenkoN}:
\begin{equation}
\label{p2}
\tilde \Delta^{ij}_{m}(x,y) := m^{-4+\zeta_{i}+\zeta_{j}} \Delta^{ij}(x m^{-\zeta_{x}},y m^{-\zeta_{y}}) \ ,
\end{equation}
and supposing that $\tilde \Delta_{m} \to \tilde \Delta $ for $m\to0$.
We find that for $\Delta_{xx}(x,0)$,  $\Delta_{yy}(x,0)$,  $\Delta_{xx}(0,y)$,  $\Delta_{yy}(0,y)$ and $\Delta_{xy}(0,y)$ {\it separately} such a scaling collapse is possible. There is no doubt that $\zeta_{y}=1$, with consistently rather small errors: the scatter from the different estimations is $\zeta_{y} = 1.009\pm 0.015.$


However the results for $\zeta_{x}$ are less consistent.  We find different values, depending on which quantity we consider. There is a clean data-collapse on figure\ \ref{f:D2} for $\tilde \Delta_{yy}$ with $\zeta_{x}=2$; however on figures \ref{f:D1} and \ref{f:D3} for $\tilde \Delta_{xx}$ the best collapse is  with $\zeta_{xx}=1.6$; finally on figure \ref{f:D5} for $\tilde\Delta_{xy}$, the best scaling collapse is for $\zeta_{x}= 1.25$, where however some of the data are noisy (but note that at least the data for the second-smallest mass, which are already very difficult to obtain, and for some of the larger masses show only little noise, s.t.\ noise does not seem to be an issue here.) Let us recall for comparison that for $N=1$ a particle driven in a random energy landscape with a box distribution belongs to universality class III with $\alpha=3$, i.e.\ $\zeta_x=4/3$, see sections \ref{WCfc} and \ref{KN1}. (For a box random {\it force} it is $\alpha=1$ and $\zeta=1$).

To conclude, we have shown that the unique attractor trajectory can be defined for the particle with $N=2$, and that the disorder correlator $\Delta_{ij}(x,y)$ can be measured.


\section{Conclusion}

To conclude we have shown how the renormalized disorder correlator $\Delta(w)$, central to the Functional RG
theory of depinning, can be measured for a manifold of internal dimension $d$ driven by a spring in a $N=1$
random landscape. This correlator contains information about the stick slip motion of the interface.

We have solved analytically the case $d=0$ of a particle in a short-range correlated pinning-force landscape,
finding three universality classes. In each case we have obtained the 
universal fixed-point forms for
$\Delta(w)$ for quasi-static driving, i.e.\ the depinning fixed point. In all cases it exhibits the
famous cusp at $w=0^+$. We have also obtained the distribution of critical forces, avalanche sizes and 
waiting times, and checked the general relations conjectured to hold between their moments and the cusp
$\Delta'(0^+)$. While the exponent $\zeta$ can take various values depending on the class, the avalanche-size
exponent was found to be $\tau=0$ in all cases, which invalidates, at least at a naive level, the conjecture
$\tau=2 - 2/(d+\zeta)$. We also found that the distribution of avalanche sizes and waiting times are {\it
identical} in the scaling limit.

We have extended our results to a particle driven in a force landscapes with the correlations of a random walk.
In the Brownian case, known as the ABBM model for interface motion and Barkhausen noise, it is possible to solve
for any driving velocity and check the quasi-static limit. Remarkably this model has much in common with the
mean field theory of avalanches and recent FRG results for avalanche distributions in $d=4$. Since $\zeta=4$,
the avalanche exponent for $v=0^+$ obeys, in that case, the conjecture $\tau=2 - 2/(d+\zeta)$.

In each case we have emphasized the connections between the depinning problem for a particle and the extremal
statistics of records, with and without drifts.

These exact results in $d=0$ provide interesting checks and interpretations of the Functional RG theory, and
help us understand what we should expect for manifolds.

Finally, we started addressing the problem of depinning for $N>1$, mostly numerically;  even for a particle it is  quite non-trivial. We have checked numerically the consistency of the method, based on the
ergodicity in presence of driving via a quadratic well. Numerous open problems remain.

\acknowledgements It is a pleasure to thank Michel Bauer, Jean Philippe Bouchaud, Andrei Fedorenko, Satya
Majumdar and Alberto Rosso for stimulating discussions, some in the context of the Beg-Rohu school ``Manifolds in random media, random matrices and extreme value statistics'', organized by Gulio Biroli and Claude Godreche, 
whom we gladly acknowledge.

This work was supported by ANR under program 05-BLAN-0099-01, and in part through NSF under grant number PHY05-51164 during the program SLE06 at KITP.

\appendix

\section{Dynamical action and observables}
\label{app:dynamical action}

In this Section we establish the relation between the effective action
and observables; this is an extension to the dynamics of the proof given in
\cite{LeDoussal2006b,LeDoussalinprep} for the statics. It. This relation allows to measure the correlator of the dynamical FRG as explained in the text.
Since the dynamical field theory is significantly more complicated than the static replica field theory, the
arguments presented here may be slightly less general and rely on further assumptions about the nature of the
quasi-static limit and its ergodic properties, some remaining to be demonstrated, as e.g.\  extensions to
$N>1$ components is less straightforward than in the statics \cite{LeDoussal2006b,LeDoussalinprep}. At a formal
level however the arguments are rather similar.

We use notations of the text and consider the equation of motion:
\begin{eqnarray}
 \eta \partial_t u(x,t) &=& F_x[u(t);w(t)] + \xi(x,t) \label{eq:eqmo2} \\
 F_x[u;w] &=& m^2 (w(x) - u(x)) +  \nabla_x^2 u(x) + F(x,u(x)) \nonumber
\end{eqnarray}
where $w(t)$ is given, and $\xi(x,t)$ the thermal noise. We denote in general implicitly
$u^s_{xt}=u(x,t;u_0(x,t_0),F,\xi)$ the solution for given initial condition, disorder and thermal noise.

\subsection{Definition  of functionals}
Let us first recall the definition of the useful functional of the dynamical field theory. We write the
dynamical (MSR) action ${\cal S}$ in compact notation as:
\begin{eqnarray}\label{lf7}
{\cal S}[u,\hat u] &=& \hat u \cdot g^{-1} \cdot u + \hat u \cdot A^{(0)}[u]
- \frac{1}{2} \hat u \cdot B^{(0)}[u] \cdot \hat u \nonumber\ . \\ 
&&+ O(\hat u^3)
\end{eqnarray}
For any ``vectors'' $u,v$ we denote $u \cdot v := \int_{xt} u_{xt} v_{xt}$ (and additional index
contraction for $N>1$), $A$ and $B$ are respectively vector and matrix functionals. $g_{xy}$ can be an
arbitrary (time independent) symmetric matrix but the usual choice is (in Fourier) $g^{-1}_q = q^2 + m^2$. The
functionals defining the bare action are
\begin{eqnarray}\label{lf8}
 A^{(0)}[u]_{xt} &=& \eta \partial_{t} u_{xt} \\
 B^{(0)}[u]_{xt,x't'} &=& 2 \eta T \delta_{xx'} \delta_{tt'} + \Delta_0(u_{xt}-u_{x't'}) \delta_{xx'}\qquad
\end{eqnarray}
which is the standard MSR action averaged over disorder. The statistical tilt symmetry (STS) of the bare action
states that $A^{(0)}[u]$ and $B^{(0)}[u]$ are invariant under the change $u_{xt} \to u_{xt} + \phi_x$. It
implies the same symmetry for $A[u]$ and $B[u]$. The generating function of connected correlations is
\begin{align}\label{lf9}
 \exp(W[w,\hat w]) &:= \int {\cal D}[u]\, {\cal D}[\hat u]\, \exp\Big[ -{\cal
S}[u,\hat u]\nonumber \\
& \qquad+ \hat u \cdot g^{-1} \cdot w + \hat w \cdot g^{-1} \cdot u \Big]\ ,
\end{align}
where the sources have been redefined, following \cite{LeDoussal2006b,LeDoussalinprep}, in a convenient way for
the following. It also admits an expansion similar to the action:
\begin{eqnarray}\label{lf11}
 W[w,\hat w] &=& \hat w \cdot g^{-1} \cdot w - \hat w \cdot \hat A[w]
+ \frac{1}{2} \hat w \cdot \hat B[w] \cdot \hat w \nonumber \\
&&+ O(\hat w^3)
\end{eqnarray}
where the STS implies that $\hat A[w]$ and $\hat B[w]$ are invariant under the change $w_{xt} \to w_{xt} +
\phi_x$. Finally the effective action functional $\Gamma$ is defined as usual as the Legendre transform of $W$:
\begin{equation}\label{lf10}
W[w,\hat w] + \Gamma[u,\hat u] = \hat u \cdot g^{-1} \cdot w + \hat w \cdot g^{-1} \cdot u\ .
\end{equation}
It admits the expansion:
\begin{eqnarray}
\label{lf7a} \Gamma[u,\hat u] &=& \hat u \cdot g^{-1} \cdot u + \hat u \cdot A[u]
- \frac{1}{2} \hat u \cdot B[u] \cdot \hat u \nonumber \\
&& + O(\hat u^3)\ .
\end{eqnarray}
STS implies the same symmetry for $A[u]$ and $B[u]$ as for $A^{(0)}[u]$ and $B^{(0)}[u]$.

\subsection{Relations to observables}

The functional $W$ is directly related to observables, i.e.\ correlation functions, in the following way:
Consider the average over solutions of the equation of motion,
\begin{eqnarray}\nn
&& \!\!\!\overline{ \langle \rme^{\int_{xt} \hat w \cdot g^{-1} \cdot u^s}
\rangle_{\xi,u_0} } = \int {\cal D} [u] {\cal D}[\hat u]\, \rme^{- {\cal S}_w[u,\hat u] +
 \hat w \cdot g^{-1} \cdot  u_{xt}} \\
&& = \int{\cal D} [u] {\cal D}[\hat u]\, u \rme^{- {\cal S}[u,\hat u] +  \hat u \cdot g^{-1} \cdot w  +  \hat w
\cdot g^{-1} \cdot u} = \rme^{W[w,\hat w]}\qquad
\end{eqnarray}
Here ${\cal S}_w$ is the MSR action in presence of $w$, while ${\cal S} \equiv {\cal S}_{w=0}$ is the action
defined above, in the absence of $w$. As usual, to take into account the initial conditions (if necessary) all
time integrals start at $t_0$ and an additional integral ${\cal D}[u_0]\, P[u_0]$ fixes its probability at
$t=t_0$. Expanding the above average, one finds:
\begin{eqnarray}
W[w,\hat w] &=& \hat w \cdot g^{-1} \cdot \overline{\langle u
\rangle }_w \nn\\
&&+ \frac{1}{2} \hat w \cdot g^{-1} \cdot  ( \overline{\langle u   u \rangle }_w - \overline{\langle u
\rangle}_w ~~
\overline{\langle u \rangle}_w  ) \cdot g^{-1} \cdot \hat w \nn\\
&& + O(\hat
w^3)
\end{eqnarray}
Note that the (matrix) average $\overline{\langle u u \rangle } - \overline{\langle u \rangle}~
\overline{\langle u \rangle}$ is not the connected thermal average but the connected double (disorder+thermal)
average. Note the index $w$ which indicates that the above averages, e.g. $\overline{\langle u_{xt} \rangle_w }$,
are averages w.r.t. ${\cal S}_w$, i.e.\ in presence of the (given) driving $w_{xt}$. Comparing with (\ref{lf11})
we obtain:
\begin{eqnarray}
 \label{rel} \overline{\langle u_{xt} \rangle_w } &=& w_{xt} - g_{xy} \hat A_{yt}[w] \\
 \overline{\langle u_{xt} u_{x't'} \rangle_w } - \overline{\langle
u_{xt} \rangle_w}~~ \overline{\langle u_{x't'} \rangle_w} &=& g_{xy}
g_{x'y'} \hat B_{yt,y't'}[w]\ ,\qquad\ 
\end{eqnarray}
where summation (integration) over repeated indices is implicit. Until now $w_{xt}$ is arbitrary. For a uniform
driving $w_t$ one has:
\begin{eqnarray}
&& \overline{\langle (w_t - \bar u_t) \rangle } = m^{-2} \frac{1}{L^d} \int_y \hat A_{yt}[w] \\
&& \overline{\langle (w_t - \bar u_t) (w_{t'} - \bar u_{t'}) \rangle
} - \overline{\langle (w_t - \bar u_t) \rangle } ~~
\overline{\langle (w_{t'} - \bar u_{t'}) \rangle }\qquad  \nn\\
&& \qquad = m^{-4}
\frac{1}{L^{2 d}} \int_{yy'} \hat B_{yt,y't'}[w]
\end{eqnarray}
These are the (spatially) local parts of the $\hat A$ and $\hat B$ functionals, and we expect:
\begin{eqnarray}
&&\lim_{L \to \infty} \lim_{\partial_t w(t) \to 0^+} \lim_{T \to 0}
\frac{1}{L^d} \int_y \hat A_{yt}[w] = f_c \qquad \\
 &&\lim_{L \to \infty} \lim_{\partial_t w(t) \to 0^+} \lim_{T \to 0}
\frac{1}{L^{2 d}} \int_{yy'} \hat B_{yt,y't'}[w] \nn\\
&&\qquad= L^{-d} \hat \Delta(w_t - w_{t'})
\end{eqnarray}
More relations can be derived e.g.\ by considering the formal expansion (symbolically):
\begin{eqnarray}
&&\!\!\! W[w+\delta w,\hat w]\nn \\
&&= \sum_{n_1 n_2}  \frac{1}{n_1! n_2!}
\overline{\langle \hat u..\hat u u .. u \rangle_w }^c(g^{-1} \delta
w)^{n_1} (g^{-1} \hat w)^{n_2}\qquad\qquad
\end{eqnarray}
Hence one has:
\begin{equation}
 \overline{\langle \hat u_{x't'} u_{xt} \rangle_w } = g^{-1}_{xx'} \delta_{tt'}
 - (\nabla_w \hat A[w])_{x't',xt}
\end{equation}
Let us also recall that the quadratic parts of $W$ and $\Gamma$ are (always):
\begin{eqnarray}
W_{\mathrm{quad}}[w,\hat w] &=& \hat w \cdot g^{-1} \cdot R \cdot g^{-1}
\cdot w \nn\\&&+ \frac{1}{2} \hat w \cdot g^{-1} \cdot C \cdot g^{-1} \cdot \hat w \\
\Gamma_{\mathrm{quad}}[u, \hat u]&=& \hat u \cdot R^{-1} \cdot u -
\frac{1}{2} \hat u \cdot D \cdot \hat u\ ,\qquad
\end{eqnarray}
where $R$ and $C$ are the exact response and correlation functions
(in the absence of $w$) and $C=R^t D R$.

Finally it is useful to mention the terms  without disorder:
\begin{eqnarray}
 A[u]&=& a \cdot u\\
  \nabla_u A[u] &=& a^t\\
   g^{-1} + a &=& R^{-1} \\
 \hat A[w] &=& \hat a \cdot w \\
  \nabla_w \hat A[w] &=& \hat a^t \\
  g^{-1} - \hat a &=& g^{-1} \cdot R \cdot g^{-1}
\end{eqnarray}
with:
\begin{eqnarray}
&& \hat a \cdot (1 + g \cdot a) = a \\
&& \hat a = a \cdot (1 + g \cdot a)^{-1}\qquad
\end{eqnarray}

\subsection{Legendre transform}

Our aim is to relate the functionals $\hat A$ and $\hat B$, which are observables as established above, to the
functionals $A$ and $B$ associated to the effective action.

Let us thus perform the Legendre transform. In this transformation one defines the functionals $w[u,\hat u]$ and
$\hat w[u, \hat u]$ which allow to compute $\Gamma$ from $W$ using (\ref{lf10}). One has
\begin{eqnarray}\label{lf12}
 w &=&  g \cdot \frac{\delta \Gamma}{\delta \hat u} = u + g \cdot A[u]
- g \cdot B[u] \cdot \hat u + O(\hat u^2)
\\
\label{d6}
 \hat w &=& g \cdot \frac{\delta \Gamma}{\delta u} \\
 &=& \hat u + g \cdot
\nabla_u A[u] \cdot \hat u -\frac{1}{2} g \cdot \nabla_u \hat u \cdot
B[u] \cdot \hat u + O(\hat u^3)  \nonumber
\end{eqnarray}
The inverse relations are useful as well:
\begin{eqnarray}\label{d7}
u &=&  g \cdot \frac{\delta W}{\delta \hat w} = w - g \cdot \hat A[w]
+ g \cdot \hat B[w] \cdot \hat w + O(\hat w^2)
\\
\label{d8}
 \hat u &=& g \cdot \frac{\delta W}{\delta w}\\ \nonumber
 & =& \hat w - g \cdot
\nabla_w \hat A[w] \cdot \hat w +\frac{1}{2} g \cdot \nabla_w \hat w \cdot
\hat B[w] \cdot \hat w + O(\hat w^3)
\end{eqnarray} From (\ref{lf12}) and (\ref{d7}) one deduces that
\begin{equation}\label{d22}
w-u = g\cdot A[u] +O (\hat{u}) = g \cdot \hat{A}[w] + O (\hat{w})\ .
\end{equation}
This implies that
\begin{equation}\label{d23}
\mbox{$\displaystyle A[u] = \hat{A}[w]\ts_{\hat{u}=0 } =
\hat{A}[u+g\cdot A[u]]$} \ .
\end{equation}
which allows in principle to compute one functional from the other. One has the equivalent relation:
\begin{equation}\label{d232}
\mbox{$\displaystyle \hat A[w] = A[w-g\cdot \hat A[w]]$} \ .
\end{equation}
Moreover
\begin{equation}\label{d24}
B[u] = - g^{-1}\cdot {\rmd w}/{\rmd \hat{u} } \ ,
\end{equation}
where $\hat{u}$ should be set to zero at the end. We have used a notation which makes the position of the
indices clear. From (\ref{d6}):
\begin{equation}\label{d25}
{\rmd \hat{w}}/{\rmd \hat{u}} = 1 +g \cdot \nabla_u A[u]
- g \cdot \nabla_u \hat u \cdot B[u] + O(\hat
u^2)
\end{equation}
Note the subtle difference with
\begin{equation}\label{d27}
{\rmd {w}}/{\rmd {u}} = 1 +g \cdot A[u] \stackrel{\leftarrow}{\nabla}_{u}
- g  \cdot B[u]\cdot\hat{u}  \stackrel{\leftarrow}\nabla_u  + O(\hat
u^2)\ .
\end{equation}
This means that (at least at order $O(\hat u)=O(\hat w)$, but eventually even exact)
\begin{equation}\label{d28}
g^{-1} \cdot \rmd w /\rmd u = \left[g^{-1} \cdot \rmd \hat w /\rmd
\hat u  \right]^{t}
\end{equation}
From (\ref{d7}) follows
\begin{eqnarray}\label{d26}
B[u] &=& - g^{-1}\cdot {\rmd w}/{\rmd \hat{u} } \nonumber \\
&=& - g^{-1}
{\rmd}\left( u + g \cdot \hat A[w] - g \cdot \hat B[w] \cdot \hat w +
O(\hat w^2) \right)\Big/{\rmd \hat{u} } \nonumber\\
&=& {\rmd} \left(
-\hat A[w] + \hat B[w] \cdot \hat w + O(\hat
w^2)\right)\Big/{\rmd \hat{u} }\nonumber \\
&=&  \hat{A} [w] \stackrel{\leftarrow}{\nabla} _{w} \cdot g \cdot B[u] + \hat{ B} [w] \cdot \rmd \hat{w}/\rmd
\hat u\ ,
\end{eqnarray}
where it is implicit that $\hat{u}$ is set to zero at the end. It implies
\begin{equation}\label{d29}
\left( 1-[\nabla_{w} \hat{A} [w]]^{t} \cdot g\right) \cdot B[u] = \hat{B}[w]\left(1+ g \cdot \nabla_{u}A [u]
\right) +O (\hat{u}) \ .
\end{equation}
This can also be written as:
\begin{equation}\label{d30}
\left(\rmd \hat{u}/ \rmd \hat{w} \right)^{t}\cdot B[u] = \hat{B}[w]\cdot \rmd
\hat{w}/\rmd \hat{u}
\end{equation}
Equivalently:
\begin{equation}\label{d31}
 B[u] = \left ( \rmd \hat{w}/\rmd \hat{u}\right)^{t}\cdot
\hat{B}[w]\cdot \rmd \hat{w}/\rmd \hat{u}
\end{equation}
Finally, the relation between the $\hat B$ and $B$ functionals can be written as:
\begin{equation}\label{d32}
B[u] = \left(1+ g \cdot \nabla_{u} A [u] \right)^{t}\cdot \hat{B}[w]\cdot \left(1+ g \cdot \nabla_{u}A [u]
\right)\ ,
\end{equation}
where in this relation $u$ and $w$ are related via:
\begin{equation}\label{d22bis}
w-u = g\cdot A[u]  = g \cdot \hat{A}[w]
\end{equation}
It can also be written equivalently as
\begin{equation}\label{d32ter}
\hat B[w] = \left(1- g \cdot \nabla_{w} \hat A [w] \right)^{t}\cdot B[u]\cdot \left(1-g \cdot \nabla_{u} \hat A
[w] \right) \ .
\end{equation}

\subsection{Evaluation of the functionals}

Let us now evaluate the functionals $\hat A_{xt}[\{w_{yt} \}]$ and $\hat B_{xt,x't'}[\{w_{yt} \}]$ in various
situations.

Let us consider first $T>0$ {\it equilibrium dynamics}, i.e.\ a driving function $w_{yt}$ which evolves
infinitely slowly between $w_{y t_1}=w_1(y)$ and $w_{y t_2}=w_2(y)$ such that the system always remains in
equilibrium (i.e.\ we consider the limit $t_2-t_1 \to \infty$ at fixed $w_1-w_2$). From (\ref{rel}) and STS it is
clear that $\hat A[w]=\hat A[0]=0$ in that limit. This implies $u=w$ in (\ref{d22bis}) and also $A[u]=0$. It
then implies that $\hat B[w]=B[w]$ and one recovers the results of Ref. \cite{LeDoussal2006b,LeDoussalinprep}
for the statics using replicas. More precisely one expects in that limit that $\hat B_{yt,y't'}[w]=
B_{yt,y't'}[w]= \partial_{w_1(y)} \partial_{w_2(y')} R[w_1,w_2]$ where $\hat R=R$ is the two-replica functional
of the statics. Hence it is a statement {\it only} about the infinitely separated time part of the $B[w]$
functional and not about the smaller time separation part (which contains the renormalization of $\eta$ and
highly complicated activated dynamics as described in \cite{BalentsLeDoussal2003}).

Consider now $T=0$ and $w_{yt} = w(t)= v t$. From translational invariance $\hat A_{xt}[w]$ (see e.g.
(\ref{rel})) can only be a time and space independent $v$ dependent constant (assuming boundary conditions do
not break translational invariance) which we choose to call $f(m,v)$. Because of (\ref{d22bis}) one must have
the equality
\begin{equation} \label{d32ter2}
\hat A_{xt}[w] = A_{xt}[u] = f(m,v)
\end{equation}
and $u = w - m^{-2} f(m,v) = v t - m^{-2} f(m,v)$ in (\ref{d22bis}). The difference with the equilibrium statics
is that this constant is non-zero. This is allowed despite the STS symmetry because we are considering the $T=0$
limit first and the fact that $w(t)$ depends on $t$ cannot be ignored even for $v=0^+$. In that limit one
has $f_c(m,v=0^+)=f_c^{\uparrow}=-f_c(m,v=0^-)$. Of course the fact that the constant depends uniquely on $v$
assumes some ergodicity property, similarly if $w(t)$ is a more complicated adiabatic function there could be in
general some history dependence. These issues have been discussed in Section \ref{s:N-component}. For $N=1$ we
will rely on Middleton's theorem \cite{Middleton1992} which proves unicity of the solution. Note that at $T>0$ (\ref{d32ter2})
remains true with a $f(m,v,T)$ such that $f(m,0,T)=0$ in agreement with the discussion of the previous
paragraph.

Since the derivative of a constant is zero, using (\ref{d32ter2}) and (\ref{d32}) we find that at $T=0$
with the choice $w= v t$ one has:
\begin{eqnarray}
&& B_{y t, y' t'}[u= u(t)= v t - m^{-2} f(m,v)] \nn \\
&& \qquad = \hat B_{y t, y t'}[w] = \hat \Delta(v (t-t'))\ .
\end{eqnarray}
Denoting $B_{y t, y t'}[u]=\Delta(u(t)-u(t'))$ gives
\begin{equation}
\Delta(w)=\hat \Delta(w)\ ,
\end{equation}
a result on which is based our measurement of $\Delta(w)$ here
and in \cite{RossoLeDoussalWiese2006a}.

\section{Calculation of some integrals}

\label{app:integral}

Here we compute the integrals in Eq.\ (\ref{d18a}) of the main text. We need:
\begin{eqnarray}
 \int_{a}^{\infty} \rmd s \, \rme^{-s} \ln s &=& e^{-a} \ln a - \mathrm{Ei}(-a) \\
 \mathrm{Ei}(-a) &=& - \int_{a}^{\infty} dt e^{-t}/t
\end{eqnarray}
This gives:
\begin{eqnarray}
\label{d18ab}
\lefteqn{ \tilde \Delta(W)} \nn\\
& =& (1-\rme^{- W})
\int_{0}^{\infty} \rmd a \,e^{-a (1-\rme^{- W})} \ln (a)
 e^{- a\, \rme^{- W }} \ln (a\, \rme^{- W }) \nn \\
&& - (1-\rme^{- W})  \int_{0}^{\infty} \rmd a\, e^{-a (1-\rme^{- W})} \ln (a) \mbox{Ei}(- a\, \rme^{- W }) \nn \\
&& + \rme^{- W } (
\gamma_E^2 + \frac{\pi^2}{6} + X \gamma_E) - \gamma_E^2
\end{eqnarray}
Consider the integral:
\begin{eqnarray}
&-&  \int_{0}^{\infty} \rmd a\, e^{-a (1-\rme^{- W})} \ln (a) \, \mbox{Ei}(- a\, \rme^{- W }) \\ \nn
& =&  \int_{0}^{\infty}
\rmd a \,e^{-a (1-\rme^{- W})} \ln a \int_{a\, \rme^{- W }}^{\infty} \rmd t\, e^{-t}/t \nn\\
& =&  \int_1^\infty \rmd b \int_{0}^{\infty} \rmd a e^{-a (1-\rme^{- W})} \ln a \int_{a\, \rme^{- W }}^{\infty} \rmd t\,
e^{-t b} \nn\\
& =& \int_1^\infty \frac{\rmd b}{b} \int_{0}^{\infty}
\rmd a  \ln a e^{-a (1-\rme^{- W}) - a\, \rme^{- W } b} \nn\\
& =& \int_1^\infty \frac{\rmd b}{b} \frac{1}{(1-\rme^{- W}) + b \rme^{- W }
} \nn\\
&& \times \int_{0}^{\infty} \rmd a\, e^{-a} (\ln a - \ln((1-\rme^{- W}) + b \rme^{- W })) \nn\\ \nn
& =& - \gamma_E \frac{W}{1-e^{-W}} - \frac{1}{6(1-e^{-W})}\times \\
&&  \times\left[\pi^2 - 3 W^2 + 3 \ln^2(e^W-1) + 6 \text{Li}_2\left(\frac{1}{1-e^X}\right) \right] \nn
\end{eqnarray}
This yields
\begin{equation}\label{c96}
\tilde \Delta (x) = \frac{x^2}{2}-\frac{1}{2} \log
^2\left(e^x-1\right)-\text{Li}_2\left(\frac{1}{1-e^x}\right)
\end{equation}
which can be rewritten as (\ref{c97}) in the main text.

\section{Avalanche process and Markov chain}
\label{app:markov}

It is useful to recast the avalanche process for the discrete model of uncorrelated forces
as a Markov chain, and define an algorithm for easy use in the numerics.

Let us index jumps by $n$, they occur at positions $u_n^-:=u_n$ (integer), where the force is $F_n$ (real). Note
that the $w$-position of the jump us $w_n=u_n - m^{-2} F_n$. Given $(u_n,F_n)$ one finds the next jump
$(u_{n+1},F_{n+1})$ by the following algorithm:
\begin{align}
 u_{n+1}&=u_n+s_n \quad , \quad F_{n+1} = f_{s_n} \\
 s_n&=\min(p=1,2,\ldots \text{such that} ~~ f_p < F_n + m^2 p )
\end{align}
where $f_1,f_2,..$ are a sequence of i.i.d.\ random variables of distribution $P_0(f)$. The $s_n$ (integers
greater or equal to one) are the size of the avalanche and are determined at the same time. The variables
$(s_n,F_{n+1})$ form a Markov chain with conditional probability $P(s_n,F_{n+1}|F_{n})$.
\begin{align}
& P(s,F'|F) = P_0(F') \theta(F+m^2 s > F')  \prod_{k=1}^{s-1} H(F+m^2 k)  \\
& P(s|F) = (1-H(F+m^2 s) \prod_{k=1}^{s-1} H(F+m^2 k)\\
& H(F)=\int_F^\infty df P_0(f)
\end{align}
which is normalized $\sum_{s \geq 1} P(s|F)=1$ using that $1-H_1 + (1-H_2) H_1 + (1-H_3) H_1 H_2 + ..= 1 - H_1
H_2 H_3 ...$ and the fact that the $H_k$ tend to zero as $k$ increases.

Starting from, for instance, $P_0(F_0)$, the distribution for $F_n$ is given by:
\begin{align}
& P_n(F_n) = P(F_n|F_{n-1})P(F_{n-1}|F_{n-2}).. P(F_1|F_0) P_0(F_0) \\
& P(F'|F)= \sum_{s \geq 1} P(s,F'|F)
\end{align}
It converges to a stationary probability, noted $\tilde P(F)$, which satisfies:
\begin{eqnarray}
&& \sum_{s \geq 1} P(s,F'|F) \tilde P(F) = \tilde P(F') \label{steady}
\end{eqnarray}
Once we find $\tilde P(F)$ the joint distribution of $(s_n,F_{n+1},F_n)$ is known:
\begin{eqnarray}
P(s_n,F_{n+1},F_n) = P(s_n F_{n+1}|F_{n}) \tilde P(F_n)
\end{eqnarray}
Hence the avalanche size distribution is:
\begin{eqnarray}
P(s) = \int dF P(s|F) \tilde P(F)
\end{eqnarray}
The sequence of waiting times is such that:
\begin{eqnarray}
w_{n+1}-w_n=W_n= s_n - m^{-2}(F_{n+1}-F_n)
\end{eqnarray}
Hence the joint size and waiting time distribution is:
\begin{align}
P(W,s) =& \int dF dF' \delta(W-s + m^{-2}(F'-F)) \nn\\
&\qquad \times P(s,F'|F) \tilde P(F)
\end{align}
The problem is thus to determine the solution of (\ref{steady}). One can formally write it as an
infinite product:
\begin{align}
& \tilde P(F_0) = \int dF_1 dF_2 .. \prod_{k=1}^\infty P(F_{k-1}|F_k)  \label{infinite} \\
& P(F'|F) = P_0(F') \sum_{s \geq 1} \theta(F+m^2 s > F')  \prod_{k=1}^{s-1} H(F+m^2 k)
\end{align}
which however also contains an infinite number of integrations. This method does not seem very practical (see
however \cite{math}) and in the text we obtain the result by another method.

\section{General results for uncorrelated disorder}
\label{app:details}

\subsection{Proof that $P(w)=P(s)$}

Suppose, that the following variable transformation holds between $j-w$, and the corresponding
\begin{equation}\label{175}
f(j-w)= a_{w}(j)
\end{equation}
up to a constant shift and a rescaling. We recall that $f(y)=\exp(y)$ (Gumbel), $f(y)=y^{\alpha}$, $y>0$
(Weibull), and $f(y)= (- y)^{-\alpha}$, $y<0$, (Frechet).

Then
\begin{equation}\label{176}
P_{w}(w',s>S) = \int \rmd y f''(y) \rme^{-f(y+W)-f(y+s)+f(y)}\ ,
\end{equation}
where integration bounds depend on the class (real axis for Gumbel, positive axis for Weibul and negative axis
for frechet). Using the relations in the text one finds that the joint waiting-time and avalanche size
distribution is
\begin{equation}\label{177}
P(w>W,s>S) =  \frac{ \displaystyle\int_{0}^{\infty}\rmd y\, f''(y) \rme^{-f(y+W)-f(y+S)+f(y)}}{
\displaystyle\int_{0}^{\infty}\rmd y \,f''(y) \rme^{-f(y)}}\ .
\end{equation}
where the denominator is such that the distribution is properly normalized. Also note that for some choice of
variables, one has to be careful with the bounds of integration, see the Fr\'echet classe, eq.\ (\ref{173}).

This formula shows that avalanche-size and waiting-time distribution are equal for all microscopic
(uncorrelated) disorder: $P(S)=P(W)$.

\subsubsection{Distribution of local area $a=w s$ }

The distribution of the local area $a=w s$ defined in the text can be obtained as follows:
\begin{align}\label{PA}
&P(a>A)=  \int\limits_{0}^{\infty}\! \rmd W \int\limits_{0}^{\infty}\!\rmd S \left[\partial_{w}\partial_{s} P(w>W,s>S) \right] \theta(WS-A) \nn \\
&= \int\limits_{0}^{\infty}\! \rmd W \int\limits_{0}^{\infty}\!\rmd S \, P(w>W,s>S) \partial_{W}\partial_{S}\theta(WS-A)\nn\\
&= \int\limits_{0}^{\infty}\! \rmd W \int\limits_{0}^{\infty}\!\rmd S \,  P(w>W,s>S) \nn\\
& \qquad \qquad \qquad \times[ \delta(A-WS)-WS\, \delta'(A-WS)]\nn\\
&= -A\frac{\partial}{\partial A} \int\limits_{0}^{\infty}\! \rmd W \int\limits_{0}^{\infty}\!\rmd S \,
P(w>W,s>S) \delta(A-WS)
\end{align}
We note a subtle point that when writing the last term as the derivative w.r.t.\ $A$ of $- A \int P(w>W,s>S)
\delta(A-WS)$, the explicit derivative of $A$ cancels the first term.

Inserting the integral representation (\ref{177}) for $ P(w>W,s>S)$ yields
\begin{align}\label{PA3a}
&P(a> A)= \nn\\
&\ \quad  \frac{\displaystyle -\frac {A \partial}{\partial A} \int\limits_{0}^{\infty}\! \frac{\rmd W}{W}
\int\limits_{0}^{\infty}\!\rmd y \,  f''(y) \,\rme^{-f(y+W)-f(y+A/W)+f(y)}} { \displaystyle\int_{0}^{\infty}\rmd
y \,f''(y) \rme^{-f(y)}}
\end{align}
Note that boundary terms from the partial integration in (\ref{PA}) vanish except for $w=0$ or $s=0$, in which
case $A=0$. Thus we could possibly have a term $\sim \delta(A)$. However we know that for $w=0$ we do not have a
diverging probability for an avalanche, and vice versa for $w$ and $s$ exchanged. The above result can also be
written as
\begin{align}\label{PA3b}
&P(a> A)= \nn\\
&\ \quad  \frac{\displaystyle -\frac {A \partial}{\partial A}\int\limits_{0}^{\infty}\!{\rmd w}
\int\limits_{0}^{\infty}\!\rmd y \,  f''(y) \,\rme^{-f(y+\rme^{w})-f(y+A \rme^{-w})+f(y)}} {
\displaystyle\int_{0}^{\infty}\rmd y \,f''(y) \rme^{-f(y)}}
\end{align}
For the Gumbel class, $f(y)=\rme^{y}$, and this yields the formula given in the text.

\section{First-passage times and avalanche distributions}

\label{app:passage}

In this Appendix we recall the basic method of the backward diffusion equation to compute distributions of first
passage times on a half line and an interval, and then extend it to compute the first-passage-time
distribution in the velocity-diffusion equation of the ABBM model.

\subsection{First-passage-time distribution on a half line}

Let us call $T(w';w)$ the first-passage time at $w'$ of a BM starting at position $w<w'$ at time 0. Let us
recall that the generating function
\begin{eqnarray}
G(w',p;w) = \langle e^{-p T(w';w)} \rangle
\end{eqnarray}
satisfies the diffusion equation and boundary conditions
\begin{eqnarray}\label{E2}
&& D \partial_{w}^2 G + b \partial_{w} G = p G \label{genfirst} \\
&& G(w',p;w')=1 \quad , \quad G(w',p=0;w=-\infty)=0 \nn
\end{eqnarray}
for $b>0$. To see that, one introduces the diffusion kernel on $x \in [-\infty, w']$ in presence of an absorbing boundary at $x=w'$, which satisfies:
\begin{eqnarray}
&& \partial_t P = D \partial_{x}^2 P - b \partial_{x} P  = D \partial_{w}^2 P + b \partial_{w} P \\
&& P(x, t=0|w,0)=\delta(x-w) \quad , \quad P(w',t|w,0)=0 \nn
\end{eqnarray}
By definition of the exit time one has for $p>0$:
\begin{eqnarray}
 G(w',p;w)& =& - \int_0^\infty \rmd t\, e^{-p t} \partial_t \int_{x<w'} P(x ,t|w,0)\nn \\
&=& 1-p \int_{x<w'}\hat P(x ,p|w,0)\ ,
\end{eqnarray}
with $\hat P$ the Laplace transform of $P$. The latter satisfies:
\begin{equation}
 p \hat P(x ,p|w,0) - \delta(x-w) = (D \partial_{w}^2  + b \partial_{w})\hat  P(x ,p|w,0)\ .
\end{equation}
Multiplying with $-p$ and integrating over $x$ from $-\infty$ to $w'$ yields (\ref{genfirst}).

The solution of (\ref{E2}) including the boundary conditions is
\begin{equation}
 G(w',p;w)= \exp\left(\frac{b - \sqrt{4 p D + b^2}}{2 D} (w'-w)\right)\ .  \label{laplace}
\end{equation}
For $b \geq 0$ it satisfies $\lim_{p \to 0} G(w,p;w')= 1$. It is then inverted into the probability
$P(u;w'-w) du$ that $T(w';w) \in [u,u+du[$:
\begin{equation}
P(u;W) = \frac{W}{\sqrt{4 \pi D}} u^{-3/2} \exp \left( - \frac{(b u- W )^2}{4 D u}  \right)
\theta(u) \label{firstp}\ ,
\end{equation}
where $W=w'-w$. Note that for negative drift $b<0$ one has:
\begin{eqnarray}
&& \lim_{p \to 0} G(w,p;w')= 1 - q   \\
&& q := \mbox{Prob}( T(w';w) =  \infty) = 1 - \exp\left(- \frac{|b| (w'-w)}{D}\right) \nn
\end{eqnarray}
since in that case there is a finite probability $q$ that the walk starting at $w$ never hits $x=w'$.

The Laplace transform of the probability to be at $w$ at time $t$ can be written as the probability to arrive there for the first time, and then repeatedly going with and against the drift, coming always back to $w$: 
\begin{eqnarray}
&&\!\!\mbox{LT}_{t \to p} \frac{w_0}{\sqrt{4 \pi D t}} e^{-(w-b t)^2/(4 D t)}
=
\frac{w_0}{\sqrt{4 p D + b^2}} e^{\frac{b - \sqrt{4 p D + b^2}}{2 D} w}
\nn\\
 && \approx  \frac{w_0^2}{2 D} G(0,p;-w) \frac{1+ \Pi_0^{-}(p)}{1 - \Pi_0^{+}(p) \Pi_0^{-}(p)} \label{geom}\ ,
\end{eqnarray}
where $\Pi_0^+(p) = G_{b=|b|}(w_0,p;0)$ is the return probability going along the drift, and  $\Pi_0^-(p) =
G_{b=-|b|}(w_0,p;0)$ going against the drift (with $w_0>0$ a small cutoff which allows to cross $0$ in the
microscopic model). Eq. (\ref{geom}) expresses that the probability to be near $x=0$ is a sum of $n$-th
passage time events; the factor $w_0^2/2 D = dt = dw^2/(2 D)$ is the change of measure from time to space.

Similarly consider the problem of the last passage time $t_0$ of a Brownian at $w=0$, with initial condition
$w=0$ at $t=0$. Its Laplace transform can again be expressed as a geometric series \begin{eqnarray}\nn
 \Pi_{\mathrm{last}}(p) &=& 
 \left[1- \Pi_0^-(0)\right] \frac{\Pi_0^{+}(p)}{1 - \Pi_0^{+}(p) \Pi_0^{-}(p)}\\
&& \approx \frac{1}{\sqrt{1 + \frac{4 D p}{b^2}}}
\end{eqnarray}
Hence:
\begin{eqnarray}
&& \Pi_{\mathrm{last}}(t_0) = \frac{|b|}{\sqrt{4 \pi D t_0}} e^{-b^2 t_0/(4 D)} \\
&& \langle t_0 \rangle = \frac{2 D}{b^2} \quad , \quad \langle t_0^2 \rangle =   \frac{12 D^2}{b^4}
\end{eqnarray}

\subsection{First-passage-time distribution on an interval}

Consider now a Brownian starting at $w$ in an interval $[w_a,w_b]$. Consider the functions $G_a(w,p)$,
$G_b(w,p)$ and $G(w,p)=G_a(w,p)+G_b(w,p)$ which satisfy the same differential equation (\ref{genfirst}) but with boundary
conditions:
\begin{eqnarray}
 G_a(w_a,p)&=&1 \quad , \quad G_a(w_b,p)=0\nn \\
 G_b(w_b,p)&=&1 \quad , \quad G_b(w_a,p)=0 
 \end{eqnarray}
Then $G(w,p) = \langle e^{-p T_{ab}(w)} \rangle$ is the generating function for the first exit time $T_{ab}(w)$
of the interval $[w_a,w_b]$ by a walker starting at $w \in [w_a,w_b]$ at time zero. It satisfies
(\ref{genfirst}) with boundary conditions $G(w_a,p)=1$ and $G(w_b,p)=1$. One finds:
\begin{eqnarray}
 G_a(w,p) &=& e^{\frac{b}{2 D} (w_a-w)} \frac{\sinh(\mu (w_b-w))}{\sinh(\mu (w_b-w_a))} \\
  G_b(w,p) &=& e^{\frac{b}{2 D} (w_b-w)} \frac{\sinh(\mu (w-w_a))}{\sinh(\mu (w_b-w_a))} \\
 \mu &=& \frac{1}{2 D} \sqrt{4 D p + b^2}
\end{eqnarray}
One checks that $G(w,p=0)$ irrespective of the sign of $b$ as the walk is certain to exit the interval. The
function $G_a(w,p)$ (resp. $G_b(w,p)$) is the same generating function restricted to walks exiting in $w_a$
(resp. $w_b$), with normalizations:
\begin{eqnarray}
&& p_a = G_a(w,p=0) = \frac{\exp(\frac{b}{D} (w_b-w))-1}{\exp(\frac{b}{D} (w_b-w_a))-1} \\
&& p_b = G_b(w,p=0) = 1-p_a
\end{eqnarray}
for $b>0$.

One way to derive these results is to introduce the diffusion kernel $P(x, t|w,0)$ with absorbing boundary
conditions at $x=w_a$ and $x=w_b$. Its expression reads in Laplace:
\begin{eqnarray}
&&\!\!\! P(x ,p|w,0) = \nn\\
&&= \left\{  \begin{array}{l}\displaystyle
 e^{\frac{b}{2 D}(x-w)} \frac{\sinh(\mu (x-w_a)) \sinh(\mu (w_b-w_0)) }{D \mu \sinh(\mu (w_b-w_a))} \nn \\
  \hspace{0.5\columnwidth}\mbox{ for } w_a < x < w \qquad  \\
  \displaystyle
  e^{\frac{b}{2 D}(x-w)} \frac{\sinh(\mu (w_b-x)) \sinh(\mu (w-w_a)) }{D \mu \sinh(\mu (w_b-w_a))} \nn \\
  \hspace{0.5\columnwidth}\mbox{ for }  w < x < w_b\qquad 
\end{array}\right.\\
\end{eqnarray}
It satisfies the diffusion equation with the proper boundary conditions. This is obtained as follows
\begin{eqnarray}\nn
 G(p,w) &=& - \int_0^\infty \rmd t\, e^{-p t} \partial_t \int_{w_a}^{w_b} \rmd x\, P(x, t|w,0) \\
& =& D (\partial_x P(x ,p|w,0) |_{x=w_a} - \partial_x P(x, p|w,0) |_{x=w_b})\nn\\
\end{eqnarray}
where the first term is $G_a(p,w)$ and the second $G_b(p,w)$. They represent the flux from each boundary.

\subsection{Avalanche distribution for the ABBM model}
\label{app:avalanchedistrib}

Let $s=T({\sf v}',{\sf v})$ be the first passage ``time'' from ${\sf v}$ at ``time'' $u=0$ to ${\sf v}'<{\sf v}$ at
``time'' $s$, the avalanche size (here defined with some velocity cutoff ${\sf v}'$). The generating function
$G({\sf v}',p;{\sf v})= \overline{\exp(-s p)}$ satisfies the backward diffusion equation:
\begin{eqnarray}
&& \sigma \partial_{\sf v}^2 G + (\frac{m^2 v}{{\sf v}} - m^2) \partial_{\sf v} G = p G  \label{eqGa} \\
&& G({\sf v}',p;{\sf v}')=1 \quad G({\sf v}',p;+\infty)=0\qquad 
\end{eqnarray}

\subsubsection{Solution without the drift term}
From now on we denote:
\begin{eqnarray}
  x = \frac{m^2 v}{\sigma}
\end{eqnarray}
The solution, if one first drops the drift term is:
\begin{eqnarray}
&& G({\sf v}',p;{\sf v})= \left(\frac{{\sf v}}{{\sf v}'}\right)^{\frac{1-x}{2}} \frac{K_{\frac{|1-x|}{2}}({\sf
v}\sqrt{p/\sigma})}{K_{\frac{|1-x|}{2}}({\sf v}'\sqrt{p/\sigma})} \label{fullform}
\end{eqnarray}
One expects that it describes correctly small avalanches.

There are two cases. For $x \leq 1$ one has $G(p=0)=1$, i.e.\ the velocity is certain to reach any fixed ${\sf
v}'>0$. The leading behaviour at small ${\sf v}$,${\sf v}'$  is then for $x<1$:
\begin{eqnarray}
 G-1 &=& - C (p/4 \sigma)^{(1-x)/2} \label{simplified} \\
 C &=& - \frac{\Gamma(\frac{x-1}{2})}{\Gamma(\frac{1-x}{2})} ({\sf v}^{1 - x} - ({\sf v}')^{1 -x })
\end{eqnarray}
Inverse Laplace transform from $p$ to $s$ yields a distribution $1/s^\tau$ with the value of $\tau=(3-x)/2$
given in the text. The small avalanche cutoff $s_0$, necessary since $1/s^\tau$ is not normalizable at small $s$
for $\tau>1$, is provided by ${\sf v}$,${\sf v}'$ and its full form can in principle be obtained by inverse
Laplace transform from $p$ to $s$ of (\ref{fullform}). Its order of magnitude is easy to read from
(\ref{fullform}) as $s_0 \sim {\sf v}^2/\sigma$ (see below a more precise estimate).

One notes that the limit ${\sf v}' \to 0$ can be taken for $x<1$ for any ${\sf v}$. It is then easy to inverse
Laplace transform (\ref{fullform}) in that limit and to obtain the probability that if the velocity at $u=0$ is
${\sf v}$, then the next stopping point $u'$ is in the interval $u' \in [s,s+ds]$ as:
\begin{eqnarray}
&& G(0,s;{\sf v}) ds = \frac{1}{\Gamma(\mu)} \frac{ds}{s} (\frac{s_0}{s})^\mu e^{-s_0/s} \\
&& s_0 = {\sf v}^2/(4 \sigma)\ , \qquad \mu = (1 - x)/2
\end{eqnarray}
Next one can use the stationary distribution to find the probability that choosing a $u=0$ the next stopping
point is at $s$:
\begin{eqnarray}
 \int_0^\infty d{\sf v} G(0,s;{\sf v}) P_{\mathrm{eq}}({\sf v}) = \frac{2 \sigma}{\Gamma(\mu) \Gamma(x) v s^\mu}
(\frac{m^2}{\sigma})^x (4 \sigma)^{-\mu}  \nonumber
\end{eqnarray}
where to be consistent we have assumed $s \ll s_m=\sigma/m^4$.

For $x > 1$, the probability of ever reaching ${\sf v}'>0$ is $G(p=0)=(\frac{{\sf v}'}{{\sf v}})^{x-1} < 1$,
from (\ref{fullform}). At small ${\sf v}$,${\sf v}'$ (and of the same order) the following expansion holds for
$1<x<3$:
\begin{equation}
G = \left(\frac{{\sf v}'}{{\sf v}}\right)^{x-1} \left[ 1 + \frac{\Gamma(\frac{1-x}{2})}{\Gamma(\frac{x-1}{2})} \left({p \over 4
\sigma}\right)^{\frac{x-1}2} ({\sf v}^{x-1} - ({\sf v}')^{x-1 }) \right]
\label{xgt1}
\end{equation}
This shows that, {\it conditioned to returning near zero velocity} the avalanches size distribution has again a
tail $1/s^\tau$ for $s \gg s_0$, with $\tau=(1+x)/2 > 1$.

\subsubsection{Solution with the drift term}

Taking into account the drift term in (\ref{eqGa}) reintroduces the large-size cutoff for avalanches at
$s_m=\sigma/m^4$.

It is also possible to solve the full problem, with the drift. One finds for the Laplace transform of the first-passage time:
\begin{eqnarray}
&&\!\!\! G({\sf v}',p;{\sf v})= \left(\frac{{\sf v}}{{\sf v}'}\right)^{1- x} e^{\frac{m^2 - \sqrt{m^4+ 4 p \sigma}}{2 \sigma}
({\sf v}-{\sf v}')} \nn\\
&&\qquad \times \frac{U \left(1 - {\sf v} \frac{m^2}{2 \sigma} (1 + \frac{m^2}{\sqrt{m^4+ 4 p \sigma}}) , 2- x,
\frac{\sqrt{m^4+ 4 p \sigma}}{\sigma} {\sf v}\right)}{ U\left(1 - {\sf v}' \frac{m^2}{2 \sigma} (1 + \frac{m^2}{\sqrt{m^4+
4 p \sigma}}) , 2- x, \frac{\sqrt{m^4+ 4 p \sigma}}{\sigma} {\sf v}'\right)}\nn\\
\end{eqnarray}
which reduces, for $v=0^+$ to the expression obtained for quasi-static avalanches:
\begin{eqnarray}
 G({\sf v}',p;{\sf v})= e^{\frac{m^2 - \sqrt{m^4+ 4 p \sigma}}{2 \sigma} ({\sf v}-{\sf v}')}
\end{eqnarray}
from which, after inverse Laplace, (\ref{distribava}) was obtained.

One can now check that $G(p=0)=1$ for all $x$ and ${\bf v}, {\bf v}'> 0$; hence thanks to the drift the walk
comes back infinitely often, as announced in the text.

Let us consider the leading behaviour at small ${\sf v}$,${\sf v}'$. The expansion has the form:
\begin{eqnarray}
 G = \frac{A (1 + O({\sf v}^2)) + B {\sf v}^{1 - \frac{m^2 v}{ \sigma}} (1 + O({\sf v}))}{A (1 + O(({\sf
v}')^2)) + B ({\sf v}')^{1 - \frac{m^2 v}{ \sigma}} (1 + O({\sf v}'))} \nonumber
\end{eqnarray}
where $A$ and $B$ are complicated functions of $p$. Here we focus on the case $v<\sigma/m^2$, then:
\begin{eqnarray}
G - 1 = \frac{B}{A} \left({\sf v}^{1 - \frac{m^2 v}{ \sigma}} - ({\sf v}')^{1 - \frac{m^2 v}{ \sigma}} \right)\ .
\end{eqnarray}
This yields
\begin{eqnarray}
 G-1 &=& \left(\frac{\sqrt{m^4+ 4 p \sigma}}{\sigma}\right)^{1 - \frac{m^2 v}{ \sigma}} \frac{\Gamma(-1 + \frac{m^2
v}{\sigma})}{\Gamma(1- \frac{m^2 v}{\sigma})}\nn\\
&&\times \frac{\Gamma(1 -  \frac{m^2 v}{2 \sigma} (1 +
\frac{m^2}{\sqrt{m^4+ 4 p \sigma}})) }{ \Gamma(\frac{m^2 v}{2 \sigma} (1 - \frac{m^2}{\sqrt{m^4+ 4 p \sigma}}))}\nn\\
&& \times
\left({\sf v}^{1 - \frac{m^2 v}{\sigma}} - ({\sf v}')^{1 - \frac{m^2 v}{\sigma}}\right) \ .
\end{eqnarray}
For $p \gg s_m^{-1}=m^4/\sigma$ one can check that this expression reproduces (\ref{simplified}) above hence in
that case the distribution of small avalanches can be found neglecting the drift. From this expression Laplace
inversion allows, in principle, to obtain the full avalanche distribution. We will not attempt to perform it but
note that there is an additional pole structures for $p=p_n$:
\begin{eqnarray}
&& p_n = - \frac{m^4}{4 \sigma} \frac{n (n-x)}{(n-\frac{x}{2})^2} \quad , \quad n=1,2,\ldots\ ,
\end{eqnarray}
which implies a decay
\begin{eqnarray}
P(s) \sim e^{- \frac{(1-x)}{(2-x)^2} s/s_m }
\end{eqnarray}
at large $s \gg s_m$.

\section{Some 1-point observables for the Brownian force landscape}
\label{app:E}
We present here a few partial results for 1-point observables for the Brownian force landscape, deferring a more complete study to the future.

Note that since $F(u)$ is an unbounded Brownian landscape, it has an infinite threshold force $2
f_d=\max F(u) - \min F(u)$. However the model studied here of a particle dragged by a parabolic well is
well defined, and from it one defines an average critical force:
\begin{eqnarray}
f_c(m) := m^2 \overline{[w- u(w)]} \sim m^{2 - \zeta}
\end{eqnarray}
which diverges \footnote{Note that in fact this was already the case for some members of class I.} as $m \to 0$
for $\zeta>2$. As discussed in the main text, for $\gamma=1$ one has $\zeta=4$.

Let us now examine the 1-point probability of the process $u(w)$. Since it is a long-range correlated landscape, there is
a subtlety linked to the choice of boundary conditions.

\subsection{Special boundary conditions}
If we first fix $u(w=0)=0$, the probability $P_w(u)$ that $u(w)=u$ for $u>0$ is equal
to the probability that the first passage time of $\phi(u')$ at $w$ is $u$, starting at zero.
Hence we find
\begin{eqnarray}
 P_w(u) = P(u;w) \ ,   \label{res0}
\end{eqnarray}
where $P(u;w)$ is defined in (\ref{firstp}). In fact, it follows from the Markov property that,
if we impose
$u(w=0)=0$ and leave the future unconstrained, the $n$-point probability is:
\begin{eqnarray} \label{resmarkov-b}
 P_{w_1,..w_n}(u_1,..,u_n) &=& P(u_1;w_1) P(u_2-u_1;w_2-w_1) \nn \\
&& \times \ldots\times P(u_n-u_{n-1};w_n-w_{n-1}) \nn\\
\end{eqnarray}
for $0 \leq w_1 \leq w_2 ... \leq w_n$. Computing the moments of (\ref{res0}) one finds:
\begin{eqnarray}
 \overline{u(w)-w} &=& 0 \\
 \overline{(u(w)-w)^2} &=& 2 D w =  \frac{2 \sigma}{m^4} w   \label{second-b} \ ,
\end{eqnarray}
hence one finds that the critical force $f_c(m)$, which is proportional to the average extension of the spring
pulling the particle is {\em zero}! On the other hand, the total area of the hysteresis loops per unit length
cannot vanish - in fact from (\ref{second-b}) we can guess that it should grow as $\sqrt{w}$. Hence by contrast
with the case of the uncorrelated force landscape, these two quantities cannot be equal.

To understand this apparent paradox let us note that to insure $u(w=0)=0$ one needs to impose rather strong
conditions, e.g. $F(u)=0$ for all $u \leq 0$. Otherwise, there is a non-zero probability that the BM has taken
values $\phi(u)>w=0$ in the past, i.e.\ for $u<0$, which is in contradiction with $u(w=0)=0$. If we now want to
use the hysteresis loop argument in a symmetric way, it would require a similar choice at some prescribed
$u(w=W)=0$. But then (\ref{res0}) and (\ref{resmarkov}) do   not hold anymore (it holds for a Brownian
unconstrained in the future). Since that procedure produces a non-stationary result, we do not discuss it
further.

\subsection{A more generic situation}

First note that shifting $F(u)$ by a constant leaves (\ref{corr1}) unchanged, but also shifts
$\overline{w-u(w)}$ while it does not change the area of the hysteresis loop. Hence comparing $f_c(m)=m^2
\overline{[w-u(w)]}$ with the area of the hysteresis loop makes sense only for a statistically symmetric
landscape. To get rid of this unimportant shift we can consider a distribution of forces symmetric around zero by
 setting $F(u=0)=0$. This is one way to fix the problematic zero mode of the Brownian landscape. This does not
mean that $u_1:=u(w=0)$ is necessarily at zero. $u_1$ is the position of the first crossing of $\phi(u)=w=0$ by
the BM. The probability distribution of $u_1 \leq 0$ is given by (see Appendix \ref{app:passage}):
\begin{eqnarray}
 \Pi_{\mathrm{last}}(u_1) = \frac{m^2}{\sqrt{4 \pi \sigma |u_1|}} e^{- m^4 |u_1|/(4 \sigma)}
\theta(-u_1)
\end{eqnarray}
This yields a critical force:
\begin{eqnarray}
&& f_c(m) = - m^2 \overline{u_1} =  \frac{2 \sigma}{m^2} \label{fcnew}
\end{eqnarray}
which makes more sense, i.e.\ it is positive and obeys the expected scaling.

However, to be a bit more general, what we have just computed is the conditional probability
$P_{w_1}(u_1|F(u_a)=f_a)$ with the choice $u_a=0$ and $f_a=0$ (which can be realized using shifts of the axis)
{\it and} the additional choice $w_1=u_a-f_a/m^2$ (hence $w_1=0$). One wonders whether the critical force
depends on that choice.

To answer this question, we need to compute $P_{w_1}(u_1|F(0)=0)$ as a function of $w_1$. There are two cases: If $w_1<0$ then
$u_1 \geq 0$ and one has
\begin{eqnarray}
&& \!\!\! P_{w_1 \leq 0}(u_1|F(0)=0) \nn \\
&& = \theta(-u_1) \mbox{LT}^{-1}_{p \to -u_1} G_{b=|b|}(|w_1|,p;0) \Pi_{\mathrm{last}}(p)  \ .\qquad 
\end{eqnarray}
This gives
\begin{equation}
P_{w_1 \leq 0}(u_1|F(0)=0) = \theta(-u_1) \frac{|b|}{\sqrt{4 \pi D |u_1|}}
e^{-\frac{(|b| |u_1|-|w_1|)^2}{4 D |u_1|}}\ . \nn
\end{equation}
One finds
\begin{equation}
 \int \rmd u_1 u_1 P_{w_1 \leq 0}(u_1|F(0)=0) = - \frac{2 D}{b^2} - \frac{|w_1|}{|b|}
\end{equation}
If $w_1>0$ there are two subcases, and the total reads:
\begin{eqnarray}
&&\!\!\!\!\!\!\! \!\! \!\! P_{w_1 \geq 0}(u_1|F(0)=0) \nn \\
&=& \theta(-u_1) \mbox{LT}^{-1}_{p \to -u_1} G_{b=- |b|}(w_1,p;0) \Pi_{\mathrm{last}}(p)  \nn \\
&& + \theta(u_1) \mbox{LT}^{-1}_{p \to u_1} G_{b=|b|}(w_1,p;0) \nn\\
&& \qquad \times (1-G_{b=- |b|}(w_1,p=0;0)) 
\end{eqnarray}
This gives:
\begin{eqnarray}\nn
&& \!\!\!\!\!\!\! \!\! \!\! P_{w_1 \geq 0}(u_1|F(0)=0) \\
& =& \theta(-u_1) \frac{|b|}{\sqrt{4 \pi D |u_1|}}
e^{-\frac{(- |b| |u_1|-w_1)^2}{4 D |u_1|}} \nn \\
&& + \theta(u_1) \frac{w_1}{\sqrt{4 \pi D}}
u_1^{-3/2} e^{-\frac{ (|b| u_1- w_1 )^2}{4 D u_1}} (1-e^{-\frac{|b|}{D} w_1})\ ,\qquad
\end{eqnarray}
which can be checked to be correctly normalized. One gets:
\begin{align}
& \int_{-\infty}^0 \rmd u_1 u_1 P_{w_1 \geq 0}(u_1|F(0)=0) = -
\left(\frac{2 D}{b^2} + \frac{w_1}{|b|}\right) e^{- \frac{|b|}{D} w_1} \nn \\
& \int^{+\infty}_0\rmd u_1 u_1 P_{w_1 \geq 0}(u_1|F(0)=0) = \frac{w_1}{|b|} 
\left(1 - e^{- \frac{|b|}{D} w_1}\right)
\nonumber
\end{align}
Hence we find, setting $|b|=1$ and $D=\sigma/m^4$, conditioned to $F(0)=0$:
\begin{align}
& m^2 \overline{[w_1-u(w_1)]} =  2 \sigma/m^2 , \quad w_1  \leq 0 \\
& m^2 \overline{[w_1-u(w_1)]} =  2 (m^2 w_1 + \sigma/m^2 ) e^{-m^4 w_1/\sigma}, \quad w_1  \geq 0 \nn
\end{align}
For the past ($w_1 \leq 0$) one recovers the previous result, while for the future it decreases to zero at
large $w_1$.

We could try to generalize further by picking a $u_a$ and averaging over $f_a$ with some distribution. Using the
condition $w=u-F(u)/m^2$ one can check that:
\begin{eqnarray}
 P_{w_1}(u_1|F(u_a)=f_a) = P_{w_1-u_a+f_a/m^2}(u_1-u_a|F(0)=0) \nn
\end{eqnarray}
Hence we can use the previous calculation. For fixed $w_1$, $u_a$ and $f_a$ one has:
\begin{align}\nn
& m^2 \int du_1 (w_1 - u_1) P_{w_1}(u_1|F(u_a)=f_a) \\
& = m^2 \int du_1 (w_1 - u_1) P_{w_1-u_a+f_a/m^2}(u_1-u_a|F(0)=0) \nn \\
& = m^2 \int du'_1 (w'_1 - f_a/m^2 - u'_1) P_{w'_1}(u'_1|F(0)=0) \nn \\
&= \left\{
\begin{array}{ll}
2 \sigma/m^2 - f_a , \qquad \quad\mbox{for}\quad w_1-u_a+f_a/m^2 \leq 0 \\
 2 (m^2 w'_1 + \sigma/m^2 ) e^{-m^4 w'_1/\sigma} - f_a \\
 \qquad \qquad\qquad \qquad \mbox{for}\quad w'_1=w_1-u_a+f_a/m^2 \geq 0
\end{array}
\right. \nn \\
\end{align}
which can, in principle, be averaged on $f_a$. It is not clear however at this stage which distribution to choose
and how to relate these quantities to the area of the hysteresis loop. Further work is needed to clarify these
issues.

\section{First return probabilities in $d$ dimensions}
\label{app:G}
In this appendix, we recall standard methods to derive the first return probability in $d$ dimensions, using a regular hypercubic lattice. 

Consider a random walk on the  integers $\mathbb{Z}$. 
The probability to return to the origin after $t$ steps ($t=2 m$ even) is \cite{GrinsteadSnellProba}
\begin{equation}\label{G1}
u_t = \left( {t \atop t/2}\right) 2^{-t}\ . 
\end{equation}
This is seen by noting that giving weight $a$ for a step to the right, and $b=1-a$ for a step to the left,  the probability to be at $x$ after $t$ steps is the same as taking $i$ steps to the right and $t-i$ to the left, with $x=2 i -t$ and equal to $p_x = a^i b^{t-i}  \left( {t \atop i}\right)$, from which the above result is obtained for $a=b=1/2$, $i=t/2$.
We note $u_0=1$. 

The probability to return to the origin for the first time after $t$ steps is noted $f_t$, and we note $f_0=0$. In order to be at the origin at time $t$, we must return for the first time no later than $t$, and can then make a new excursion. Therefore we have (for $t>0$ even)
\begin{equation}\label{G2}
u_t = f_2 u_{t-2} + f_4 u_{t-4} + \ldots + f_t u_0
\end{equation}
Introducing the generating functions
\begin{equation}
u(x):= \sum_{m=0} ^\infty  u_{2 m} x^m \ , \qquad f(x):= \sum_{m=0} ^\infty  f_{2 m} x^m \ ,
\end{equation}
(\ref{G2}) can be written as
\begin{equation}\label{G6}
u(x) =1+ f(x) u(x) \ .
\end{equation}
This gives for the probabilities (\ref{G1})
\begin{equation}
u(x) = \frac{1}{\sqrt{1-x}}\ , \qquad f(x)= 1-\sqrt{1-x}\ ,
\end{equation}
thus the first return probability in $d=1$ at time $t$ is 
\begin{equation}
f_t = \frac{u_t}{t-1}\ . 
\end{equation}
At large times, 
\begin{equation}
u_{2 m} \approx \frac 1{\sqrt{\pi m}}\ , \qquad f_{2 m} \approx \frac1{2 \sqrt {\pi}  m^{3/2}}\ .
\end{equation}
We now want to calculate the same quantities in $d$ dimensions, noted $u_t^d$ and $f_t^d$. $u_t^d$ is simply:
\begin{equation}
u_t^d = (u_t)^d\ .
\end{equation}
As an example, in $d=2$
\begin{equation}
u^{d=2}(x) = \frac{2 K(x)}{\pi }\ , \qquad f^{d=2}(x) = 1-\frac{\pi }{2 K(x)}\ ,
\end{equation}
with $K$ the elliptic $K$ function, and in $d=3$:
\begin{equation}
u(x) = \left[\frac{2 K(\frac{1}{2} (1-\sqrt{1-x}))}{\pi }\right]^2
\end{equation}
In general, an analytic solution for the coefficients is not possible. We therefore give the asymptotic behavior for large $t$, i.e.\ small $1-x$, s.t.\ the series picks up contributions at large times. Noting $s:=- \ln x$, we get for $d<2$
\begin{eqnarray}
u^d(x)& =& \sum_{m=0}^\infty u_{2 m}^d x^m \approx  \int_0^\infty \rmd m \frac1{(\pi m)^{d/2}}\exp(-s m) \nn\\
&=& \pi ^{-d/2} s^{\frac{d}{2}-1} \Gamma
   \left(1-\frac{d}{2}\right)\ .
\end{eqnarray}
Note that for $d\ge 2$, a UV cutoff is needed, which we discuss below. Using (\ref{G6}) yields
\begin{equation}
f^d(x) = 1-\frac1{u^d(x)} \approx 1-\frac{\pi ^{d/2} s^{1-\frac{d}{2}}}{\Gamma
   \left(1-\frac{d}{2}\right)}\ .
\end{equation}
Transforming inverse Laplace gives
\begin{equation}\label{G12}
f_{2 m}^d\approx \frac{2-d}{2}  m^{\frac{d}{2}-2} \pi ^{\frac{d}{2}-1}
   \sin \left(\frac{d \pi }{2}\right)\ \mbox{\ \ for\ \ }d< 2\ .
\end{equation}
Let us now consider $d>2$. Then 
\begin{eqnarray}
 u^d(x) &\approx& u^d(1) +  \pi ^{-d/2}   s^{\frac{d}{2}-1} \Gamma
   \left(1-\frac{d}{2}\right) \\
  u^d(1) &\approx& 1+ \pi ^{-d/2} \zeta \left(\frac{d}{2}\right) \quad , \quad d < 2 \\
    u^{d=3}(1) &=& \frac{\pi}{\Gamma(3/4)^4}= 1.3932 \ldots
    \end{eqnarray}
where the approximation for $u^d(1)$ valid for $d$ near 2 was obtained summing using $u_{2 m}^d\approx (\pi m)^{-d/2}$ (it gives $1.469$ for $d=3$). Then 
\begin{eqnarray}
 f^d(x) &=& 1-\frac1{u^d(x)} \nn\\
 &\approx& 1 - \frac1{u^d(1)} +  \frac1{u^d(1)^2} \pi ^{-d/2}   s^{\frac{d}{2}-1} \Gamma
   \left(1-\frac{d}{2}\right)\nn\\
&&\qquad +O\left(s^{d-2}\right)\ .
\end{eqnarray}
Note that $1-1/u^d(1)$ is the probability that the walk never returns (which equals $0.2822\ldots$ in $d=3$). Conditioned to returning, the probability that the first return occurs at  $t=2m$ steps decays at large-$t$ as:
\begin{equation}\label{G15}
u^d(1) f_{t=2 m}^d \approx \frac1{u^d(1)} \pi ^{-d/2}  m^{-d/2} 
\mbox{\ \ for\ \ }d>2\ .
\end{equation}
We see that at $d=2$ the exponent for the first-return probability changes direction, and that the amplitudes of both (\ref{G12}) and (\ref{G15}) go to 0.


\begin{thebibliography}{100}
\bibitem{foot2} STS implies that $\hat A[w]$ and $\hat B[w]$
are invariant under the change $w_{xt} \to w_{xt} + \phi_x$,
and same for $A$,$B$,$A^0$,$B^0$ functionals.
\bibitem{footnoteII}
Distributions with unbounded tails $p(f) \sim A \alpha (\alpha+1)
f^{2+\alpha}$ at large negative $f$ correspond to $\alpha \to -
\alpha$, i.e.\ $\zeta=2+2/\alpha$, $f_c^0= 0$, $m^2
\rho_m=(A/m^2)^{1/\alpha}$ (and $c=\Gamma(1 - \frac{1}{\alpha})$).
However they are dominated by rare events with no finite continuum
limit for cumulants of order larger than $\alpha$.
\end{thebibliography}

\tableofcontents

\end{document}